\documentclass[aps,superscriptaddress,nofootinbib,notitlepage,twocolumn]{revtex4-1}

\usepackage[latin1]{inputenc}
\usepackage{graphicx}
\newcommand\aap{{\rm A\&A}}%     % Astronomy and Astrophysics 
\newcommand\mnras{{\rm MNRAS}}%   % Monthly Notices of the RAS 
\usepackage{float}
\usepackage{latexsym}
\usepackage{graphicx}
\usepackage{amssymb}
\usepackage{amsmath}
\usepackage{amsfonts}
\usepackage{hyperref}
\usepackage{cool}
\usepackage{dcolumn}
\usepackage{textcomp}
\usepackage{color}
\usepackage{xfrac}
\usepackage{slashed}
\usepackage{multirow}
\usepackage{comment}

\def\be{\begin{equation}}
\def\ee{\end{equation}}
\def\figs/B{B}
\def\bea{\begin{eqnarray}}
\def\eea{\end{eqnarray}}
\def\bg{\begin{eqnarray}}
\def\nd{\end{eqnarray}}

\def\cos{{\rm cos}}

\def\log{{\rm log}}
\def\ln{{\rm log}}

\newcommand\jcap{{\rm JCAP}}%  % Journal of Cosmology and Astroparticle Physics

\newcommand{\Mpl}{M_{\rm pl}}
\newcommand{\eq}{{\rm eq}}

\def\be{\begin{equation}}
\def\ee{\end{equation}}

\definecolor{orange}{rgb}{1,0.5,0}

\begin{document}

\title{The Early Dark Sector, the Hubble Tension, and the Swampland}

\begin{abstract}
We consider the interplay of the Early Dark Energy (EDE) model, the Swampland Distance Conjecture (SDC), and cosmological parameter tensions.  EDE is a proposed resolution of the Hubble tension relying upon a near-Planckian scalar field excursion, while the SDC predicts an exponential sensitivity of masses of other fields to such an excursion, $m \propto e^{- c |\Delta \phi|/\Mpl}$ with $c \sim {\cal O}(1)$. Meanwhile, EDE is in tension with large-scale structure (LSS) data, due to shifts in the standard $\Lambda$CDM parameters necessary to fit the cosmic microwave background (CMB). One might hope that a proper treatment of the model, e.g., accounting for the SDC, may ameliorate the tension with LSS. Motivated by these considerations, we introduce the Early Dark Sector (EDS) model, wherein the mass of dark matter is exponentially sensitive to super-Planckian field excursions of the EDE scalar. The EDS model exhibits new phenomenology in both the early and late universe, the latter due to an EDE-mediated dark matter self-interaction, which manifests as an enhanced gravitational constant on small scales.  This EDE-induced dark matter-philic ``fifth force'', while constrained to be small, remains active in the late universe and is not screened in virialized halos. We find that the new interaction with dark matter partially resolves the LSS tension. However, the marginalized posteriors are nonetheless consistent with $f_{\rm EDE}=0$ at 95\% CL once the Dark Energy Survey Year 3 measurement of $S_8$ is included.  We additionally study constraints on the model from Atacama Cosmology Telescope data, and find a factor of two improvement on the error bar on the SDC parameter $c$, along with an increased preference for the EDE component. We discuss the implications of these constraints for the SDC, and find the tightest observational constraints to date on a swampland parameter, suggesting that an EDE description of cosmological data is in tension with the SDC. 
\end{abstract}

\author{Evan McDonough}
\affiliation{Kavli Institute for Cosmological Physics and Enrico Fermi Institute, The University of Chicago, Chicago, IL 60637, USA}
\affiliation{Department of Physics, University of Winnipeg, Winnipeg, MB R3B 2E9 Canada}

\author{Meng-Xiang Lin}
\affiliation{Kavli Institute for Cosmological Physics and Enrico Fermi Institute, The University of Chicago, Chicago, IL 60637, USA}

\author{J.~Colin Hill}
\affiliation{Department of Physics, Columbia University, New York, NY, USA 10027}
\affiliation{Center for Computational Astrophysics, Flatiron Institute, New York, NY, USA 10010}

\author{Wayne Hu}
\affiliation{Kavli Institute for Cosmological Physics and Enrico Fermi Institute, The University of Chicago, Chicago, IL 60637, USA}

\author{Shengjia Zhou}

\affiliation{Department of Physics, Columbia University, New York, NY, USA 10027}

%\date{\today}

%\pacs{}
\maketitle

\tableofcontents

%%%%%%%%%%%%%%%%%%%%%%%%%%%%%%%%%%%%%%
%%%%%%%%%%%%%%%%%%%%%%%%%%%%%%%%%%%%%%
\section{Introduction}
\label{sec:intro}
%%%%%%%%%%%%%%%%%%%%%%%%%%%%%%%%%%%%%%
%%%%%%%%%%%%%%%%%%%%%%%%%%%%%%%%%%%%%%

The Early Dark Energy (EDE) model \cite{Poulin:2018cxd} is a prominent candidate to resolve the Hubble tension \cite{Verde:2019ivm}. However, this model faces challenges both from data, in the form of exacerbated tensions with large-scale structure observations \cite{Hill:2020osr,Ivanov:2020ril,DAmico:2020ods}, and from theory, namely whether the model can be self-consistently described as a low-energy limit of a high-energy theory including gravity. To understand the interplay of these challenges, in this work we take guidance from the {\it Swampland Distance Conjecture} \cite{Ooguri:2006in} (and its extension to axions \cite{Baume:2016psm,Klaewer:2016kiy,Blumenhagen:2017cxt,Scalisi:2018eaz}), and extend EDE to an Early Dark {\it Sector}.

The Hubble tension, namely, the discrepancy in the value of the Hubble constant $H_0$ measured locally via the cosmic distance ladder using Type Ia supernovae (SNIa) \cite{Riess:2019cxk,Riess:2020fzl} and the value inferred from the cosmic microwave background (CMB) \cite{Aghanim:2018eyx}, from large-scale structure (LSS) \cite{2016ApJ...830..148C,Aubourg:2014yra,Cuceu:2019for,Schoneberg:2019wmt,Abbott:2017smn,Philcox:2020vvt}, and from other probes \cite{Verde:2019ivm}, presents a challenge to the standard $\Lambda$CDM cosmological model. In particular, the disagreement between ${\it Planck}$ 2018 CMB observations and the SH0ES 2020 cosmic distance ladder measurement  stands at $5.0\sigma$ statistical significance \cite{Riess:2021jrx}, with the two values given by $H_0 = 67.37 \pm 0.54$ km/s/Mpc  \cite{Aghanim:2018eyx} and $H_0 = 73.04 \pm 1.04 \, {\rm km/s/Mpc}$~\cite{Riess:2021jrx}, respectively.  While some local measurements have yielded $H_0$ values that are not in statistical disagreement with the $\Lambda$CDM-predicted value from CMB and LSS data (e.g.,~\cite{Freedman:2019jwv,Birrer:2020tax}), it is generally true that local $H_0$ probes have yielded higher values than expected in $\Lambda$CDM.

A plethora of cosmological models have been proposed to bring these data sets into concordance, and resolve the Hubble tension. For a recent review see, e.g., \cite{Shah:2021onj}. These range from modifications to the early (pre-recombination) universe, to the late universe, and to the theory of gravity in the local universe. However, all approaches face severe challenges: For example, late universe models that leave the sound horizon at the drag epoch unchanged are heavily constrained by the inverse cosmic distance ladder and generally cannot explain the SH0ES measurement \cite{Efstathiou:2021ocp}. Early universe models that reduce the the sound horizon at recombination can successfully raise the Hubble constant while maintaining consistency with CMB observations, but are often in tension with LSS observations, namely the galaxy clustering and cosmic shear auto- and cross-correlation two-point functions from the Dark Energy Survey Year 1 \cite{Hill:2020osr} and BOSS full-shape anisotropic galaxy clustering \cite{ Ivanov:2020ril}; see also \cite{Jedamzik:2020zmd} and \cite{Lin:2021sfs}. Nonetheless, the relative success of early universe models at raising the inferred $H_0$ motivates the search for an embedding into a more complete and yet still well motivated model that is consistent with all data sets. Several recent models have been proposed along these lines, e.g., \cite{Clark:2021hlo,Allali:2021azp,Karwal:2021vpk}.

An interesting case study is Early Dark Energy \cite{Poulin:2018cxd}. In this class of models, the expansion rate is increased near matter-radiation equality, so as to reduce the sound horizon at recombination, and thereby raise the $H_0$ value inferred from the angular scale of the sound horizon. The model can accommodate larger values of $H_0$ than $\Lambda$CDM whilst not degrading the fit to the CMB, and is thereby compatible with both SH0ES and {\it Planck}. However, the larger $H_0$ is accompanied by shifts in other $\Lambda$CDM parameters, such as the dark matter density $\Omega_c h^2$, the scalar spectral index $n_s$, and the amplitude of density perturbations $\sigma_8$. This brings the model into tension with LSS data \cite{Hill:2020osr,Ivanov:2020ril,DAmico:2020ods}. Accordingly, when additional LSS data are included in the analysis, e.g., from the Dark Energy Survey, the Kilo-Degree Survey (KiDS) \cite{Hildebrandt:2016iqg,2020A&A...633A..69H}, and the Subaru Hyper Suprime-Cam (HSC) survey \cite{Hikage:2018qbn}, or from BOSS \cite{Ivanov:2020ril}, the evidence for an EDE component is significantly diminished~\cite{Hill:2020osr,Ivanov:2020ril,DAmico:2020ods} (see~\cite{Smith:2020rxx} for an alternative viewpoint).

The minimal EDE model is comprised of a scalar field $\phi$ with potential $V(\phi) = V_0 \left[ 1 - \cos(\phi/f)\right]^n$. This potential, first proposed in \cite{Kamionkowski:2014zda}, is a generalization of the usual axion potential (see \cite{Marsh:2015xka} for a review). In this model, the relative energy density in $\phi$ is peaked at a critical redshift $z_c$, at which point the scalar field constitutes a fraction $f_{\rm EDE} \equiv \rho_\phi (z_c)/\rho_{tot} (z_c)$ of the energy density of the universe. The parameters of the model follow from simple considerations: $n\geq2$ so as to have the EDE field's energy density rapidly redshift away following $z_c$, $V_0 ^{1/4} \sim {\rm eV}$ so as to constitute $\approx 10\%$ of the universe at $z_{\rm eq}$, and $f\lesssim M_{pl}$ so as to endow the scalar with a mass $m \sim H(z_{\rm eq})$ and thereby set $z_c \sim z_{\rm eq}$.

This model is, at best, a phenomenological description of a more complicated theory. The conventional origin of periodic axion potentials is instantons. A complete model would need to explain why a tower of instantons $V(\phi) \sim \sum_n c_n e^{-S_n}\cos(n \phi/f)$, with $S_n$ the instanton action,
conspires to take the required form, despite the Planckian decay constant $f\sim M_{\rm pl}$, which would conventionally be associated with a total breakdown of the instanton expansion (see, e.g., \cite{Banks:2003sx,Rudelius:2014wla,Stout:2020uaf}). One might presuppose that the model exists as a low-energy limit of a UV-complete theory, e.g., that EDE is in the landscape of string theory \cite{Bousso:2000xa,Susskind:2003kw}, and that the low-energy parameter fine-tunings are sensible from the UV perspective. However, it might equally well be the case that the EDE model is in the swampland \cite{Vafa:2005ui,Ooguri:2006in}. So-called ``swampland conjectures'' (for a review, see \cite{Palti:2019pca,Brennan:2017rbf,vanBeest:2021lhn}) attempt to delineate the boundaries of the landscape, and identify those properties that low-energy theories inherit from the high-energy theory. In particular, the Swampland Distance Conjecture \cite{Ooguri:2006in} holds that any Planckian field excursion $|\Delta \phi| \sim M_{\rm pl}$, such as that in EDE, causes an exponential suppression of the mass of other fields in the theory, $m \propto e^{- c |\Delta \phi|/M_{\rm pl}}$, with $c>0$ a number of $\mathcal{O}(1)$.

In this work we study the interplay of the swampland and the EDE model. We consider the impact of the Swampland Distance Conjecture (SDC) \cite{Ooguri:2006in} (and its extension to axions \cite{Baume:2016psm,Klaewer:2016kiy,Blumenhagen:2017cxt,Scalisi:2018eaz}) on the EDE inference of $H_0$ and on the tension of EDE and LSS data \cite{Hill:2020osr,Ivanov:2020ril,DAmico:2020ods}. To this end, we promote EDE to an {\it Early Dark Sector} (EDS). We consider an EDE-dependence of the mass of dark matter, given by,
\be
m_{\rm DM}(\phi) = m_0 e^{ c \phi / M_{pl}},
\ee
where $\phi$ is initially $\phi_i \in [0, \pi f]$, and is zero in the present universe. We assume for simplicity that the above applies to all of the dark matter (as also considered in, e.g., the ``Fading Dark Matter'' model \cite{Agrawal:2019dlm,Anchordoqui:2019amx}). The SDC prediction is that $c$ is positive and order-1, such that the dark matter is exponentially lightened when $\phi$ rolls from $\phi_i \sim \Mpl$ to $\phi \sim 0$. We perform data analysis allowing $c$ to vary, and allow the data to decide both the magnitude and sign of $c$.

We find that positive $c$ ($c>0$), which is the sign of $c$ predicted by the SDC, raises $S_8$ and exacerbates the tension with LSS data in this model. On the other hand, we find that a small but negative $c$ can lower $S_8$ without decreasing $H_0$, while simultaneously improving the fit to the CMB. This occurs due to an interplay of imprints on the cosmic microwave background, both at high-$\ell$ and on scales that enter the horizon around $z_c$, and imprints on the growth of structure, caused by a relative shift in the redshift of matter-radiation equality and by an induced attractive dark matter self-interaction (a dark-matter-philic ``fifth-force'') \footnote{This is related but distinct from the ``cosmic axion force'' \cite{Kim:2021eye}; in that work an ultra-light scalar mediates an interaction with the Standard Model, whereas in the in the EDS model the interaction is confined to the dark sector.}.

We perform a Markov Chain Monte Carlo (MCMC) analysis of a (`baseline') combined data set comprised of {\it Planck} 2018 primary CMB and CMB lensing data \cite{Planck2018likelihood,Aghanim:2018eyx,2018arXiv180706210P}; BAO distances from the SDSS DR7 main galaxy sample~\cite{Ross:2014qpa}, the 6dF galaxy survey~\cite{2011MNRAS.416.3017B}, and SDSS BOSS DR12~\cite{Alam:2016hwk}; the Pantheon supernovae data set \cite{Scolnic:2017caz}; and the SH0ES $H_0$ measurement. We find a modest overall preference for $c<0$, with the best-fit value $c=-5 \times 10^{-3}$.

We find that the EDS model is able to accommodate a lower $S_8$ than in EDE, and thereby lessen the tension with LSS data. To substantiate this, we supplement our baseline data set with Dark Energy Survey Year-3 data (DES-Y3) \cite{DES:2021wwk}, approximated as a prior on $S_8\equiv \sigma_8 (\Omega_{\rm m} / 0.3)^{0.5}$, and we repeat the MCMC analysis. We find that the best-fit EDS is better able to accommodate the DES-Y3 measurement than is EDE, with a relative reduction in $\chi^2 _{\rm DES-Y3}$ of $1.1$. However, like previous analyses \cite{Hill:2020osr,Ivanov:2020ril,DAmico:2020ods}, we find that the combined data set including DES-Y3 is statistically consistent with $f_{\rm EDE}=0$, indicating that there is little Bayesian justification for this 4-parameter extension of $\Lambda$CDM.

Finally, we study the impact of recent CMB temperature and polarization data from the Atacama Cosmology Telescope (ACT) \cite{ACT:2020gnv, ACT:2020frw}.  The ACT DR4 data significantly improve upon the precision of {\it Planck} on small angular scales. The ACT collaboration analysis of the EDE model \cite{Hill:2021yec} found a moderate preference for $f_{\rm EDE}>0$, in contrast to results from \emph{Planck}. 
We perform an MCMC analysis of the EDS model fit to the baseline data set supplemented with ACT DR4 temperature and polarization spectra. Analogous to the EDE analysis of \cite{Hill:2021yec}, we find that the inclusion of ACT data increases the preference for $f_{\rm EDE}>0$, and significantly constrains the timing $z_c$. We find a factor of two improvement on the constraint on $c$ relative to the baseline data set.

Turning these analyses on their head, we may ask what the data, when analyzed in the context of the EDE model, have to say about the Swampland Distance Conjecture. We find a 95\% CL upper limit on $c$ given by $c<0.068$ for the baseline data set, and $c<0.035$ and $c<0.042$ at 95\% CL when DES-Y3 or ACT are included, respectively.  We interpret this as a modest tension between the Swampland Distance Conjecture and the EDE model, at the level of a $4-7\%$ fine-tuning.

The structure of this paper is as follows.  In Sec.~\ref{sec:EDS} we introduce the Early Dark Sector model, the dynamics, and the physics behind it. In Sec.~\ref{sec:pheno} we detail the imprint on the cosmic microwave background and on large-scale structure. In Sec.~\ref{sec:constraints} we discuss the data sets that will be used in our analyses, and perform MCMC analyses of the model fit to varying data set combinations. We detail the implications of this for the Swampland Distance Conjecture in  Sec.~\ref{sec:swamp}, and conclude in Sec.~\ref{sec:discussion}.

We work in natural units, where the speed of light is unity. The parameter $c$ refers exclusively to the coupling parameter of the EDS model, and not the speed of light.  We denote 
the reduced Planck mass $M_{\rm pl} (=2.435 \times 10^{18}$ GeV). Unless otherwise stated, values for $H_0$ are given in units of km/s/Mpc.

%%%%%%%%%%%%%%%%%%%%%%%%%%%%%%%%%%%%%%
%%%%%%%%%%%%%%%%%%%%%%%%%%%%%%%%%%%%%%
\section{From Early Dark Energy to The Early Dark Sector}
\label{sec:EDS}
%%%%%%%%%%%%%%%%%%%%%%%%%%%%%%%%%%%%%%
%%%%%%%%%%%%%%%%%%%%%%%%%%%%%%%%%%%%%%

The idea underlying the EDE model \cite{Poulin:2018cxd} is to shrink the comoving sound horizon at last scattering, $r_s$, defined by
\begin{equation}
\label{eq:rs}
   r_s(z_*) = \int _{z_*} ^\infty \frac{{\rm d} z}{H(z)} c_s(z) ,
\end{equation}
with $z_*$ the redshift of last scattering and $c_s$ the sound speed of the photon-baryon plasma, through the inclusion of an additional source of energy density, namely the EDE. The reduced sound horizon allows an increased $H_0$ while remaining consistent with CMB observations of the {\it angular scale} of the sound horizon, $\theta_s$, defined by,
\begin{equation}
\label{eq:thetas}
    \theta_s = \frac{r_s (z_*)}{D_A(z_*)},
\end{equation}
where $D_A$ is the angular diameter distance to last scattering. By adjusting the redshift dependence of the EDE component, the CMB damping scale can simultaneously be adjusted to match observations, albeit at the expense of introducing a tuning or coincidence into the EDE model.

The baseline EDE model \cite{Poulin:2018cxd} is  described by a canonical scalar field, with potential energy given by
\be
\label{eq:EDE_V}
V(\phi) = m^2 f^2 \left[ 1- \cos \frac{\phi}{f}\right]^3.
\ee
This potential, of the form first proposed in \cite{Kamionkowski:2014zda}, is a generalization of the usual axion potential, corresponding to a careful fine-tuning of an instanton expansion or of other non-perturbative effects (see, e.g.,  the discussion in \cite{Hill:2020osr}).  Alternative realizations and variations on the EDE model abound, see, e.g.,\ \cite{Poulin:2018cxd,Smith:2019ihp,Agrawal:2019lmo,Alexander:2019rsc,Lin:2019qug,Sakstein:2019fmf,Niedermann:2019olb,Niedermann:2020dwg,Kaloper:2019lpl,Berghaus:2019cls}. 

The common feature of these models is that the energy density transitions between redshifting slower than ordinary matter to redshifting faster across a critical redshift. 
In the baseline EDE model this is achieved as follows. At early times the scalar is frozen in place by Hubble friction, and effectively behaves at dark energy. The scalar is released from Hubble friction when  $H\simeq m$,  for 
a typical value of the initial field 
$\phi_i = \mathcal{O}(f)$. Around this time,
the scalar field makes its maximal contribution to the energy density of the universe, i.e., the ratio of energy densities
\begin{equation}
    f_{\rm EDE}(z)  \equiv \frac{\rho_{\rm EDE}(z)}{ \rho_{\rm tot}(z)},
\end{equation}
where $\rho_{\rm tot}$ is the total energy density, is maximal when $z=z_c$. As a shorthand, we will denote $f_{\rm EDE} \equiv f_{\rm EDE}(z_c)$, and will explicitly specify $f_{\rm EDE}(z)$ when referring to the above. At times after $z_c$, i.e., at lower redshifts, the field rolls down the potential $V(\phi)$ and undergoes damped oscillations. The energy density of the scalar rapidly redshifts away, naively leaving no trace in the post-recombination  universe.

One can easily estimate the model parameters necessary to resolve the Hubble tension. 
The sound horizon and damping scale are most sensitive to dynamics that occur in the decade of redshift preceding last scattering \cite{Knox:2019rjx}.
This effectively imposes $z_c \sim z_\eq$, which in turn determines the mass parameter $m$
as
\begin{equation}
    m \sim 10^{-27} {\rm eV} .
\end{equation}
Meanwhile, the discrepancy in the Hubble constant $H_0$ is roughly $10\%$, which, combined with $\phi_i = \mathcal{O}(f)$ by standard arguments (see, e.g., \cite{Marsh:2015xka}), implies that
\begin{equation}
    V(z\sim z_{c}) \sim 0.1\, H_{\eq}^2 \Mpl^2 ,
\end{equation}
and hence,
\begin{equation}
    f \sim \Mpl .
\end{equation}
Thus we see the EDE scalar field, insofar as it is relevant to the Hubble tension, naturally undergoes a field excursion $|\Delta \phi| \sim f \sim M_{pl}$.

Little is known about field theories near the Planck scale. At these scales one can reasonably expect quantum gravity effects, e.g., from string theory, to become relevant. When assessing models, in lieu of a concrete string theory construction, one approach is to take guidance from known calculable string theory examples, distilled into a simple set of conjectures -- so-called ``Swampland'' conjectures \cite{Vafa:2005ui} (for a review, see \cite{Palti:2019pca,Brennan:2017rbf,vanBeest:2021lhn}). The Swampland conjectures collectively aim to delineate the boundary between effective field theories that are inconsistent once gravity is quantized (or more precisely, EFTs that do not admit a UV completion into quantum gravity \cite{Palti:2019pca}), and those that are consistent with quantum gravity (and hence do admit UV completion). 

Of particular relevance to EDE is the SDC~\cite{Ooguri:2006in}. The SDC holds that any low-energy effective field theory is only valid in a region of field space bounded by the Planck scale, and the breakdown of effective field theory that occurs at Planckian field excursions is encoded in an exponential sensitivity of the mass spectrum of the effective theory. This can be expressed as, for the mass of at least one such field in the spectrum,
\begin{equation}
\label{eq:m_general}
    M \sim M_0 e^{- \alpha |\Delta|/\Mpl} ,
\end{equation}
where $\Delta$ is the distance traversed in field space, and $\alpha$ is an order-1 parameter. There are numerous concrete examples that support the SDC. For example, consider a universe with an extra dimension that is a circle of radius $R$. Dimensional reduction on the circle yields a tower of massive Kaluza-Klein excitations, with masses given by
\begin{equation}
    m_n ^2 \simeq  n^2 \Mpl^2 e^{- 2 \varphi/\Mpl },
\end{equation}
where $\varphi \equiv \Mpl\log( \Mpl R)$ is the canonically normalized radius of the circle.  At large field values $\varphi \gtrsim \Mpl$, the Kaluza-Klein fields become exponentially light and a priori cannot be neglected. For other examples of the scaling in Eq.~\eqref{eq:m_general}, see, e.g., the review in~\cite{vanBeest:2021lhn}.

The EDE scenario is precisely the sort of model that the SDC is designed to address, namely a model with Planckian field excursions. While this is not unique to EDE, and is exhibited also in late-universe dark energy models, such as quintessence \cite{Tsujikawa:2013fta}, the EDE model is unique in that this exponential sensitivity is activated in the high-redshift universe. Thus one might hope that cosmological observables such as the CMB and LSS may be powerful probes of the couplings predicted by the SDC, e.g., of the form in Eq.~\eqref{eq:m_general}, in the EDE model.

\begin{figure*}
    \centering
    \includegraphics[width=0.32\textwidth]{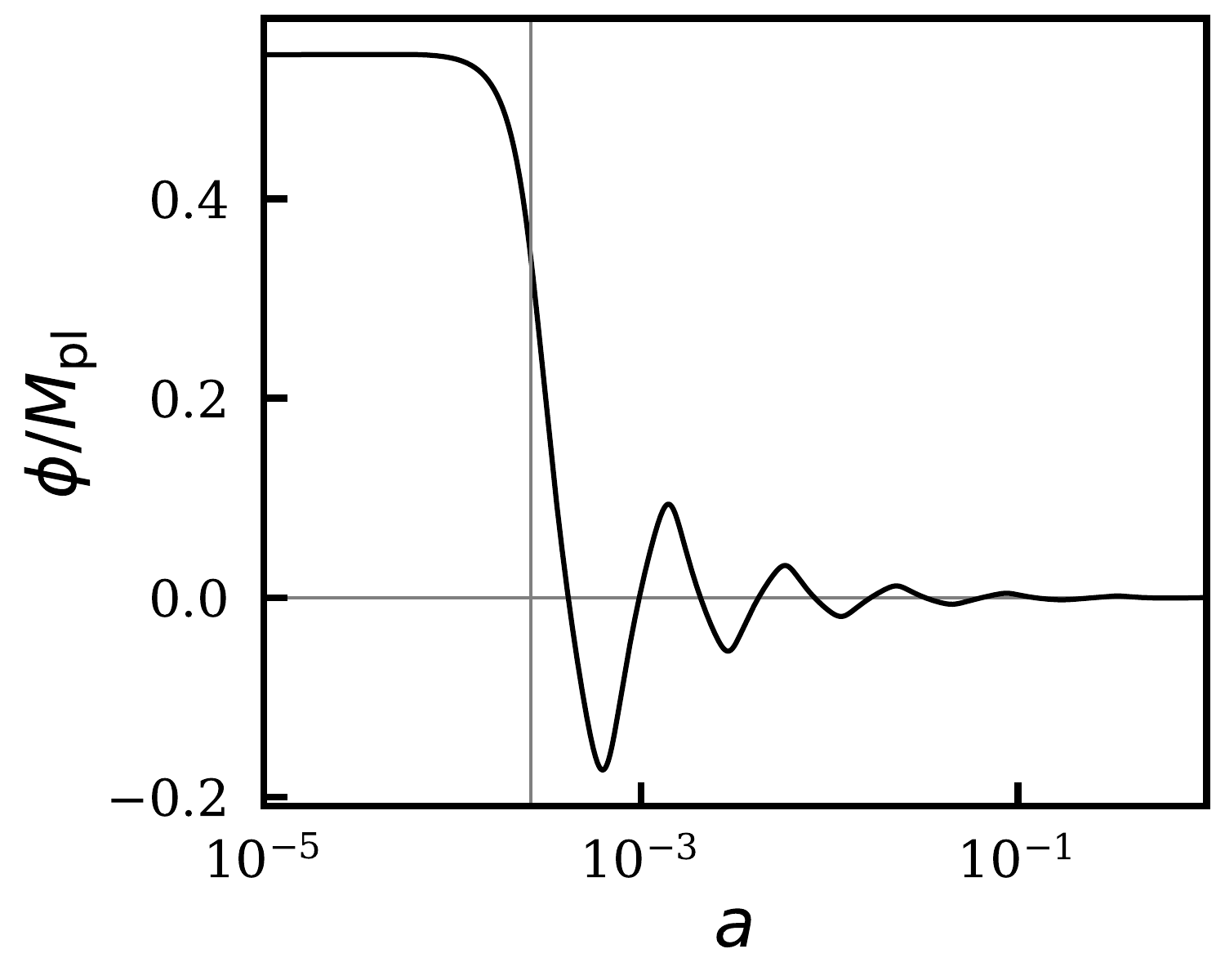}
    \includegraphics[width=0.32\textwidth]{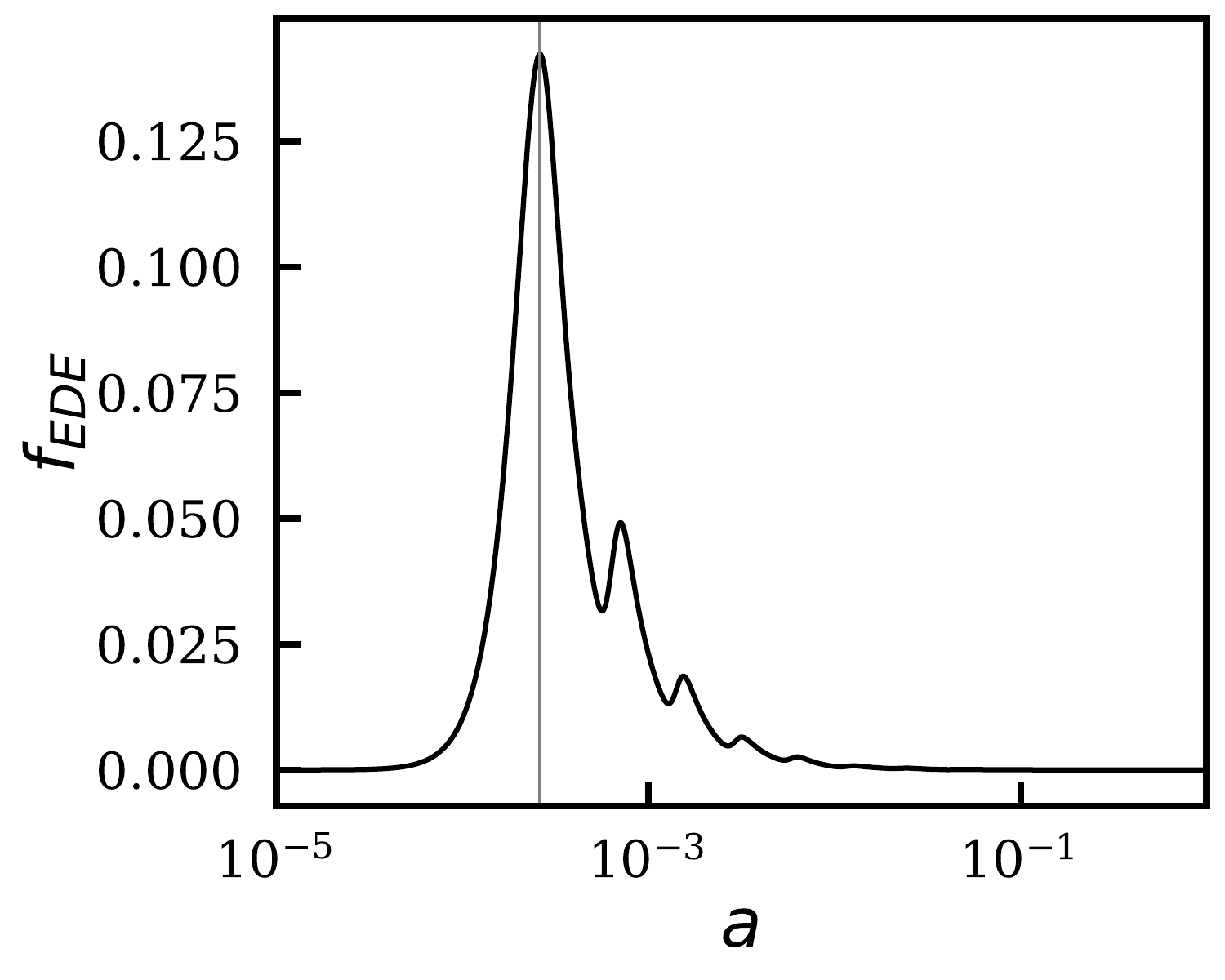}
    \includegraphics[width=0.32\textwidth]{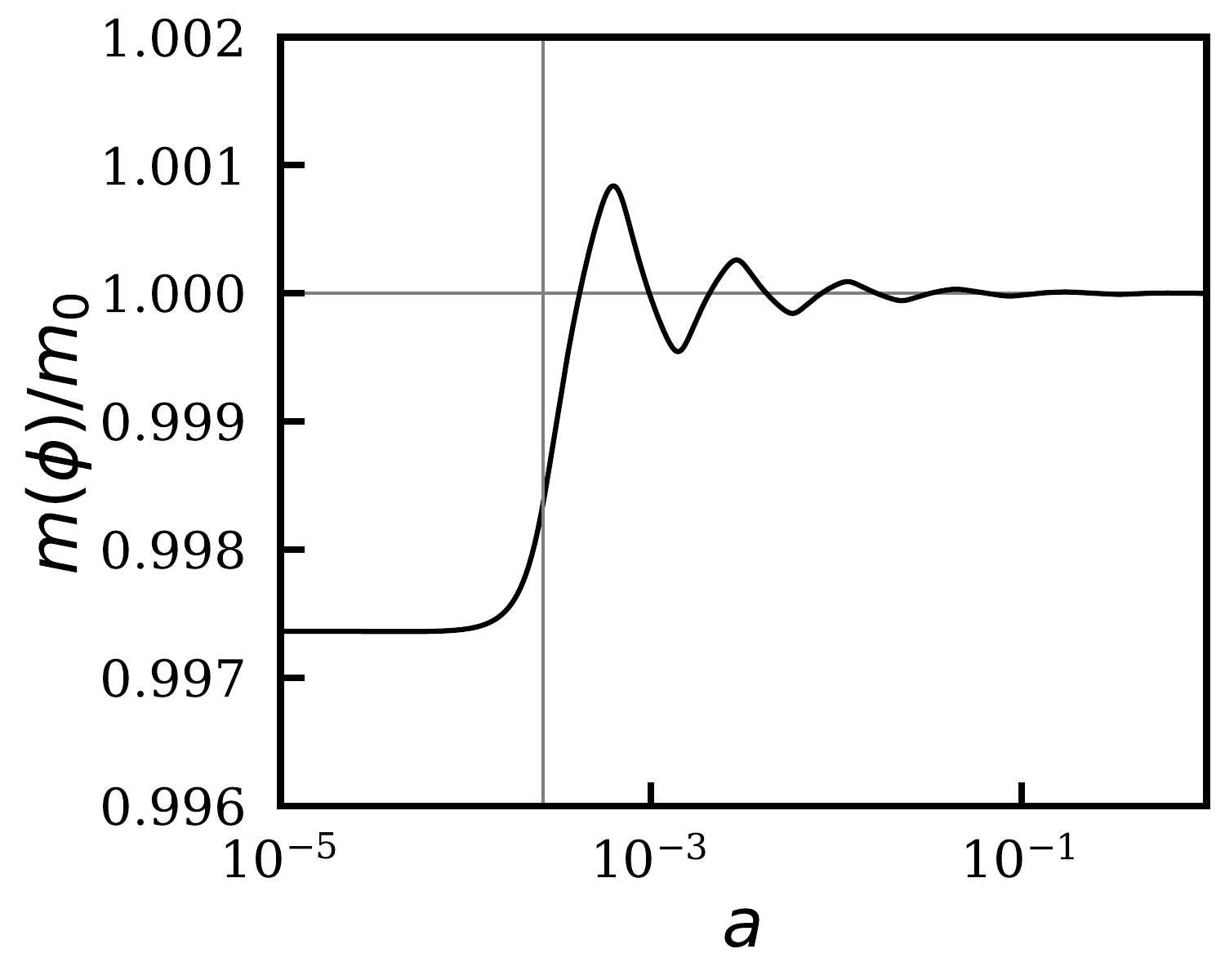}
    \caption{  Fiducial example background evolution of the scalar field, the energy density fraction $f_{\rm EDE}$, and the dark matter mass $m_{\rm DM}({\phi})$. The vertical lines indicate the location of $z_c$. The scalar field indeed undergoes a Planckian field excursion (up to an order-1 factor), leading to a $\approx 0.3\%$ change to $m_{\rm DM}$ around $z_c$.  See Eq.~\eqref{eq:bestfitparams} for parameters.}
    \label{fig:EDSexample}
\end{figure*}

With all this in mind, in this work we consider a simple model that implements these ideas. We extend the EDE model to the Early Dark {\it Sector} (EDS), and consider a coupling of the EDE field to dark matter of the form predicted by the SDC. While fields that exhibit the mass scaling in Eq.~\eqref{eq:m_general} could in principle be an arbitrary fraction of the total dark matter, for simplicity we assume $\phi$ couples to all dark matter. As a concrete model, we consider the following Lagrangian:
\begin{equation}
    \mathcal{L} = \frac{1}{2} (\partial \phi)^2+ i\bar{\psi} \slashed{D}\psi  - V(\phi) - m_{\rm DM}(\phi)\bar{\psi}\psi ,
    \label{eq:Lmodel}
\end{equation}
where $\phi$ is the EDE scalar with potential $V(\phi)$ and ${\psi}$ is a fermionic cold dark matter candidate with $\phi$-dependent mass $m_{\rm DM}(\phi)$. We consider the specific form of the potential $V(\phi)$ given by Eq.~\eqref{eq:EDE_V}, and a field-dependent mass $m_{\rm DM}(\phi)$ given by
\be
\label{eq:mDMc}
m_{\rm DM}(\phi) = m_0 e^{ c \phi / \Mpl},
\ee
as motivated by the SDC, and in particular the extension of the SDC to axions \cite{Baume:2016psm,Klaewer:2016kiy,Blumenhagen:2017cxt,Scalisi:2018eaz}. In our work we fix the convention that  $\phi$ decreases over the course of cosmic evolution, i.e., $\phi$ evolves from $\phi_i >0$ in the early universe to $\phi_f \sim 0$ in the present universe. The SDC then predicts that $c$ defined by Eq.~\eqref{eq:mDMc} is positive ($c>0$), such that the dark matter mass is decreased by a Planckian field excursion of $\phi$.  In what follows, we refer to the
system  defined by Eqs.~\eqref{eq:Lmodel}, \eqref{eq:mDMc}, and \eqref{eq:EDE_V},  as the EDS model.

The background cosmology of the EDS model Eq.~\eqref{eq:Lmodel} is specified by the Friedmann equations, along with the scalar field equation of motion,
\begin{equation}
\ddot{\phi}+2aH\dot{\phi}+a^2\frac{dV}{d\phi}=-a^2\frac{c}{M_{\rm pl}} \rho_{\rm DM} ,
\ee
where dot denotes a derivative with respect to conformal time and $H=(1/a){\rm d}a/{\rm d}t$ where $t$ is cosmic time, and the conservation equation for  the joint stress-energy of the dark matter and scalar 
field. The latter leads to the modified continuity equation for the dark matter density,
\begin{equation}
\label{eq:rhoDMEOM}
\dot{\rho}_{\rm DM} +3aH\rho_{\rm DM}=\frac{c}{M_{\rm pl}} \dot{\phi}\rho_{\rm DM}.
\end{equation}
A full derivation of the equations of motion at the background and linear-perturbation level is given in App.~\ref{app:EOM}.  We may understand the background cosmology in a relatively straightforward way. On the dark matter side, Eq.~\eqref{eq:rhoDMEOM} may be solved exactly, to give the evolution of the dark matter density at all times. We find
\begin{equation}
\rho_{\rm DM}(a)
= \frac{3 \Mpl^2 H_0^2 \Omega_{\rm DM}}{ a^3} \frac{m_{\rm DM}(\phi)}{m_{\rm DM}(\phi_0)} ,
\label{eq:rhoDM}
\end{equation}
with $m_{\rm DM}(\phi_0)$ the present-day dark matter mass. This is consistent with the conservation of the comoving DM number density, $a^3 n_{\rm DM}(a) =3 \Mpl^2 H_0^2 \Omega_{\rm DM}/m_{\rm DM}(\phi_0) $. Meanwhile, the scalar field may be understood as evolving in a time-dependent effective potential, which can be read off from Eq.~\eqref{eq:phiEOM} as
\begin{equation}
V_{\rm eff} (\phi,a) \equiv V(\phi) + \rho_{\rm DM}(a) ,
\label{eq:Veff}
\end{equation}
where $\rho_{\rm DM}(a)$ is given by Eq.~\eqref{eq:rhoDM}.

As a fiducial numerical example, we consider the best-fit model in the fit to primary CMB, CMB lensing, BAO, SNIa, and SH0ES data, to be presented later in this work (see Tab.~\ref{tab:parameters-basline}). We will refer to this example throughout; the parameters (to be varied in Sec.~\ref{sec:pheno} and sampled in our MCMC analysis) are given by,  for the EDS parameters,
\begin{alignat}{3}
     \label{eq:bestfitparams}
      f_{\rm EDE}&=0.142, \qquad &\log_{10}(z_c)&=3.58 , \\
     \theta_i \equiv \frac{\phi_i}{f}&= 2.72, & c_\theta \equiv  c\cdot \frac{f}{M_{\rm pl}}&=-0.0010, \nonumber
\end{alignat}
where we have defined $c_\theta$ as $c$ in units of $f$, analogous to the rescaling of $\phi$ into $\theta$, and 
\begin{alignat}{3}
\label{eq:bestfitparams-2}
     100 \theta_s &= 1.04114  , \qquad & \Omega_b h^2 &=0.02284, \\
     \Omega_c h^2&=0.13043 ,&\log(10^{10}A_s) &= 3.079, \nonumber \\  
     n_s&=0.9931 ,&\tau &=0.0600 ,\;\;\;  \nonumber 
\end{alignat}
 for the $\Lambda$CDM parameters. The corresponding particle physics parameters are given by
 \begin{alignat}{3}
     c&=-0.0049 , \qquad & \phi_i &= 0.55 \Mpl ,\nonumber\\  
     f &= 0.20 \Mpl , & m&= 5.4 \times 10^{-28}\, {\rm eV},
 \end{alignat}
 implying a change in the dark matter mass,
 \begin{equation}
     \frac{\Delta m_{\rm DM}}{m_{\rm DM}} \equiv \frac{m(\phi_i) - m_0}{m_0} = -0.003.
 \end{equation}
The tension-related derived cosmological parameters are given by
  \begin{alignat}{3}
     \label{eq:bestfitparams-derived}
     H_0&=72.52 \, 
     ,\qquad & S_8& =0.848, \\
     \sigma_8 &=0.848 ,& \Omega_m &= 0.3000, \nonumber 
 \end{alignat}
 which can be compared with the SH0ES 2020 measurement $H_0 = 73.2 \pm 1.3$~\cite{Riess:2020fzl}, and the DES-Y3 measurements \cite{DES:2021wwk} $S_8 = 0.776 \pm 0.017$, $\Omega_m=0.339 ^{+0.032} _{-0.031}$, and $\sigma_8 =  0.733 ^{+0.039} _{-0.049}$.  Note that SH0ES has been included in the data sets that are used in this fit, while DES-Y3 has not.  We will discuss in detail the tension with and interplay between these data sets in Sec.~\ref{sec:constraints}.
 
 The cosmological evolution of the EDE scalar field, the fractional energy density $f_{\rm EDE}(z)$, and the dark matter mass $m_{\rm DM}(\phi)$, for the above parameters, are shown in Fig.~\ref{fig:EDSexample}. The scalar field undergoes an $\mathcal{O}(\Mpl)$ excursion, and near $z=z_c=3801$ comprises 14\% of the energy density of the universe. This energy density is rapidly dissipated as the field rolls down the potential and begins to oscillate, and at $z= 10^3$ its contribution is less than 2\% of the energy density of the universe. The dark matter mass undergoes a fractional change corresponding to a mass that is $0.3\%$ lighter in the early universe than in the late universe.

The equations of motion for linear perturbations of the scalar field and dark matter may be derived following the same procedure as for the background evolution, namely, from the variation of the action with respect to the scalar field perturbations and the conservation of the perturbed joint stress-energy tensor (see App.~\ref{app:EOM}). In the synchronous gauge, we find for the scalar field perturbation, 
\begin{equation}
\ddot{\delta\phi}+ 2aH \dot{\delta\phi}+\left(k^{2} + a^{2}\frac{d^{2}V}{d\phi^{2}}\right)\delta \phi+\frac{\dot{h}}{2}
\dot{\phi}  = -a^2 \frac{c \rho_{\rm DM}}{\Mpl}  \delta_c,
\end{equation} 
and for the dark matter,
\begin{equation}\label{denscont}
\dot{\delta}_c  + \theta + \frac{\dot{h}}{2}=\frac{1}{\Mpl}c \dot{\delta\phi},
\end{equation}
\begin{equation}
\label{eq:theta}
\dot{\theta}  + aH\theta = \frac{1}{\Mpl}c k^{2} \delta \phi-\frac{1}{\Mpl}c\dot{\phi}\theta ,
\end{equation}
where $\theta \equiv \partial_i v^i$ and $h$ is the trace of the spatial metric perturbation.  These results are specific to the choice of SDC-inspired dark matter mass dependence in Eq.~\eqref{eq:mDMc}; the equations of motion for a general $\phi$-dependent dark matter mass $m(\phi)$ are given in App.~\ref{app:EOM}. The phenomenology of perturbations will be discussed in detail in Sec.~\ref{sec:pheno}.

Finally, we note the model we consider here is similar to, but distinct from, the modified gravity implementation of coupled EDE in \cite{Karwal:2021vpk}. While both setups include a field-dependent dark matter mass, here we consider an axion-like sinusoidal $V(\phi)$, Eq.~\eqref{eq:EDE_V}, whereas \cite{Karwal:2021vpk} considered a monomial $V(\phi)=\lambda \phi^4$. These two choices for $V(\phi)$ are known to exhibit different phenomenology; see, e.g., the discussion in \cite{Agrawal:2019lmo, Smith:2019ihp}.

%%%%%%%%%%%%%%%%%%%%%%%%%%%%%%%%%%%%%%
%%%%%%%%%%%%%%%%%%%%%%%%%%%%%%%%%%%%%%
\section{Phenomenology: The CMB and the Growth of Structure}
\label{sec:pheno}

Here we investigate the novel EDS impact of the coupling between the scalar field and dark matter on the CMB and large-scale structure of the Universe.

\subsection{CMB}

\begin{figure}[!htb]
\centering
\includegraphics[width=\columnwidth]{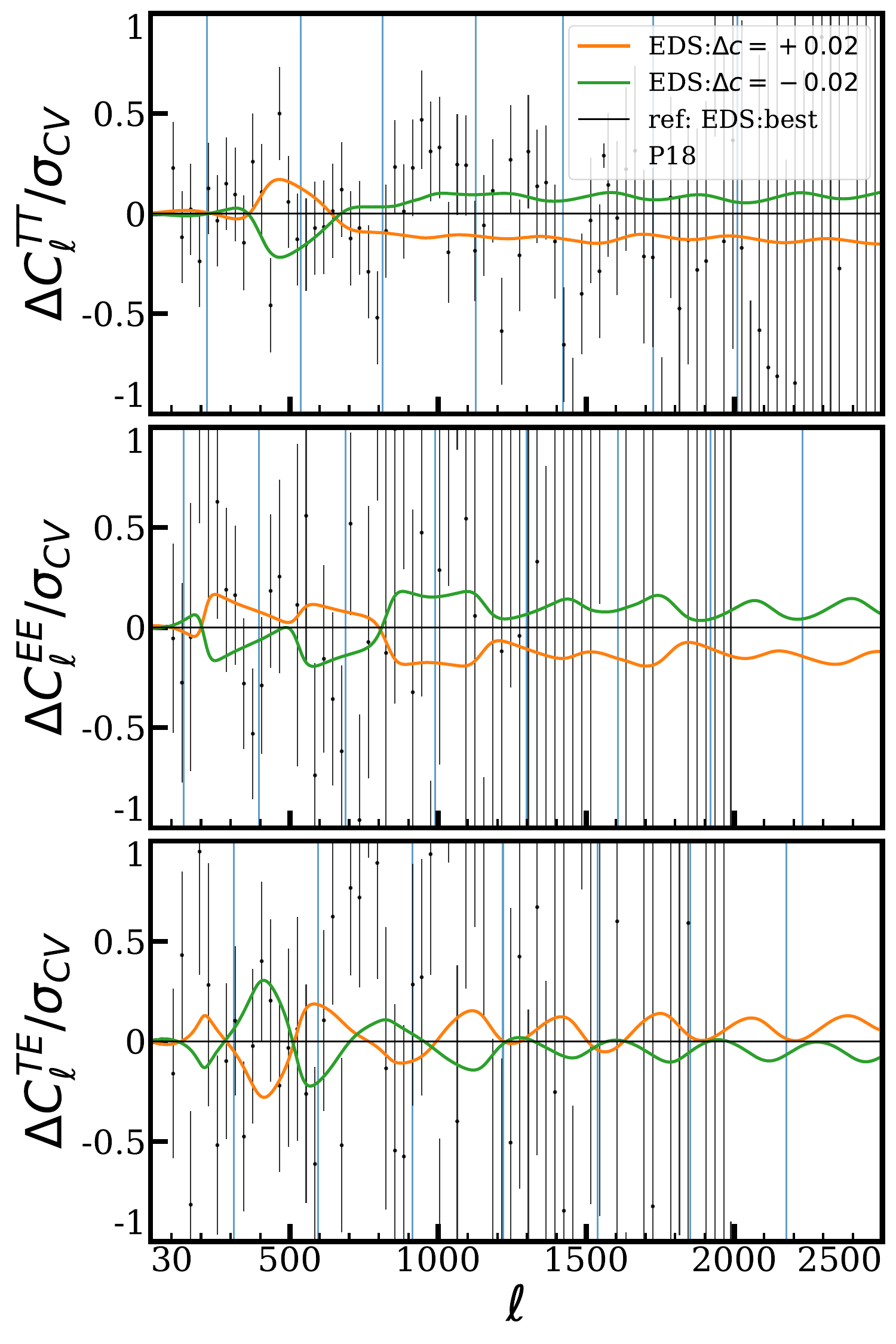}
\caption{\label{fig:fix_thetas_EDS_TTTEEE}
Planck 2018 data residuals relative to the
EDS best-fit model to the baseline data set. Models with $\Delta c=\pm0.02$ around the best-fit $-0.005$ with all other parameters fixed to their values in Eqs.~\eqref{eq:bestfitparams} and \eqref{eq:bestfitparams-2} are shown for comparison. 
The blue vertical lines indicate the positions of the acoustic peaks in the best-fit EDS model. }
\end{figure}

\begin{figure}[!htb]
\centering
\includegraphics[width=\columnwidth]{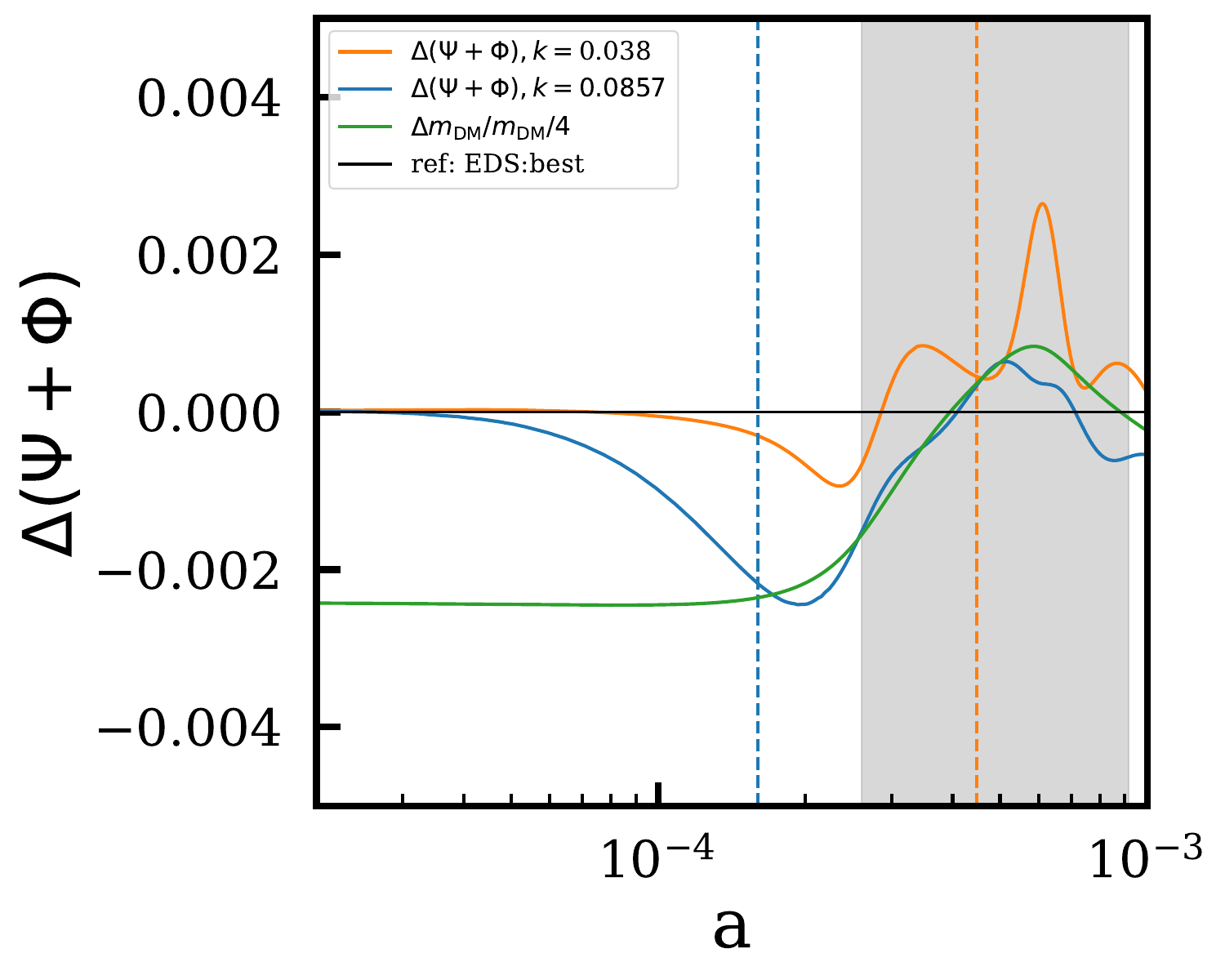}
\caption{\label{fig:EDS_ks_Weyl}
Time evolution of Weyl potential, in units of the initial comoving curvature pertubation, and the dark matter mass for $\Delta c=-0.02$ with respect to the best-fit EDS model. All the other parameters are fixed to their values in Eq.~(\ref{eq:bestfitparams}) and \eqref{eq:bestfitparams-2}. 
The dashed vertical lines indicate locations of $kr_s(a)=\pi$ for each $k$ mode with the same color where $r_s$ is the comoving sound horizon.
The shaded area indicates the epoch between $z_c$ and recombination. 
} 
\end{figure}

\begin{figure}[!htb]
\centering
\includegraphics[width=\columnwidth]{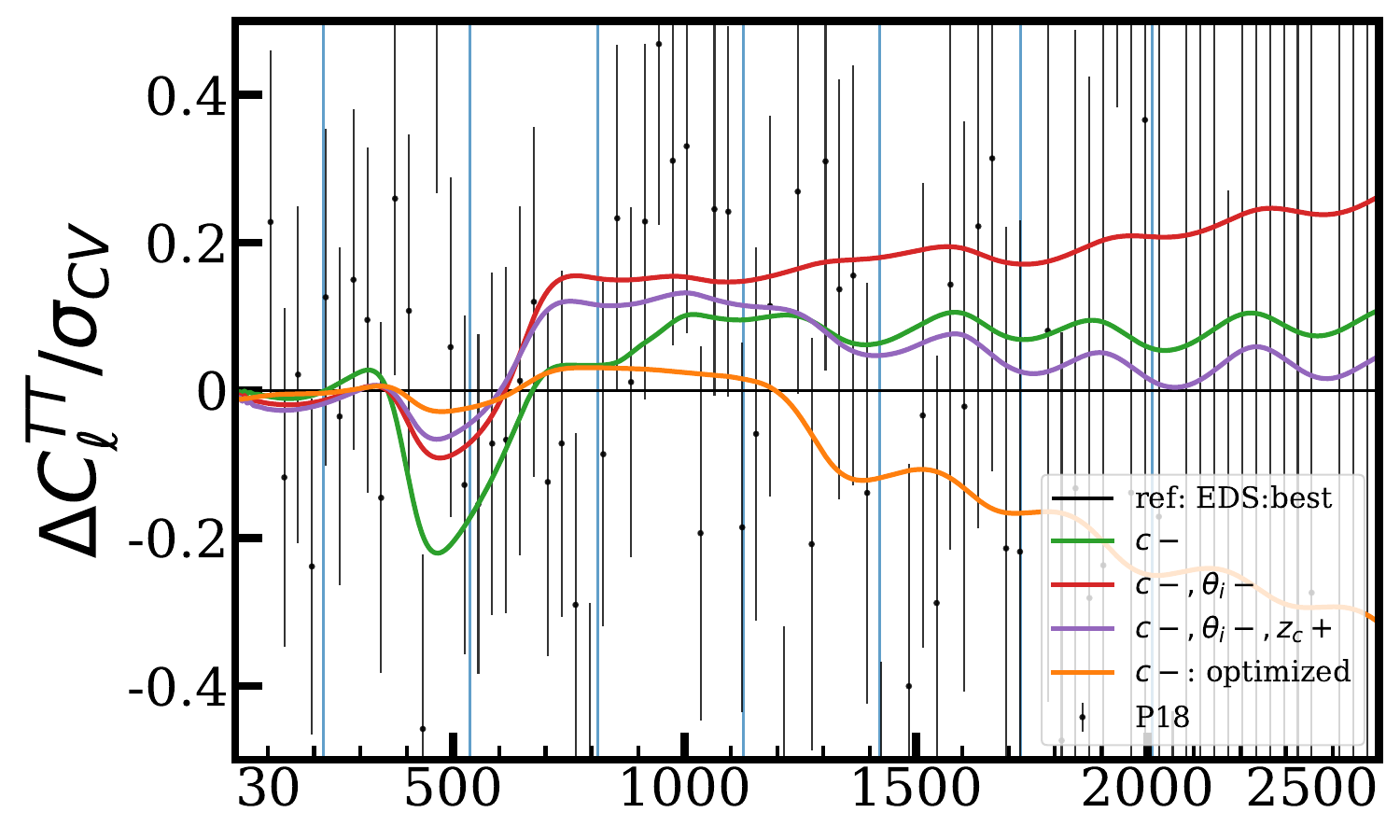}
\caption{\label{fig:EDS_minima_TT}
Comparison  between the Planck $TT$ data and both the global best-fit EDS model where $c=-0.005$ (black line) and the best model for the baseline data set with $c=-0.025$ fixed ($c-$ optimized, orange line). The other curves show the effect of varying the EDS parameters $c$, $\theta_i$, and $z_c$ from  the former to the latter in the direction indicated by  the $+$ and $-$ with the remaining parameters  fixed to the global best-fit model.}
\end{figure}

In Fig.~\ref{fig:fix_thetas_EDS_TTTEEE}, we show the impact of varying $c$ with the other parameters fixed to their values in Eqs.~\eqref{eq:bestfitparams} and \eqref{eq:bestfitparams-2} compared with the Planck $TT$, $EE$, $TE$ data. The various models are plotted as differences with respect to the best-fit model to baseline data set in units of the cosmic variance per multipole,
\begin{equation}
\sigma_{\rm CV} =
\begin{cases}
\sqrt{\frac{2}{2\ell+1}} C_\ell^{TT}, & {TT} \,;\\
\sqrt{\frac{1}{2\ell+1}}  \sqrt{ C_\ell^{TT} C_\ell^{EE} + (C_\ell^{TE})^2}, & {TE} \,;\\
\sqrt{\frac{2}{2\ell+1}}  C_\ell^{EE}, & {EE}, \, \\
\end{cases}
\end{equation}
of the best-fit model. From the 
$\Delta c=\pm 0.02$ parameter variations around the best-fit $c=-0.005$, which is comparable to the scale of its observational errors, we can see that the main effects on the $TT$ power spectrum of decreasing $c$ is a localized decrease in power near $\ell\sim500$ and an increase in power at high multipole moments $\ell \gtrsim 700$.  

These effects are induced by the gravitational effects of  the change of dark matter mass on the CMB acoustic oscillations. These gravitational effects come through the Newtonian gauge Weyl potential $\Psi+\Phi$; see, e.g., \cite{Lin:2018nxe}.   The change in the Weyl potential drives acoustic oscillations, especially around the epoch that the oscillations reach their first extrema $k r_s(z)=\pi$ where $r_s$ is the comoving sound horizon. In Fig.~\ref{fig:EDS_ks_Weyl}, we show the time evolution of the Weyl potential and dark matter mass for the $\Delta c=-0.02$ model with respect to the best-fit EDS model. The Weyl potential is shown in blue and orange curves for $k=0.038$ and $0.0857\,{\rm Mpc^{-1}}$, which correspond to $\ell\sim500$ and $1100$, respectively. The dashed vertical lines indicate locations where $k r_s(z)=\pi$ for each $k$ mode with the same color, and the shaded area indicates the epoch between $z_c$ and recombination.
We see that the Weyl potential change follows the dark matter mass change, which oscillates with time. For a negative $c$, the dark matter mass is smaller before $z_c$ and larger during an epoch between $z_c$ and recombination.
For modes that cross $k r_s=\pi$ well before $z_c$, 
the decrease in the dark matter mass at that time causes a larger relative decay in the Weyl potential and a corresponding increase in the amplitude of the acoustic peaks at the corresponding multipoles $\ell \gtrsim 700$.
On the other hand, for modes that cross right around
$z_c$, the change in the Weyl potential flips sign at the critical phase for driving the acoustic mode, leading to a local decrement in the power around $\ell \sim 500$.

As we can see from  Fig.~\ref{fig:fix_thetas_EDS_TTTEEE}, these effects for variations of $\Delta c =\pm 0.02$  with other parameters fixed are too large to be accommodated by the data and must be compensated by other parameters. This can be done largely within the EDS sector itself, without substantially modifying the other $\Lambda$CDM parameters
of Eq.~(\ref{eq:bestfitparams-2}), especially $\Omega_c h^2$.  We study these compensating effects in Fig.~\ref{fig:EDS_minima_TT}, where we show $TT$ power spectra for both the global best-fit EDS model where $c=-0.005$ (black line) and the best model with  $c=-0.025$ fixed (orange line). We then iteratively perform the parameter shifts from the former to the latter so as to understand the compensations and hence the expected parameter degeneracies in the fit to data. One may appreciate from Fig.~\ref{fig:EDS_minima_TT} that lowering c from $c=-0.005$ to $c=-0.025$ generates a significant dip in $C_{\ell} ^{\rm TT} $ around $\ell\sim500$. This can be compensated by lowering the initial phase $\theta_i$, however this comes at the expense of significant residuals at somewhat higher multipoles.  Next, tuning $z_c$ changes the damping scale, and hence the high-$\ell$ amplitude. Therefore, we expect a $c-\theta_i-z_c$   degeneracy in the fit to data. This expectation is confirmed by MCMC analyses,  e.g., Fig.~\ref{fig:big-triangle-default}, to be presented in Sec.~\ref{sec:constraints}. 

In particular these compensations do not involve the
{\it present} cold dark matter density $\Omega_c h^2$, leaving a range of allowed $c$ at fixed $\Omega_c h^2$.
Indeed in the best-fit model with $c=-0.005$,
$\Omega_c h^2$ remains very close to its best-fit value for EDE (i.e.~$c=0$) but the change in the dark matter mass makes the cold dark matter density at early times smaller.    We shall see next that this delays the onset of the matter-dominated growth of density fluctuations and hence allows a smaller amplitude of structure today.

\subsection{Growth of Structure}
%%%

As we have seen, the CMB allows $c<0$ with a present dark matter density $\Omega_c h^2$ nearly fixed.    
In this context, there are two distinct effects of $c$ on the growth of structure and hence $S_8$
as can be seen in Fig.~\ref{fig:EDS_deltaEDSdm}.  The first is that for $c<0$ the dark matter mass is lighter at $z>z_c$ and the dark matter density smaller.  Therefore the start of the matter-dominated growth of density fluctuations is delayed, which leads to a smaller amplitude of fluctuations today for $c<0$, all else equal.  This can be seen in Fig.~\ref{fig:EDS_deltaEDSdm} as the negative change in density fluctuation right after $z_c$. Note that the behavior before $z_c$ is due to the Weyl potential change induced by the change of the dark matter mass, as we see in Fig.~\ref{fig:EDS_ks_Weyl}. This pre-$z_c$ effect will be suppressed for larger $k$ modes where horizon crossing occurs much earlier.  The second effect is that the $\phi$ field mediates an enhanced gravitational force for the dark matter, which increases the growth of structure for large values of $|c|$.

\begin{figure}[!htb]
\centering
\includegraphics[width=\columnwidth]{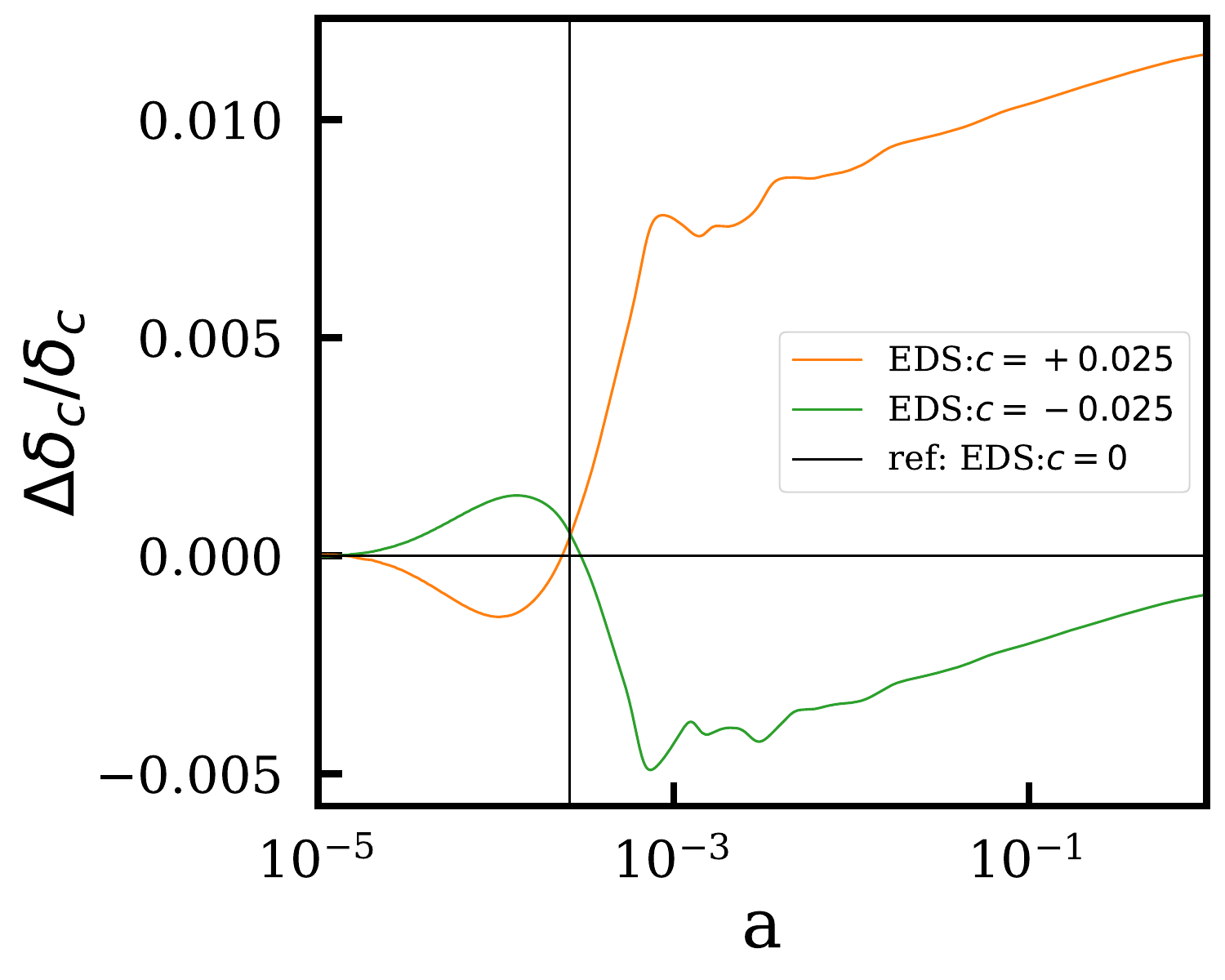}
\caption{\label{fig:EDS_deltaEDSdm} Density growth of EDS best-fit model as $c$ varied, with fixed $H_0$ and all other parameters (except $\theta_s$) fixed to their values in Eqs.~\eqref{eq:bestfitparams} and \eqref{eq:bestfitparams-2}. The vertical line indicates the location of $z_c$. Here $k=0.2 h \,{\rm Mpc^{-1}}$ .
}
\end{figure}

\begin{figure}[!htb]
\centering
\includegraphics[width=\columnwidth]{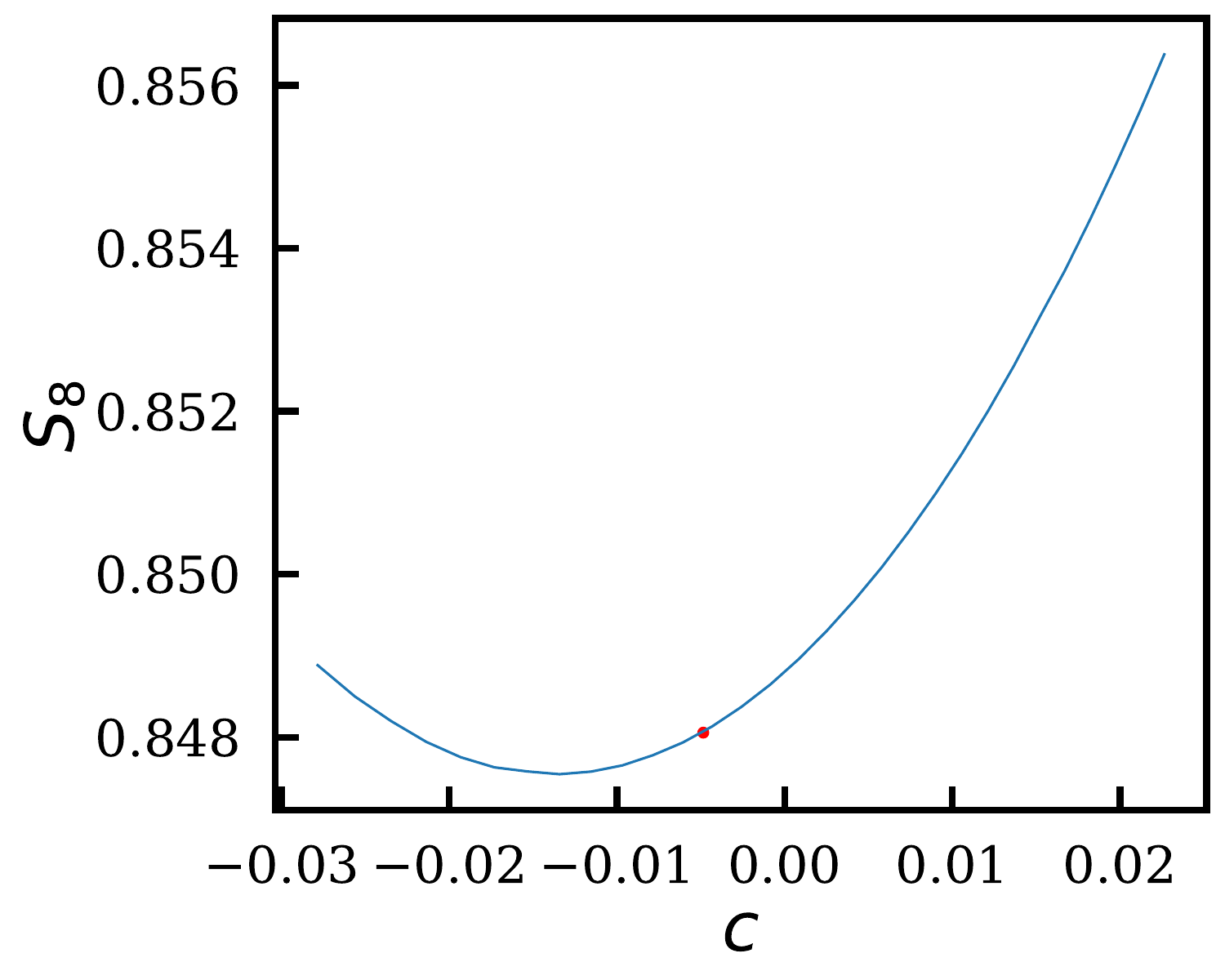}
\caption{\label{fig:S8-c}
$S_8$ value as function of $c$, with fixed $H_0$ and all other parameters (except $\theta_s$) fixed to their values in Eqs.~\eqref{eq:bestfitparams} and \eqref{eq:bestfitparams-2}. The red dot indicates the best-fit model. 
}
\end{figure}

\begin{figure}[!htb]
\centering
\includegraphics[width=\columnwidth]{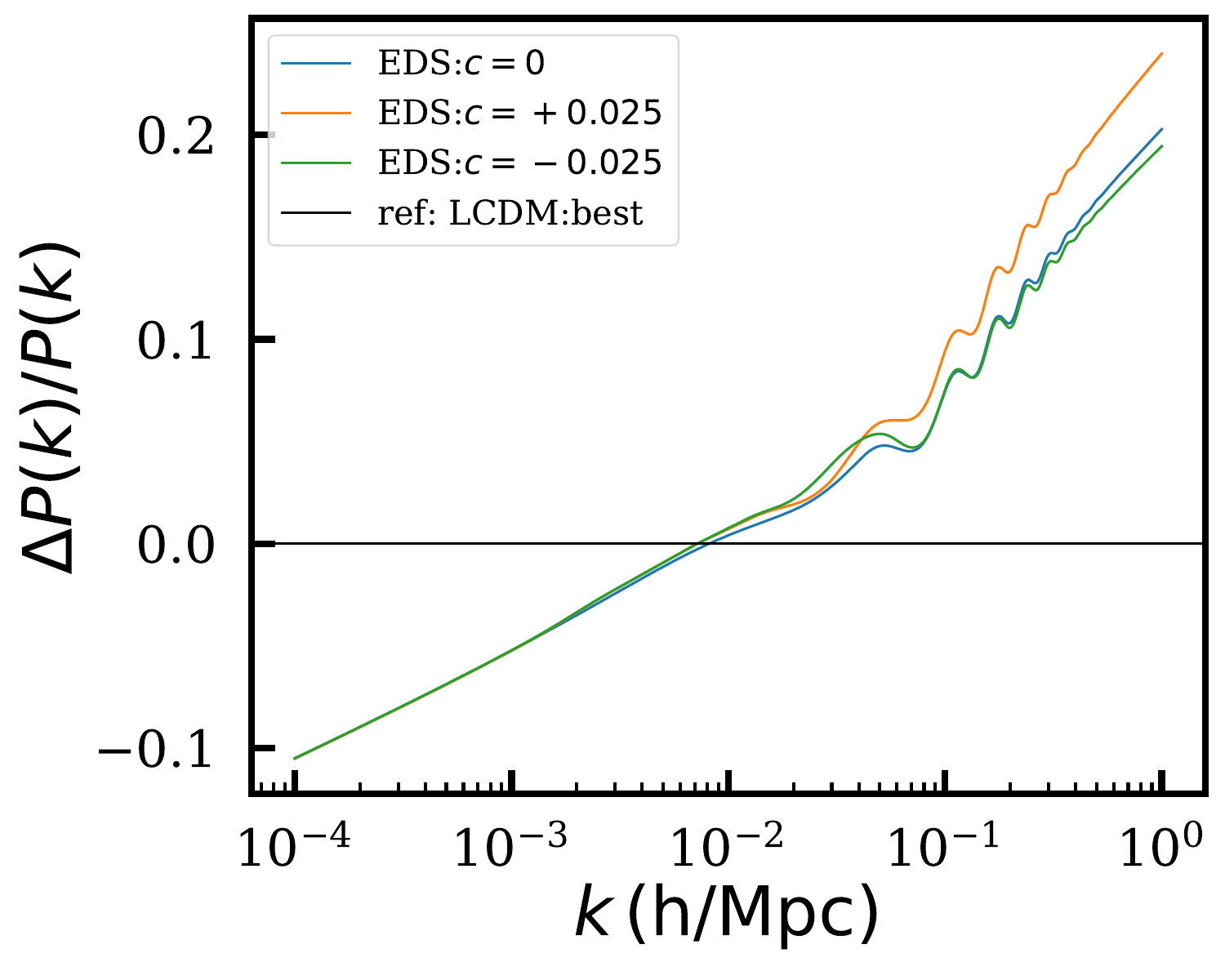}
\caption{\label{fig:Pk_EDS}
Matter power spectra of EDS best-fit model as $c$ varied, with fixed $H_0$ and all other parameters (except $\theta_s$) fixed to their values in Eqs.~\eqref{eq:bestfitparams} and \eqref{eq:bestfitparams-2}. The results are compared to the best-fit $\Lambda$CDM model.
}
\end{figure}

To understand this second effect, in App.~\ref{app:Geff} we derive the equation of the dark matter density perturbation growth at second order in $c$, under a quasistatic approximation for the sourced scalar field perturbations, namely, the assumption that spatial gradients dominate over temporal derivatives for $\delta \phi$. This is a good approximation deep inside the horizon.  In this limit, the impact of $\delta \rho_{\rm DM}$ on $\delta \phi$ takes the form of a non-oscillatory offset $\delta \phi^{(0)}\propto c \delta_c$ (see Eq.~\ref{eq:deltaphi0}). Substituting this back into the equation for $\delta_c$, the resulting effect is an $\mathcal{O}(c^2)$ self-interaction. We find
\begin{equation}
\ddot{\delta}_c +\mathcal{H}\dot{\delta}_c = 4\pi Ga^2\rho_c\delta_c \left(1+\frac{2c^2k^2}{k^2+a^2d^2V/d\phi^2} \right) \,,
\end{equation}
where $\mathcal{H}$ is the Hubble parameter defined with respect to conformal time. From this one may read off an effective gravitational constant,
\begin{equation}
G_{\rm eff} = G_N \left( 1+\frac{2c^2k^2}{k^2+a^2d^2V/d\phi^2} \right),
\end{equation}
which is independent of the sign of $c$. This expression simplifies in the high-$k$ limit, namely, for physical wavenumbers greater than the mass of the EDE scalar field, which satisfy,
\begin{equation}
\label{eq:rangeofk}
\frac{k}{a} \gg m_{\phi} \equiv \sqrt{d^2V/d\phi^2} .
\end{equation}
In this limit, we have,
\begin{equation}
\label{eq:GNeff}
G_{\rm eff} = G_N(1+2c^2),
\end{equation}
which is independent of $k$ and the scalar field potential. This enhanced gravitational constant can understood as a dark matter-philic scalar-mediated force.

The range of $k$-modes which satisfy Eq.~\eqref{eq:rangeofk} changes throughout cosmic history, as the EDE scalar evolves. Before $z_c$, for the parameters in Eq.~\eqref{eq:bestfitparams}, the field mass is $|m_{\phi}| \simeq 3.9 \times 10^{-14 } \, {\rm eV} \simeq 18 \, h/{\rm Mpc}$. After $z_c$, the field is released from Hubble friction and begins to oscillate, and the mass rapidly decreases. After this, modes come to satisfy Eq.~\eqref{eq:rangeofk}. The modes predominantly responsible for setting $S_8$, $k\approx 0.2 \, h$/Mpc, satisfy Eq.~\eqref{eq:rangeofk} shortly after $z_c$, while longer-wavelength modes begin to satisfy Eq.~\eqref{eq:rangeofk}  at later times $t_k$ as $ a(t_k)\sim k^2$. The mass eventually settles to its value at the minimum of the effective potential and quasistatically evolves with $\rho_{\rm DM}$.  We derive in App.~\ref{app:Geff} the scaling of this quasistatic mass with parameters and show that it remains negligible, even with the enhanced local $\rho_{\rm DM}$ of virialized structures. Consequently, even on nonlinear scales today, the scalar mediates an enhanced force on the dark matter.

A direct consequence of the enhanced gravitational constant in Eq.~\eqref{eq:GNeff} is that both positive and negative $c$ will increase the late-time growth of $\delta_c$. This may be appreciated from Fig.~\ref{fig:S8-c}, where we show $S_8$ as $c$ is varied (with $H_0$ held fixed). While $S_8$ may be slightly decreased by a small negative $c$, making $c$ further negative leads to a net increase in $S_8$. This may be understood analytically as follows. In the matter-dominated limit, the enhanced gravitational force on the dark matter, below the Compton scale $k \gg a m_\phi$, changes the growth rate to $\lim_{c\ll 1} \delta_c \propto a^{1+ 6 c^2/5}\simeq a(1 + \log(a) 6c^2/5)$. This determines the fractional change in $\sigma_8$ as $\Delta \sigma_8/\sigma_8 \simeq  \Delta \delta_c (z=0)/ \delta_c \simeq  -\log(a_{\rm eq})6c^2/5 \simeq 9.6 c^2$. This simple estimate captures the
qualitative behavior of $S_8$ in Fig.~\ref{fig:S8-c}; more quantitatively we find $S_8=0.8488(1 + 0.22 c + 7.93 c^2)$.

These effects are encoded in the matter power spectrum by a $c$-dependent enhancement on small scales.  The linear matter power spectrum for varying $c$ is shown in Fig.~\ref{fig:Pk_EDS}, where one may appreciate a net enhancement for both positive and negative $c$. The enhancement is lessened in the negative $c$ case, since the fifth force effect is mitigated by the delayed onset growth effect,  while the opposite occurs for $c>0$.

The imprint on the matter power spectrum is most significant on small scales. This is true for both the imprint of the shift in matter-radiation equality (from the dark matter mass variation), and of the enhanced gravitational interaction. The latter effectively `turns on' as modes come into the quasistatic approximation, and small-scale modes have had the greatest period of time spent under its influence.  In our EDS model, these two competing effects leave only a small ability to lower $S_8$ with $c$. Interestingly though, these two effects are determined by different regions of the scalar field potential: the shift in $z_{\rm eq}$ is determined by the release from Hubble friction of the axion from the hilltop of the cosine potential, while the enhanced gravitational interaction is determined by the scalar field mass in the minimum of the potential. This opens the possibility of modifying the potential in such a way as to reduce the second effect and lower $S_8$ to below its $\Lambda$CDM value, e.g., if $\phi$ becomes heavy in the late universe.  We leave the exploration of this possibility to future work.

%%%%%%%%%%%%%%%%%%%%%%%%%%%%%%%%%%%%%%
%%%%%%%%%%%%%%%%%%%%%%%%%%%%%%%%%%%%%%
\section{Constraints from  Data}
\label{sec:constraints}
%%%%%%%%%%%%%%%%%%%%%%%%%%%%%%%%%%%%%%
%%%%%%%%%%%%%%%%%%%%%%%%%%%%%%%%%%%%%%

In this work we take as our baseline data set the following combination:
\begin{enumerate}
    \item {\bf CMB}: {\it Planck} 2018 \cite{Planck2018likelihood,Aghanim:2018eyx,2018arXiv180706210P} low-$\ell$ and high-$\ell$ [\texttt{Plik}] temperature and polarization power spectra ($TT$/$TE$/$EE$), and reconstructed CMB lensing power spectrum.
    \item {\bf BAO}: distance measurements from the SDSS DR7 main galaxy sample~\cite{Ross:2014qpa}, the 6dF galaxy survey~\cite{2011MNRAS.416.3017B}, and SDSS BOSS DR12~\cite{Alam:2016hwk}, namely, the optimally combined LOWZ and CMASS galaxy samples.
    \item {\bf Supernovae}: The Pantheon supernovae data set \cite{Scolnic:2017caz}, comprised of relative luminosity distances of 1048 SNe Ia in the redshift range $0.01 < z < 2.3$ .
    \item ${\boldsymbol{H_0}}$: The 2019 SH0ES cosmic distance ladder measurement $H_0=74.03 \pm 1.42 \, {\rm km/s/Mpc}$ \cite{Riess:2019cxk}.\footnote{We use the SH0ES 2019 measurement to facilitate comparison with previous work, but note that a more recent SH0ES measurement has recently appeared, with a smaller error bar and slightly lower value ($H_0=73.04 \pm 1.04 \, {\rm km/s/Mpc}$)~\cite{Riess:2021jrx}.}
\end{enumerate}
We supplement the above baseline data set with additional LSS data from the Dark Energy Survey Year-3 (DES-Y3) analysis \cite{DES:2021wwk}:
\begin{enumerate}
    \item[5.] {\bf DES-Y3}: Dark Energy Survey Year-3 \cite{DES:2021wwk} weak lensing and galaxy clustering data, namely, galaxy-galaxy, shear-shear, and galaxy-shear two-point correlation functions, implemented as a Gaussian constraint on $S_8\equiv \sigma_8 (\Omega_{\rm m} / 0.3)^{0.5}$ corresponding to the DES-Y3 measurement $S_8 = 0.776 \pm 0.017$.
\end{enumerate}
The approximation of DES data with an $S_8$ prior procedure was validated with DES-Y1 data in the context of EDE \cite{Hill:2020osr}. In this work, in light of the significant computational expense of evaluating the full  DES 3$\times$2pt likelihood, we assume that an $S_8$ prior continues to be a good approximation in the EDS model with DES-Y3 data. As we will see, the baseline data set combination restricts the EDS model to be a small departure from EDE, and thus one expects the validation test of \cite{Hill:2020osr} to apply,  at least at the level of marginalized 1d and 2d posterior probability distributions.

Finally, we also supplement our baseline data set with CMB data from the Atacama Cosmology Telescope (ACT):
\begin{enumerate}
    \item[6.] {\bf ACT:} The ACT DR4 \cite{ACT:2020gnv, ACT:2020frw} temperature and polarization power spectra.  When combining these data with the \emph{Planck} CMB likelihood, we apply the multipole cut determined in~\cite{ACT:2020gnv} to the ACT data to avoid double-counting information, in particular setting $\ell_{\rm min, TT} = 1800$.
\end{enumerate}
The ACT collaboration analyzed the EDE model in \cite{Hill:2021yec} and found that ACT data combined with low-$\ell$ \emph{Planck} $TT$ data ($\ell < 650$, similar to \emph{WMAP}) mildly prefer a non-zero $f_{\rm EDE}$ at $\approx 3\sigma$ significance (see also \cite{Poulin:2021bjr} and \cite{Lin:2020jcb}).  When combining ACT with the full \emph{Planck} data set, this preference is no longer seen, due to the dominant statistical weight of \emph{Planck} (which does not prefer EDE on its own). In our work we consider ACT in combination with the baseline data set, including {\it Planck} 2018.  We take care in combining ACT and \emph{Planck}, and in particular we apply a multipole cut $\ell_{\rm min, TT} = 1800$ to ACT data to avoid double counting information (following \cite{ACT:2020gnv}). We additionally use increased precision settings in the theoretical computation of CMB power spectra when ACT is included in the joint data set, as emphasized in \cite{ACT:2020gnv,McCarthy:2021lfp}. 

We perform MCMC analyses of the EDS scenario using a modified version of CLASS (\cite{2011arXiv1104.2932L,2011JCAP...07..034B})\footnote{\url{http://class-code.net}} and posterior sampling with \texttt{Cobaya} \cite{torrado_lewis_2019}. We impose broad uniform priors on the $\Lambda$CDM parameters.  Following past work on Early Dark Energy (e.g., \cite{Hill:2020osr}), we impose uniform priors on the EDE parameters $f_{\rm EDE}=[0.001, 0.5]$ and $\log_{10}(z_c)=[3.1,4.3]$, and a uniform prior on the initial field displacement in units of the decay constant $f$, as $\theta_i=[0.1,3.1]$. The choice and impact of EDE priors is discussed in detail in \cite{Hill:2020osr}.  Given that the EDE physics is sensitive primarily to $\theta_i$ (and not $\phi_i$ per se), and given that $\theta_i$ is itself relatively well-constrained by data \cite{Smith:2019ihp,Hill:2020osr}, we express $m_{\rm DM}(\phi)$ as $m_{\rm DM}(\theta)=m_0 e^{c_\theta \theta}$, with $c_{\theta} \equiv c f/M_{pl}$. We impose a uniform prior  $c_\theta=[-0.08,0.08]$. Since $\theta_i$ is fairly well constrained for cases that alleviate the Hubble tension, this allows $c_\theta$ to function as a proxy for $m_{\rm DM}$. 

We follow the Planck convention for the neutrino masses, namely, we hold the sum of the neutrino masses fixed to $0.06 {\rm eV}$ with a single massive neutrino eigenstate. We analyze the MCMC chains using \texttt{GetDist} \cite{GetDist}\footnote{\url{https://github.com/cmbant/getdist}}, and consider chains to be converged when the Gelman-Rubin statistic \cite{Gelman:1992zz} satisfies $R-1 < 0.05$. To determine maximum-likelihood parameter values we use the ``BOBYQA'' likelihood maximization method implemented in \texttt{Cobaya}~\cite{Powell2009,Cartis2018a,Cartis2018b}. When handling ACT data, we use increased \texttt{CLASS} precision settings as discussed in \cite{ACT:2020gnv}, and a slightly relaxed convergence criterion $R-1<0.07$ due to the computational expense of these calculations.  In all EDS runs, we use increased \texttt{CLASS} precision setting \texttt{perturb}$\_$\texttt{sampling}$\_$\texttt{stepsize} = $0.02$.

\subsection{EDS vs EDE: The Interplay of $H_0$ and $S_8$}
\label{sec:baseline}

\begin{table*}
Constraints on the EDS scenario from Planck, BAO, SNIa, and SH0ES. 
\setlength{\tabcolsep}{12pt}
\centering
\begin{tabular}{@{}cccccc@{}}
\toprule
Model                  & $\Lambda$CDM                           & EDS        &EDE            \\
\toprule                                                                                                               
$\boldsymbol{ 100\theta_{\rm s}}$    & 1.04218 ($1.04205\pm 0.00027$)  & 1.04114 ($1.04136\pm 0.00040$) & 1.04091 ($1.04141\pm 0.00036$)  \\
$\boldsymbol{ \Omega_bh^2 }$         & 0.02249  ($0.02252\pm 0.00013$) & 0.02284 ($0.02291\pm 0.00024$) & 0.02286 ($0.02280^{+0.00020}_{-0.00022}$)  \\
$\boldsymbol{ \Omega_ch^2}$          & 0.11840 ($0.11821\pm 0.00085$)  & 0.1343 ($0.1288^{+0.0056}_{-0.0046}$) & 0.1344 ($0.1296\pm 0.0039$)  \\
$\boldsymbol{ \tau}$                 & 0.0594 ($0.0595^{+0.0068}_{-0.0078}$)     & 0.0600 ($0.0570\pm 0.0075$)     & 0.0600 ($0.0578\pm 0.0072$)    \\
$\boldsymbol{ \ln(10^{10}A_s)  }$    & 3.052  ($3.052\pm 0.015$)       & 3.079  ($3.062\pm 0.017$)      & 3.079 ($3.067\pm 0.015$)   \\
$\boldsymbol{ n_s     }  $           & 0.9686 ($0.9691\pm 0.0035$)     & 0.9931 ($0.9847\pm 0.0073$)    & 0.9930 ($0.9865\pm 0.0071$) \\
$\boldsymbol{c_\theta}$           &                                 & $-0.0010$ ($-0.0024^{+0.0091}_{-0.015}$)   &    \\
$\boldsymbol{f_{\rm EDE} }$          &                                 & 0.142  ($0.099^{+0.056}_{-0.041}$)    & 0.142 ($0.104^{+0.034}_{-0.030}$)    \\
$\boldsymbol{ \log_{10}z_c}$         &                                 & 3.58 ($3.602^{+0.071}_{-0.19}$)& 3.58  ($3.606^{+0.037}_{-0.11}$)\\
$\boldsymbol{ \theta_i }$            &                                 & 2.72 ($< 3.14$)                & 2.73  ($2.60^{+0.31}_{+0.022}$)  \\
\colrule
$c$           &                                 & $-0.005$ ($-0.011^{+0.029}_{-0.047}$)   &    \\
$\phi_i\, [\Mpl]$                  &                                 & 0.547 ($0.53^{+0.10}_{-0.15}$)            & 0.549 ($0.48\pm 0.11$) \\
$\mathrm{log}_{10}(f/{\mathrm{eV}})$ &                                 & 26.69 ($26.857^{+0.058}_{-0.37}$)    & 26.69 ($26.652^{+0.080}_{-0.14}$)\\
$\mathrm{log}_{10}(m/{\mathrm{eV}})$ &                                 & $-27.27$  ($-27.04^{+0.30}_{-0.55}$)     & $-27.28$ ($-27.195^{+0.031}_{-0.23}$) \\
%\colrule        
$\Delta m_{\rm DM}/m_{\rm DM}$       &                                 & $-0.003$ ($-0.007\pm 0.021$)        & \\

$\sigma_8$                           & 0.8093 ($0.8087\pm 0.0060$)     & 0.8481 ($0.838^{+0.011}_{-0.013}$)    & 0.8490 ($0.815\pm 0.011$) \\  
$\Omega_m$                           & 0.3047 ($0.3039\pm 0.0050$)     & 0.3000 ($0.3012\pm 0.0056$)    & 0.3003 ($0.3017\pm 0.0051$)  \\  
$S_8$                                & 0.8156 ($0.8140\pm 0.0098$)     & 0.8481 ($0.840\pm 0.014$)      & 0.8495  ($0.838\pm 0.013$) \\  
$H_0$                                & 68.16 ($68.21\pm 0.39$)         & 72.52  ($71.1\pm 1.2$)         & 72.50  ($71.2\pm 1.1$) \\
\hline
$\Delta\chi^2_{\rm tot}$             &  0                              & $-18.1 $                         & $-16.2$  \\
\botrule
\end{tabular}
\caption{ \label{tab:parameters-basline}
Maximum-likelihood (ML) parameters and 68\% CL marginalized constraints for the $\Lambda$CDM, EDS, and EDE models, in the fit to a combined data set comprised of Planck 2018 CMB, CMB lensing, BAO, SNIa, and SH0ES. Parameters in bold are sampled in the MCMC analyses. 
}
\end{table*}

\begin{figure*}%[h!]
    \centering
    \includegraphics[clip, trim=0.25cm 0cm 0.5cm 1cm,width=0.99\textwidth]{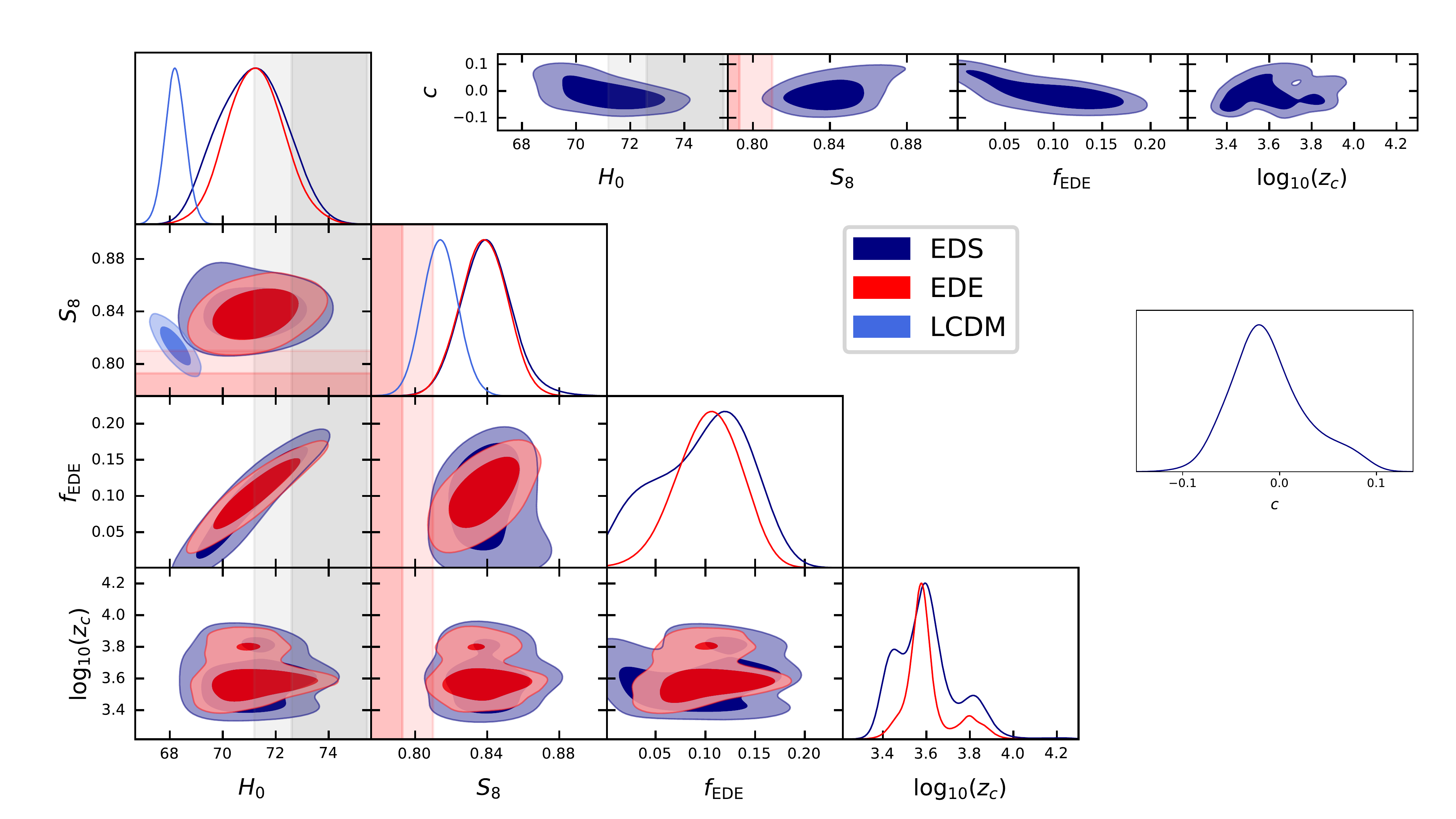}
    \caption{Interplay of the $H_0$ and $S_8$ tensions in the EDS, EDE, and $\Lambda$CDM models (as labeled). The plot shows posterior distributions for the fit to the baseline data set (CMB, CMB lensing, BAO, SNIa, and SH0ES). Shaded grey and pink bands denote the SH0ES measurement and the DES-Y3 $S_8$ constraint, respectively.}
    \label{fig:summary-default}
\end{figure*}

\begin{figure*}%[h!]
    \centering
    \includegraphics[width=0.99\textwidth]{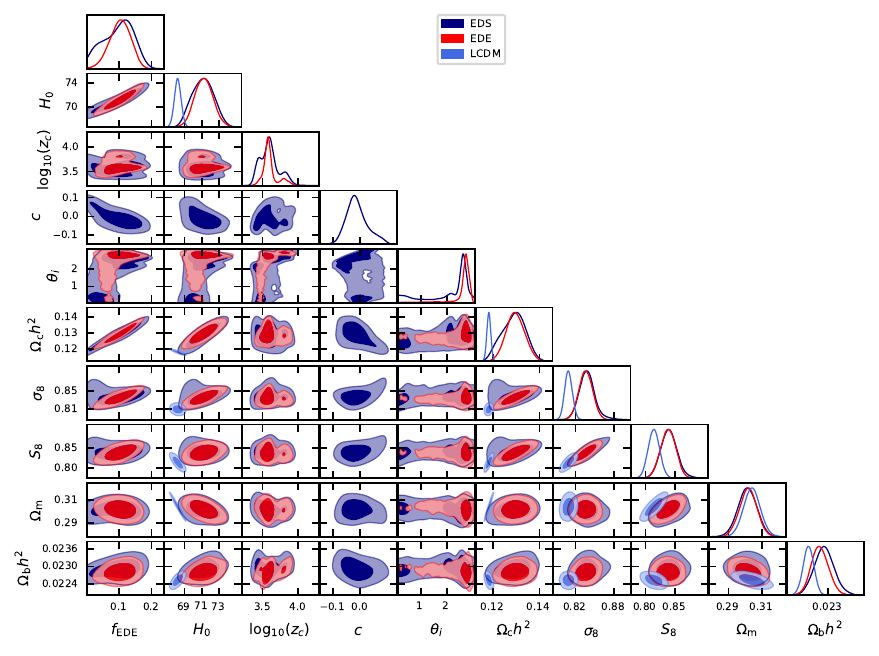}
    %\vspace{-1cm}
    \caption{Enlarged set of posterior distributions for the fit to the baseline data set (CMB, CMB lensing, BAO, SNIa, and SH0ES) for $\Lambda$CDM, EDE, and EDS. }
    \label{fig:big-triangle-default}
\end{figure*}

\begin{table}[h!]%[t]
\centering
EDS $\chi^2$ statistics \\ from the fit to {\it Planck} 2018, BAO, SNIa, SH0ES\\
%\scalebox{0.9}{
  \begin{tabular}{|l|c|c|c|}
    \hline\hline
    Datasets & $\Lambda$CDM  &  EDS  &  EDE  \\ \hline \hline
    Primary CMB: & & &\\
    \;\;\;\; \textit{Planck} 2018 low-$\ell$ TT  & 22.9   & 20.9   & 20.9\\
    \;\;\;\; \textit{Planck} 2018 low-$\ell$ EE  & 397.2  & 397.2  & 397.2\\
    \hspace{.16cm} \; \begin{tabular}[t]{@{}c@{}}\textit{Planck} 2018 high-$\ell$ \\ TT+TE+EE\end{tabular}                        & 2346.5 & 2345.1 & 2346.9 \\

    LSS:& & &\\
        \;\;\;\;\,\textit{Planck} CMB lensing        & 8.9    & 10.0   & 10.0 \\
    \;\;\;\; BAO (6dF)       & 0.00005 & 0.008  & 0.005 \\
    \;\;\;\; BAO (DR7 MGS)   & 1.7     & 2.0    & 2.0 \\
    \;\;\;\; BAO (DR12 BOSS) &  3.4    & 3.4   & 3.5 \\
    SNIa (Pantheon) & 1034.8  & 1034.7 & 1034.7 \\
    SH0ES           & 17.2    & 1.2    & 1.2 \\
    Planck prior    & 1.9     & 2.2    & 2.2 \\
    \hline
    $\Delta \chi^2 _{\rm Primary\,CMB}$   & 0   & $-3.4$   & $-1.6$ \\ 
     $\Delta \chi^2 _{\rm LSS}$   & 0   & $+1.4$   & $+1.5$ \\ 
    $\Delta \chi^2 _{\rm SH0ES}$ & 0   & $-16.0$  & $-16.0$ \\ 
     \hline 
    $\Delta \chi^2 _{\rm tot}$   & 0   & $-18.1$  & $-16.2$ \\ 
    \hline
  \end{tabular}
  %}
 \caption{$\chi^2$ statistics for the ML $\Lambda$CDM, EDS, and EDE models in the fit to the baseline data set (CMB, CMB lensing, BAO, SNIa, and SH0ES).}
  \label{tab:chi2_baseline}
\end{table}

We first perform a direct comparison of the EDE and EDS models fit to the baseline data set, namely, {\it Planck} 2018 primary CMB anisotropies, {\it Planck} 2018 CMB lensing, BAO, Pantheon, and SH0ES. The posteriors are shown in Figs.~\ref{fig:summary-default} and \ref{fig:big-triangle-default}, the best-fit parameters and parameter constraints are given in Tab.~\ref{tab:parameters-basline}, and the $\chi^2$ statistics of the best-fit models are given in Tab.~\ref{tab:chi2_baseline}.

The best-fit EDS and EDE models (Tab.~\ref{tab:parameters-basline}) have near-identical cosmological parameters. They are distinguished primarily by the parameter $c$, which is $c=- 5 \times 10^{-3}$ in EDS, while $c=0$ in EDE by definition. The models have near identical best-fit $H_0$ and $S_8$, with $H_0 = 72.50$ km/s/Mpc and $72.52$ km/s/Mpc, and $S_8= 0.8481$ and $0.8495$, for EDS and EDE respectively. Both models are a significant $\chi^2$ reduction in comparison to the best-fit $\Lambda$CDM, while the EDS model, with $c=-5\times 10^{-3}$, is a slightly better fit to the data than EDE, with a relative $\chi^2$ reduction of $\Delta \chi^2 _{\rm EDS-EDE} = -1.9$. This is driven by the high-$\ell$ CMB data, which in turn drives the mild preference for $c<0$, as discussed in Sec.~\ref{sec:pheno}.

The marginalized posterior distributions, shown in Fig.~\ref{fig:summary-default},  shed more light on the differences between the models.  From the $H_0-S_8$ panel of Fig.~\ref{fig:summary-default}, one may appreciate that the tight $H_0-S_8$ correlation in EDE is softened in EDS, evidenced by an overall flattening of the 1$\sigma$ posterior, and a slight anti-correlation of $H_0$ and $S_8$ in the 95$\%$ contour. Focusing on the SH0ES $1\sigma$ region, indicated by the dark grey band, we see that the EDS model allows a notable reduction in $S_8$ relative to EDE. This suggests that, in the high-$H_0$ context, the EDS model may allow greater compatibility with current LSS data, e.g., from the Dark Energy Survey, than the EDE model.  We return to this point in Sec.~\ref{sec:baseline-S8}.

The ability to raise $H_0$ and simultaneously lower $S_8$ in EDS relative to EDE is obscured in the 1d marginalized posteriors and the marginalized parameter constraints. This occurs due to the low-$H_0$ region of parameter space, $H_0 \lesssim 70$ km/s/Mpc, where the $95\%$ CL contour in EDS extends to significantly larger $S_8$ values than in EDE. The net effect, i.e., after marginalizing, is for the 1d $S_8$ posterior in EDS to be near-identical to that in EDE, differing only in the high-$S_8$ tail.

\begin{table}[h!]%[t]
\centering
EDS $\chi^2$ statistics \\ from the fit to {\it Planck} 2018, BAO, SNIa, SH0ES, and $S_8$ from DES-Y3 \\
  \begin{tabular}{|l|c|c|c|}
    \hline\hline
    Datasets & $\Lambda$CDM  &  EDS  &  EDE  \\ \hline \hline
    Primary CMB: & & &\\
    \;\;\;\; \textit{Planck} 2018 low-$\ell$ TT  & 22.4   & 21.0   & 20.9\\
    \;\;\;\; \textit{Planck} 2018 low-$\ell$ EE  & 396.1  &396.7   & 396.6\\
    \hspace{.16cm} \; \begin{tabular}[t]{@{}c@{}}\textit{Planck} 2018 high-$\ell$ \\ TT+TE+EE\end{tabular}                        & 2349.6 & 2344.7 &  2345.5 \\
    LSS:& & &\\
     \;\;\;\;  \textit{Planck} CMB lensing        & 9.9    &  9.9   & 9.9\\
    \;\;\;\; BAO (6dF)       & 0.011   & 0.085  & 0.078\\
    \;\;\;\; BAO (DR7 MGS)   & 2.1     & 2.7    & 2.6\\
    \;\;\;\; BAO (DR12 BOSS) &  3.4    & 4.0    & 4.0 \\
    \;\;\;\; $S_8$ (DES-Y3)  &  2.5    & 6.5    & 7.6 \\
    SNIa (Pantheon) & 1034.7  & 1034.8 &  1034.8 \\
    SH0ES           & 15.4    & 2.2    & 2.0 \\
    Planck prior    & 1.9     & 1.6    & 2.0 \\
    \hline
    $\Delta \chi^2 _{\rm Primary\, CMB}$   & 0   & -5.7   & -5.1 \\ 
    $\Delta \chi^2 _{\rm LSS}$   & 0   & +5.3   & +6.3\\ 
    $\Delta \chi^2 _{\rm SH0ES}$ & 0   & -13.2  & -13.3 \\ 
    \hline 
    $\Delta \chi^2 _{\rm tot}$   & 0   & -13.9  & -11.9 \\ 
    \hline
  \end{tabular}
  %}
 \caption{$\chi^2$ values for the ML $\Lambda$CDM, EDS, and EDE models in the fit to {\it Planck} primary CMB and CMB lensing, BAO, SNIa, SH0ES, and $S_8$ from DES-Y3.}
  \label{table:chi2_S8}
\end{table}

\begin{table*}[!ht]
Constraints on the EDS scenario from {\it Planck} 2018, BAO, SNIa, SH0ES, and $S_8$ from DES-Y3.
\setlength{\tabcolsep}{12pt}
\centering
\begin{tabular}{@{}cccccc@{}}
\toprule
Model                  & $\Lambda$CDM                           & EDS        &EDE            \\
\toprule                                                                                                               
$\boldsymbol{ 100\theta_{\rm s}}$    & 1.04202 ($1.04208\pm 0.00027$)  & 1.04143 ($1.04151\pm 0.00039$) & 1.04138  \\
$\boldsymbol{ \Omega_bh^2 }$         & 0.02258  ($0.02258\pm 0.00013$) & 0.02273 ($0.02287\pm 0.00022$)  & 0.02281  \\
$\boldsymbol{ \Omega_ch^2}$          & 0.11760 ($0.11754\pm 0.00078$)  & 0.1284 ($0.1247^{+0.0042}_{-0.0047}$) & 0.1287   \\
$\boldsymbol{ \tau}$                 & 0.0535 ($0.0577\pm 0.0071$)     & 0.0583 ($0.0557\pm 0.0074$)     & 0.0581    \\
$\boldsymbol{ \ln(10^{10}A_s)  }$    & 3.041  ($3.046\pm 0.014$)       & 3.063  ($3.051\pm 0.015$)     & 3.065   \\
$\boldsymbol{ n_s     }  $           & 0.9706 ($0.9704\pm 0.0035$)     & 0.9884 ($0.9812\pm 0.0072$)       & 0.9895  \\
$\boldsymbol{c_\theta}$           &                                 & $-0.0034$ ($-0.0044^{+0.0076}_{-0.0097}$)   &    \\
$\boldsymbol{f_{\rm EDE} }$          &                                 & 0.112  ($<0.140$)  & 0.109    \\
$\boldsymbol{ \log_{10}z_c}$         &                                 & 3.57 ({$>3.39$})    & 3.56  \\
$\boldsymbol{ \theta_i }$            &                                 & 2.69 ( {$<2.84$})               & 2.77  \\
\colrule
$c$           &                                 & $-0.020$ ($-0.020^{+0.025}_{-0.032}$)   &    \\
$\phi_i\, [\Mpl]$                  &                                 & $0.461$ ($0.46\pm 0.12$)           & 0.463 \\
$\mathrm{log}_{10}(f/{\mathrm{eV}})$ &                                 & 26.62 ($26.835^{+0.057}_{-0.43}$)    & 26.61 \\
$\mathrm{log}_{10}(m/{\mathrm{eV}})$ &                                 & $-27.29$  ($-26.90^{+0.21}_{-0.63}$)     & $-27.31$ \\
%\colrule        
$\Delta m_{\rm DM}/m_{\rm DM}$       &                                 & $-0.0009$ ($-0.0095\pm 0.014$)       &  \\

$\sigma_8$                           & 0.8024 ($0.8044\pm 0.0054$)     & 0.8287 ($0.8206\pm 0.0096$)      & 0.8320  \\  
$\Omega_m$                           & 0.3004 ($0.2999\pm 0.0046$)     & 0.2931 ($0.2961\pm 0.0052$)   & 0.2934   \\  
$S_8$                                & 0.8028 ($0.8043\pm 0.0084$)     & 0.8192 ($0.815\pm 0.010$)     & 0.8228  \\  
$H_0$                                & 68.47 ($68.51\pm 0.36$)         & 71.96  ($70.7\pm 1.2$)        & 72.02   \\
\hline
$\Delta\chi^2_{\rm tot}$             &  0                              & $-13.9$                         & $-11.9 $ \\
\botrule
\end{tabular}
\caption{ \label{tab:parameters-S8}
ML parameters and marginalized parameter constraints for $\Lambda$CDM and EDS in the fit to a combined data set comprised of {\it Planck} 2018 primary CMB and CMB lensing, BAO, SNIa, SH0ES, and $S_8$ data from DES-Y3. Parameters in bold are sampled in the MCMC analyses. For EDE we present the ML parameters, but not marginalized parameter constraints, as we do not repeat the MCMC for EDE (see~\cite{Hill:2020osr} for analysis of a similar data set combination in EDE). Upper and lower bounds are quoted at 95\% CL.}
\end{table*}

These two corners of parameter space, i.e., high-$H_0$-low-$S_8$ and low-$H_0$-high-$S_8$, correlate with the EDS parameter $c$. This can be appreciated from the $c-H_0$ and $c-S_8$ panels in Fig.~\ref{fig:summary-default},  where one may see that high-$H_0$-low-$S_8$ correlates with $c<0$, while low-$H_0$-high-$S_8$ correlates with $c>0$.  This suggests that additional $S_8$ data would prefer $c<0$; we return to this in Sec.~\ref{sec:baseline-S8}. There is an additional effect at $c<0$, which amplifies the overall preference of the baseline data set for $c<0$: the negative $c$ region includes a weak multimodality in ${\rm log}_{10}(z_c)$, and in particular at $\log_{10}(z_c) \simeq 3.8$ the 1$\sigma$ contour is contained completely within $c<0$.  These effects combine to give an overall mild asymmetry in the posterior, weighted toward $c<0$, and we find $c=-0.011^{+0.029}_{-0.047}$.

\subsection{Impact of Dark Energy Survey data}
\label{sec:baseline-S8}

We now supplement the baseline data set with DES-Y3 data \cite{DES:2021wwk}, approximated by a Gaussian constraint on $S_8\equiv \sigma_8 (\Omega_{\rm m} / 0.3)^{0.5}$ corresponding to the DES-Y3 measurement $S_8 = 0.776 \pm 0.017 $. To contextualize these results, we perform the same analysis for $\Lambda$CDM. We do not repeat the baseline+DES-Y3 analysis for EDE, in light of computational expense and given that the role of $S_8$ data in EDE was studied in detail in  \cite{Hill:2020osr}. 

The best-fit parameters and parameter constraints are given in Tab.~\ref{tab:parameters-S8} and the $\chi^2$ statistics are given in Tab.~\ref{table:chi2_S8}. Consistent with expectations from the fit to the baseline data set, Sec.~\ref{sec:baseline}, we find that when DES-Y3 is included the best-fit EDS has a lower $S_8$ than EDE, whilst having a near-identical value of $H_0$. We find $S_8=0.8192$ and $S_8=0.8228$ in EDS and EDE respectively, corresponding to a $\Delta \chi^2 _{\rm DES-Y3} = -1.1$ between the two models.  Meanwhile the $H_0$ values are respectively $71.96$ km/s/Mpc and $72.02$ km/s/Mpc for the two models, corresponding to $\Delta \chi^2 _{\rm SH0ES} = +0.1$. Comparing the total $\chi^2$, Tab.~\ref{table:chi2_S8}, we find that the best-fit EDS is an improvement over EDE of $\Delta \chi^2 _{\rm EDS-EDE} = - 2.0$.

The marginalized posterior distributions are shown in Figs.~\ref{fig:summary-default-S8} and \ref{fig:big-triangle-default-S8}. The preference for $f_{\rm EDE} > 0$ is significantly diminished when DES-Y3 is included (as expected based on previous work for EDE~\cite{Hill:2020osr,Ivanov:2020ril,DAmico:2020ods}), and in place of a detection we find only an upper bound. We find a 95\% CL upper bound $f_{\rm EDE}<0.14 $, which, while consistent with the $H_0$-resolving regime of parameter space, is also consistent with $f_{\rm EDE}=0$, similar to results in the non-interacting EDE scenario \cite{Hill:2020osr} when DES-Y1, HSC, and KV-450, are included.  However, one may appreciate from the $H_0 - S_8$ panel that EDS exhibits a substantial overlap between the 95\% CL contours of both the SH0ES measurement (grey bands) and DES-Y3 measurement (pink bands). This indicates that the EDS model fit to baseline+DES-Y3 data is statistically consistent with both SH0ES and DES-Y3, at 95\% CL. This is encoded in the marginalized parameter constraints by a broadening of the error bars in EDS relative to EDE: comparing to Tab.~VIII of \cite{Hill:2020osr}, we see that the error bar on $H_0$ in the EDE fit to a comparable combination of data sets is $ \approx \pm 1.1$, whereas in our analysis we find an error bar $\pm 1.2$.

The weighting of the posterior to $c<0$ is slightly strengthened by the inclusion of DES-Y3 data, as the additional $S_8$ data disfavors the low-$H_0$-high-$S_8$ region discussed in Sec.~\ref{sec:baseline}. We find $c=-0.020^{+0.025}_{-0.032}$ and highlight the 1d  $c$ posterior in Fig.~\ref{fig:summary-default-S8}, where the support for the $c>0$ tail of the distribution present in the fit to the baseline data set has been significantly reduced. Looking at the the $c-\log_{10}(z_c)$ panel, we again see a weak multimodality, now accompanied by a tail out to large $z_c$.

\begin{figure*}[]
    \centering
    \includegraphics[clip, trim=0.25cm 0cm 0.5cm 1cm,width=0.99\textwidth]{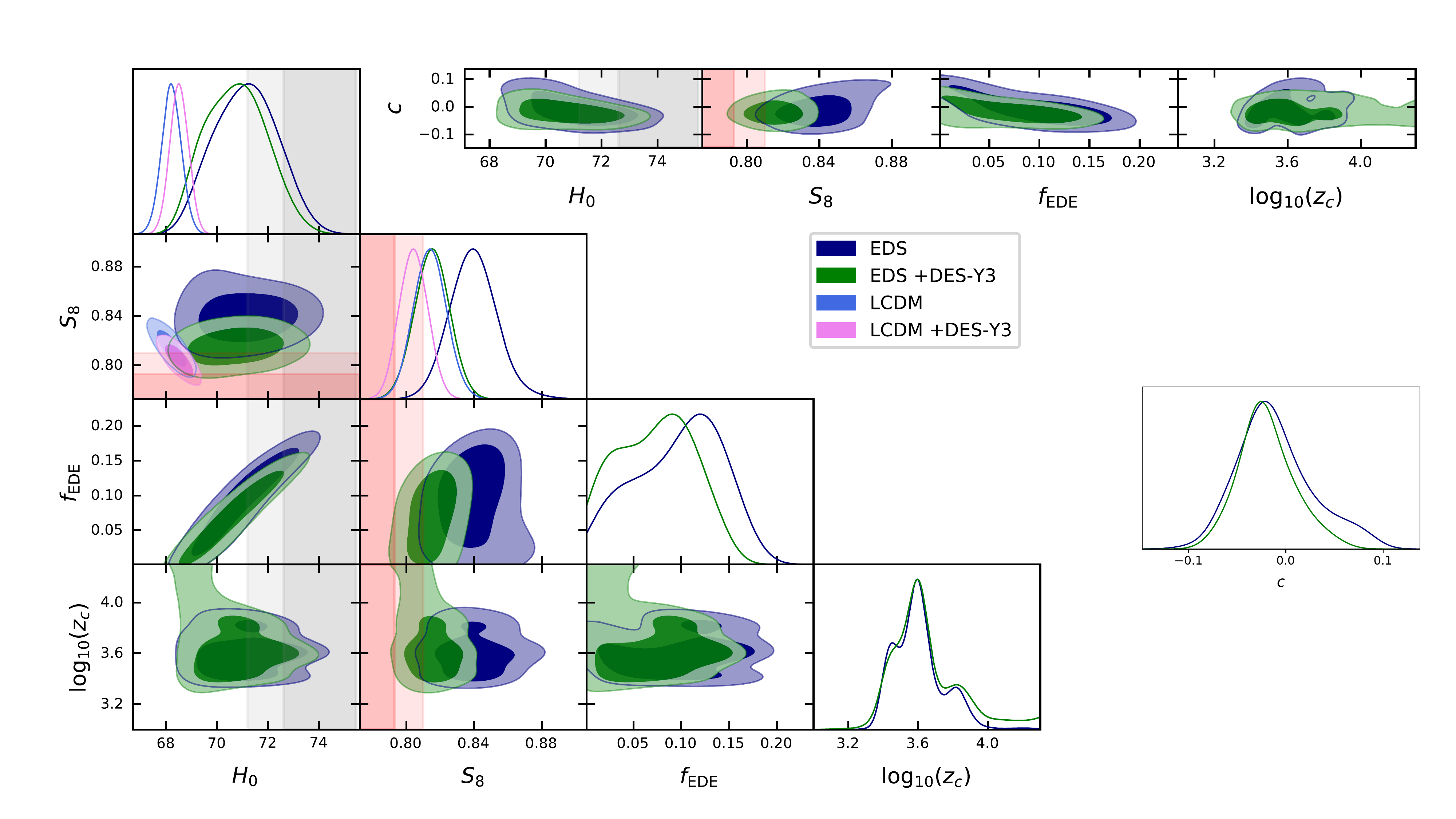}
    \caption{The impact of $S_8$ data.  The plot shows posterior distributions for the fit to the baseline data set (CMB, CMB lensing, BAO, SNIa, and SH0ES) supplemented with DES-Y3 data, approximated by a prior on $S_8$, for $\Lambda$CDM, EDE, and EDS. Shaded grey and pink bands denote the 2019 SH0ES measurement and the DES-Y3 $S_8$ constraint, respectively. }
    \label{fig:summary-default-S8}
\end{figure*}

\subsection{Constraints from ACT Data}

Finally, we consider the impact of high-precision small-scale CMB data, namely, the latest measurements from the Atacama Cosmology Telescope fourth data release (DR4)~\cite{ACT:2020frw,ACT:2020gnv}. The ACT collaboration analysis of EDE  \cite{Hill:2021yec},  in a fit to the combination of ACT, large-scale Planck $TT$, Planck CMB lensing, and BAO data, has found a moderate $\approx 3 \sigma$ preference for a non-zero EDE component, finding $f_{\rm EDE}= 0.091 _{-0.036} ^{+0.020}$. As a first look at ACT and the EDS model, we supplement our baseline data set with ACT $TT$, $TE$, and $EE$ data. We include the full Planck likelihood, including the high-$\ell$ temperature and polarization power spectra, and impose the multipole cut determined in~\cite{ACT:2020gnv} to the ACT data to avoid double-counting information, in particular setting $\ell_{\rm min, TT, ACT} = 1800$.

When using the ACT data we use enhanced precision settings in our modified version of the Boltzmann code \texttt{CLASS}. The need for this increased precision is documented in \cite{ACT:2020gnv} (see their Appendix A). This increased precision comes at the cost of additional computational expense in the MCMC analyses. In light of this, and in light of the existing ACT collaboration analyses of $\Lambda$CDM \cite{ACT:2020gnv} and EDE \cite{Hill:2021yec}, in this work we perform an MCMC analysis of only the EDS model (and not EDE or $\Lambda$CDM), and we present maximum-likelihood parameters for only EDS and $\Lambda$CDM (and not EDE).  Future optimization of the precision parameters needed for these calculations, and/or the development of emulators with which to accelerate the Boltzmann code (e.g., as in~\cite{SpurioMancini:2021ppk}), will be useful.

The best-fit parameters and parameter constraints for the analysis including ACT are given in Tab.~\ref{tab:parameters-ACT}, and $\chi^2$ statistics are given in Tab.~\ref{table:chi2_ACT}. The marginalized posterior distributions are shown in Fig.~\ref{fig:summary-default-ACT}.

Inclusion of the ACT data provides a factor of two improvement on the error on $c$.  We find $c=-0.002^{+0.015}_{-0.024}  $ in comparison with $c=-0.011^{+0.029}_{-0.047}$ from the fit to the baseline data set. This dramatic reduction is largely driven by the ability of ACT to constrain the timing of the EDE component, $z_c$. Indeed, from the posterior distribution of $\log_{10}(z_c)$ in Fig.~\ref{fig:summary-default-ACT}, one may appreciate that the inclusion of ACT data in the EDS analysis almost completely removes the multimodality exhibited in the fit of EDS to the baseline data set, as ACT removes the high-$z_c$ tail (as discussed in \cite{Hill:2021yec}). The reduced multimodality in $z_c$ propagates to the marginalized constraint on $c$, leading to an overall reduction in the error bar. 

Meanwhile, the preference for a non-zero EDE component is strengthened (as found in~\cite{Hill:2021yec,Poulin:2021bjr,Moss:2021obd}): we find the marginalized constraint $f_{\rm EDE} = 0.108^{+0.053} _{-0.023}$ when ACT is included, compared to $f_{\rm EDE} = 0.099^{+0.056} _{-0.041}$ without ACT data. However, the $f_{\rm EDE}$ posterior distribution in Fig.~\ref{fig:summary-default-ACT} is significantly broader than a Gaussian, exhibiting ample support on the boundary of the prior at $f_{\rm EDE} \approx 0$. Indeed we find the 95\% CL constraint $f_{\rm EDE} = 0.108 ^{+0.063} _{-0.095}$, which nearly reaches $f_{\rm EDE}=0$. This is reflected also in the 2d posteriors, e.g., $f_{\rm EDE}-H_0$ and $f_{\rm EDE}-S_8$, which are consistent with $f_{\rm EDE}=0$ at the 95\% confidence level. The marginalized constraints on $H_0$ and $S_8$ are consistent with those from the fit to the baseline data set, while the best-fit values of both are lower when ACT is included, with $H_0=71.79$ and $H_0=72.52$ with and without ACT respectively, and $S_8=0.8385$ and $S_8=0.8481$ with and without ACT, respectively.

Turning to the $\chi^2$ values, Tab.~\ref{table:chi2_ACT}, we find that the best-fit EDS model is an improvement over the best-fit $\Lambda$CDM model by $\Delta \chi^2 = -19.0$. This is slightly enhanced relative to that in the fit to the baseline data set ($\Delta \chi^2_{\rm tot, \,baseline} = -18$, Tab.~\ref{tab:chi2_baseline}), driven in part by $\Delta \chi^2_{\rm ACT}=-1.9$, consistent with the mild preference of ACT data for a non-zero EDE component.

\begin{figure*}
    \centering
    \includegraphics[clip, trim=0.25cm 0cm 0.5cm 1cm,width=0.99\textwidth]{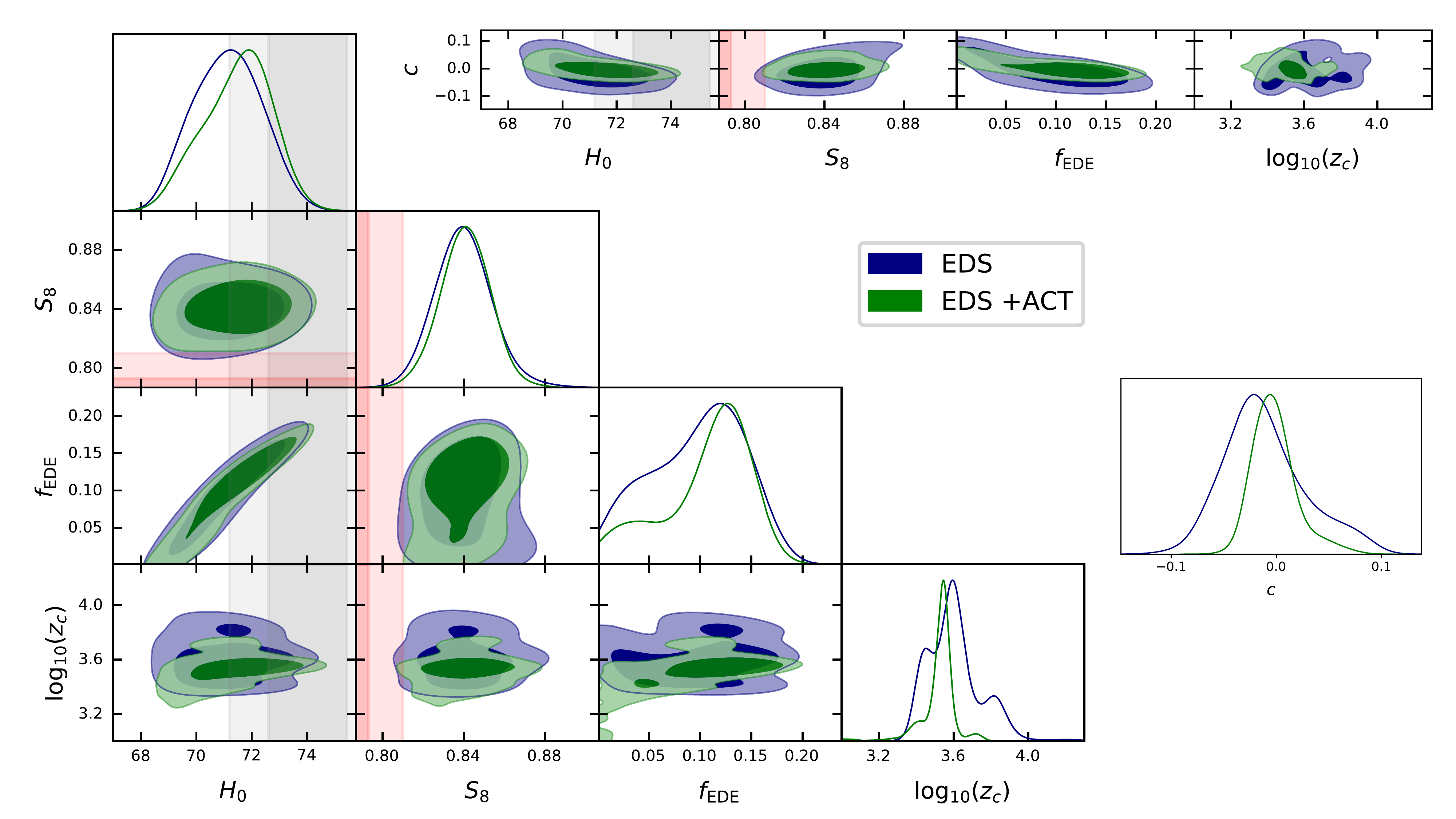}
    \caption{Constraints including ACT data.  The plot shows posterior distributions for the fit of the EDS model to the baseline data set (CMB, CMB lensing BAO, SNIa, and SH0ES) with and without the addition of ACT primary CMB data.  Shaded grey and pink bands denote the SH0ES measurement and the DES-Y3 $S_8$ constraint, respectively.}
    \label{fig:summary-default-ACT}
\end{figure*}

\begin{table*}[!ht]
Maximum Likelihood and Marginalized Parameter Constraints from the combination of the baseline data set and ACT data.
\setlength{\tabcolsep}{12pt}
\centering
\begin{tabular}{@{}cccccc@{}}
\toprule
Parameter                  & $\Lambda$CDM & EDS                  \\
\toprule                                                                                                               
$\boldsymbol{ 100\theta_{\rm s}}$    & 1.04219  &  1.04150 ($1.04151^{+0.00034}_{-0.00039}$)   \\
$\boldsymbol{ \Omega_bh^2 }$         & 0.02248  & 0.02257 ($0.02258\pm 0.00017        $)   \\
$\boldsymbol{ \Omega_ch^2}$          & 0.1181  & 0.1311 ($0.1302^{+0.0055}_{-0.0034}$)  \\
$\boldsymbol{ \tau}$                 & 0.0599       & 0.0565 ($0.0546\pm 0.0071          $)    \\
$\boldsymbol{ \ln(10^{10}A_s)  }$    & 3.059         & 3.071 ($3.068\pm 0.015            $)      \\
$\boldsymbol{ n_s     }  $           & 0.9725      & 0.9876 ($0.9865^{+0.0077}_{-0.0065}$)     \\
 $\boldsymbol{c_\theta} $ & & $ -0.0008$ ($0.0013^{+0.0013}_{-0.0065}$) \\
$\boldsymbol{f_{\rm EDE} }$          &     &   0.119  ($0.108^{+0.053}_{-0.023}   $)    \\
$\boldsymbol{ \log_{10}z_c}$         &   &  3.545  ($3.521^{+0.071}_{-0.032} $  ) \\
$\boldsymbol{ \theta_i }$            &     &  2.79     ($2.44^{+0.46}_{+0.16}      $)       \\
\colrule
$c$                     &   &  $-0.005$ ($-0.002^{+0.015}_{-0.024}  $)   \\

$\phi_i\, [\Mpl]$                  &     &       0.474 $(0.490\pm 0.093          )  $   \\
$\mathrm{log}_{10}(f/{\mathrm{eV}})$ &   &   26.61 ($26.726^{+0.011}_{-0.19}   $)   \\
$\mathrm{log}_{10}(m/{\mathrm{eV}})$ &  &  $-27.32$ ($-27.270^{+0.033}_{-0.16}  $)  \\
$\Delta m_{\rm DM}/m_{\rm DM}$       &  &      $-0.0009$ ($-0.0011^{+0.0078}_{-0.011}$)  \\

$\sigma_8$                           & 0.8128  &  0.8393 ($0.840^{+0.010}_{-0.0094}  $)   \\  
$\Omega_m$                           & 0.3003    & 0.2995 ($0.3003\pm 0.0052          $)    \\  
$S_8$                                &   0.8172 & 0.8385  ($0.841\pm 0.012            $)   \\  
$H_0$                                &     68.23    &  71.79  ($71.5^{+1.4}_{-1.1}        $)       \\
\hline
$\Delta\chi^2_{\rm tot}$             &  0                              & $-19.0$                     \\
\botrule
\end{tabular}
\caption{ \label{tab:parameters-ACT}
 Maximum likelihood (ML) parameters and marginalized parameter constraints for EDS and $\Lambda$CDM in the fit including ACT data. Parameters in bold are sampled in MCMC analyses. For $\Lambda$CDM we give only ML parameters, due to the significant computational expense of MCMC analyses at the high precision settings required to analyse ACT data. }
\end{table*}

\begin{table}[h!]%[t]
\centering
$\chi^2$ statistics \\ from the fit to {\it Planck} 2018, BAO, SNIa, SH0ES, and ACT \\
  \begin{tabular}{|l|c|c|}
    \hline\hline
    Datasets  &  $\Lambda$CDM  &  EDS   \\ \hline \hline
    Primary CMB:  & &\\
    \;\;\;\; \textit{Planck} 2018 low-$\ell$ TT  & 22.2   & 21.3   \\
    \;\;\;\; \textit{Planck} 2018 low-$\ell$ EE  & 397.2  & 396.4   \\
    \hspace{.16cm} \; \begin{tabular}[t]{@{}c@{}}\textit{Planck} 2018 high-$\ell$ \\ TT+TE+EE\end{tabular}                        & 2346.3 & 2345.9  \\
    \;\;\;\; ACT  & 243.2 & 241.2  \\
    
    LSS: & &\\
    \;\;\;\;\,\textit{Planck} CMB lensing        & 8.4    &  9.8   \\
    \;\;\;\; BAO (6dF)          & 0.0008  & 0.015\\
    \;\;\;\; BAO (DR7 MGS)     & 1.8    & 2.1\\
    \;\;\;\; BAO (DR12 BOSS)   & 3.4    & 3.4 \\
    SNIa (Pantheon)  & 1034.7 &  1034.7 \\
    SH0ES             & 16.7   & 2.5 \\
     Planck prior        & 4.1   & 2.0  \\
    \hline
    $\Delta \chi^2 _{ {\it Planck}\, {\rm primary \,CMB}}$     &  0  & $-2.1$ \\ 
        $\Delta \chi^2 _{ \rm ACT }$     &  0  & $-1.9$ \\ 
    $\Delta \chi^2 _{\rm LSS}$    &  0 & $+1.2$ \\ 
    $\Delta \chi^2 _{\rm SH0ES}$  & 0  &  $-14.2$  \\ 
    \hline 
    $\Delta \chi^2 _{\rm tot}$   & 0    & $-19.1$ \\ 
    \hline
  \end{tabular}
  %}
 \caption{$\chi^2$ statistics for the ML $\Lambda$CDM and EDS models in the fit to the baseline data set (CMB, BAO, SNIa, and SH0ES) supplemented with ACT data. }
  \label{table:chi2_ACT}
\end{table}

\subsection{The Swampland}
\label{sec:swamp}

As discussed in Sec.~\ref{sec:EDS}, the SDC states that a Planckian field excursion leads to an exponential suppression of the mass of other fields. The simple setup studied here, with the scalar field coupled to all of the dark matter, provides a minimal context within which to test the SDC.  A similar idea has been explored previously in the context of quintessence, where it was dubbed Fading Dark Matter \cite{Agrawal:2019dlm}.

The 95\% bounds on the parameter $c$ are given in Tab.~\ref{table:swamp}. The posterior distributions for swampland-related quantities (the field excursion, the axion decay constant, and the coupling $c$), along with their correlations with the Hubble parameter $H_0$, are shown in Fig.~\ref{fig:swamp}. The SH0ES measurement is shown in grey bands. From this one may appreciate that the EDE resolution of the Hubble tension scenario, namely, the ability for the EDE model to be $1-2\sigma$ consistent with SH$0$ES, indeed rests upon a Planckian field excursion $|\Delta \phi|/ \Mpl \simeq 1/2 $, and a Planckian axion decay constant $f\sim \Mpl/5$. Thus one naturally expects the parameter $c$ to play a role in this model.

However, turning to Tab.~\ref{table:swamp}, we see that the SDC parameter $c$ is constrained to be $c<0.068$ from the baseline data set, at 95\% CL, and $c<0.035$ and $c<0.042$ at 95\% CL when DES-Y3 or ACT are included, respectively. From this we infer a mild tension of the data, in the context of the EDE model, with the SDC at the level of a $4-7\%$ fine-tuning. 

While the degree of fine tuning may not be severe, it is interesting to note that these constraints are an order of magnitude stronger than constraints on other would-be $\mathcal{O}(1)$ swampland parameters. In particular, the de Sitter (dS) Swampland Conjecture \cite{Obied:2018sgi} states that scalar field potentials cannot be arbitrarily flat, and are bounded by $V'/V \geq \mathcal{O}(1)$ in Planck units. The would-be $\mathcal{O}(1)$ parameter of the dS conjecture is constrained by data to be $V'/V \lesssim 0.51$ ($2\sigma$) \cite{Raveri:2018ddi} or $V'/V \lesssim 1.35$ ($3\sigma$) \cite{Heisenberg:2018yae} (see also \cite{Akrami:2018ylq}). Compared to the constraints on SDC order-1 parameter presented in this work ($c<0.035$, $c<0.042$, and $c<0.068$ at 95\% CL), one may appreciate the latter are considerably stronger than constraints on the swampland found in previous works.

Finally, we note that a more complete analysis, which we will not pursue here, would be to allow variation in the fraction of dark matter $f_{\rm DM}$ to which the scalar field couples. This would introduce one new parameter to the already four-parameter EDS extension to $\Lambda$CDM. We expect the $\approx 5\%$ fine-tuning of $c$ in our fixed-$f_{\rm DM}$ analysis to translate to slightly lesser fine-tunings of $c$ and $f_{\rm DM}$ once $f_{\rm DM}$ is allowed to vary. 

\begin{table}[h!]%[t]
\centering
Constraints on the Swampland Distance Conjecture \\
  \begin{tabular}{|l|c|}
    \hline\hline
    Datasets & 95\% upper limit on $c$  \\ \hline \hline
    \, baseline & $c < 0.068$ \\
   ~\,baseline + $S_8$ from DES-Y3   &   $c < 0.035 $  \\  
      ~\,baseline + ACT    & $c < 0.042 $ \\   

     \hline 
  \end{tabular}
  %}
 \caption{  Constraints on the Swampland Distance Conjecture parameter $c$, defined by the early dark energy dependence of the dark matter mass $m_{\rm DM} = e^{c \phi/\Mpl}$. Upper and lower bounds are 95\% CL.}
  \label{table:swamp}
\end{table}

\begin{figure*}%[h!]
    \centering
    \includegraphics[clip, trim=8cm 0cm 7cm 1cm,width=0.99\textwidth]{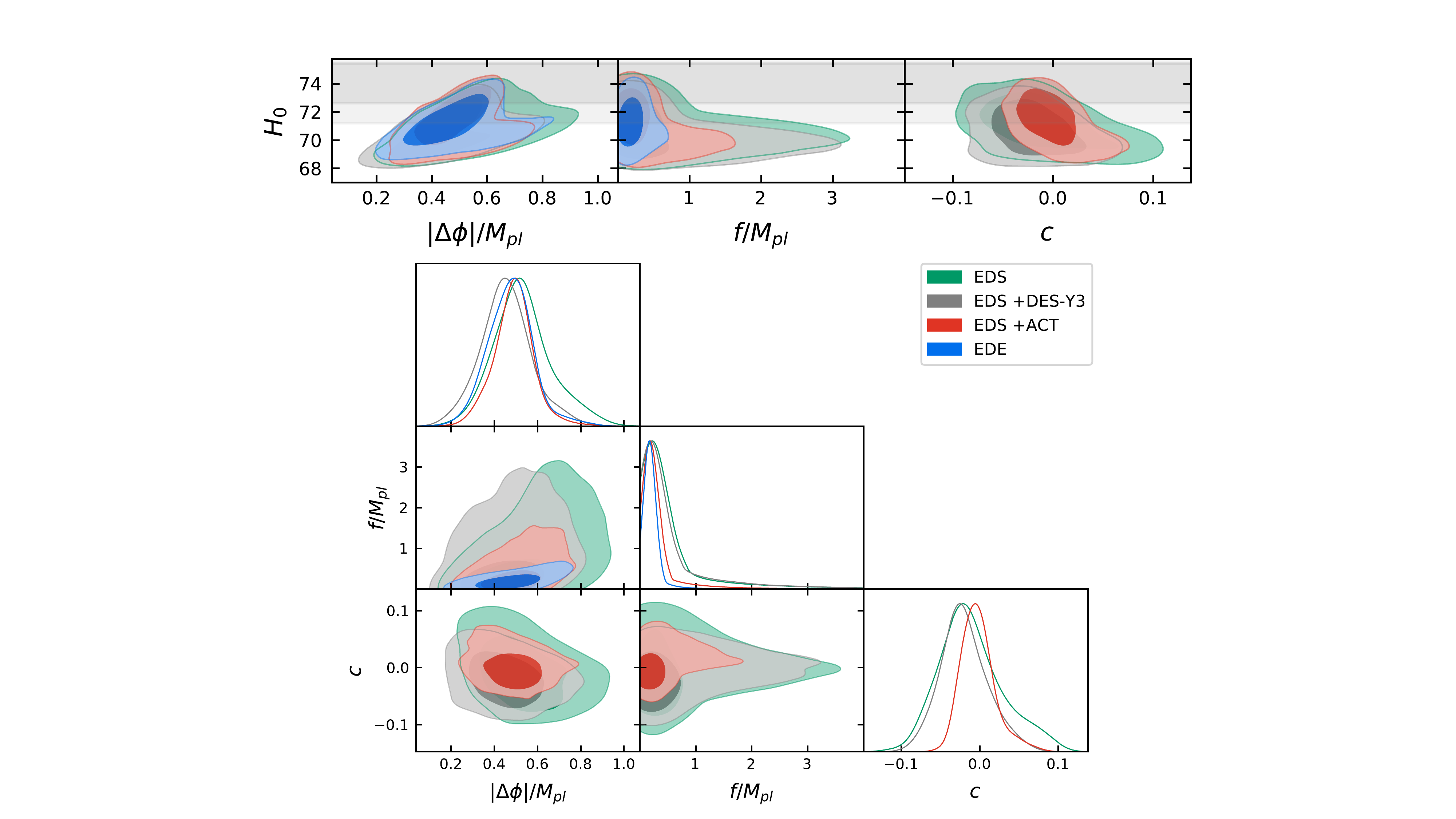}
    \caption{Early Dark Energy and the swampland conjectures. We show the posterior distributions of the field excursion, axion decay constant, and dark matter mass dependence, along with their correlation with $H_0$, in the fit to varying data sets. The swampland distance conjecture would suggest that $c=\mathcal{O}(1)>0$, while the data constrain $c < 0.068$, $0.035$, and $0.042$ at $95\%$ confidence, for the baseline data set, the baseline + DES-Y3, and baseline + ACT, respectively, and slightly prefer $c<0$.}
    \label{fig:swamp}
\end{figure*}

%%%%%%%%%%%%%%%%%%%%%%%%%%%%%%%%%%%%%%
%%%%%%%%%%%%%%%%%%%%%%%%%%%%%%%%%%%%%%
\section{Discussion}
\label{sec:discussion}
%%%%%%%%%%%%%%%%%%%%%%%%%%%%%%%%%%%%%%
%%%%%%%%%%%%%%%%%%%%%%%%%%%%%%%%%%%%%%

In this work have extended Early Dark Energy to an Early Dark {\it  Sector} (EDS). Motivated by the Swampland Distance Conjecture \cite{Ooguri:2006in} (SDC), the EDS is comprised of the EDE scalar field along with a dark matter candidate whose mass is exponentially sensitive to Planckian field excursions of the EDE scalar. The aims of this model are two-fold: (1) to understand the interplay of the $H_0$ and $S_8$ tensions, and determine whether the competition between these can be softened by embedding EDE into a larger model, and (2) to determine the extent to which EDE (namely the $H_0$-tension-resolving region of EDE parameter space) is in conflict with the SDC, and thereby determine whether the EDE resolution of the Hubble tension lies in the landscape or the swampland.

Concretely, the EDS model is a one-parameter extension of EDE, parameterized by an additional parameter $c$ corresponding to the exponent in the dark matter mass, $m_{\rm DM}(\phi)=m_0 e^{c \phi/\Mpl}$, where $\phi$ is the EDE scalar. In our sign convention, where $\phi$ is initially $>0$ and decreases over cosmic evolution, the SDC predicts that $c>0$ and $c=\mathcal{O}(1)$. The parameter $c$ has important impacts on both the CMB and on the growth of structure. In the CMB the imprint of $c$  contains a localized feature around $\ell \simeq 500$, corresponding to  modes that enter the horizon near $z_c$ and a sign reversal in its effect at much higher multipoles. This can be understood in terms of the impact of the dark matter mass on the radiation driving of acoustic oscillations, as described in Sec.~\ref{sec:pheno}.

Meanwhile, $c>0$ (at fixed $\Omega_c h^2$) leads to an enhanced growth of structure, due to the relative shift in matter radiation equality to earlier times. The growth of structure is also subject to a second effect: an effective dark matter self-interaction (a dark ``fifth force'') that is attractive, and in the limit of high $k$ has strength $c ^2 G_{N}$. This leads to enhanced structure formation on small scales for both positive and negative $c$. The combination of the two growth effects allows a small but negative $c$ to decrease $S_8$. Incidentally, this small negative $c$ also improves the fit to the CMB.

Armed with the theory motivation and understanding of the phenomenology, we have performed MCMC analyses of the EDS model fit to a baseline data set combination of {\it Planck} 2018 primary CMB and CMB lensing \cite{Planck2018likelihood,Aghanim:2018eyx,2018arXiv180706210P}; BAO from the SDSS DR7 main galaxy sample~\cite{Ross:2014qpa}, the 6dF galaxy survey~\cite{2011MNRAS.416.3017B}, and SDSS BOSS DR12~\cite{Alam:2016hwk}; the Pantheon supernovae data set \cite{Scolnic:2017caz}, and the 2019 SH0ES $H_0$ measurement \cite{Riess:2019cxk}. We have performed additional MCMC analyses of the baseline data set supplemented with Dark Energy Survey Year-3 data \cite{DES:2021wwk} and supplemented with data from ACT \cite{ACT:2020gnv, ACT:2020frw}. 

From the analysis of the baseline data set we find that EDS can accommodate lower $S_8$ values than EDE without compromising on $H_0$. The low-$S_8$-high-$H_0$ region of parameter space is correlated with small but negative $c$, and we find a mild overall preference for $c<0$ in the fit to the baseline data set. When the data set is supplemented with DES-Y3, we find that $S_8$ decreases while leaving $H_0$ nearly unchanged, while maintaining the preference for $c<0$. Compared to EDE, we find EDS is better able to accommodate the DES-Y3 data by $\Delta \chi^2 _{\rm DES-Y3,EDS-EDE}=-1.1$. This demonstrates the ability of the EDS model to at least partially resolve the tension of EDE with large scale structure data.

When ACT data are included we find a significant improvement on the constraint on $c$, driven largely by the improved constraint on $z_c$. Finally, all of these analyses constrain $c$ to be significantly less than $1$: we find $c<0.068$ from the baseline data set, at 95\% CL, and $c<0.035$ and $c<0.042$ at 95\% CL when DES-Y3 or ACT are included, respectively. Taken at face value, this indicates a tension between the EDE resolution of the Hubble tension and the SDC.

Finally, we evaluate the overall preference of the data for the EDE model vs.~EDS. To compare the EDS and EDE models we calculate the Akaike Information Criterion \cite{akaike1973information}, which for the baseline data set yields $\Delta {\rm AIC} \simeq 0$, suggesting no preference for one model over the other. A more detailed model comparison could be done by computing the Bayesian evidence for each model; we leave this for future work. 
% However, we note that, as emphasized above, constraints on the EDS parameter $c$ are consistent with $c=0$. }

We close this analysis with the following summary comments:
\begin{enumerate}
    \item The EDS extension of EDE, namely EDE with the EDE-dependent dark matter mass $m_{\rm DM}(\phi)=m_0 e^{c\phi/M_{\rm pl}}$, can partially ameliorate the tension between the EDE resolution of the Hubble tension and LSS data. However, the data are statistically consistent with $c=0$.
    \item ACT data significantly constrain both the timing $z_c$ of the EDE component and the EDS coupling parameter $c$. We find that supplementing the baseline data set with ACT data improves the constraint on $c$ by a factor of two, and nearly eliminates the preference for $c<0$.
    \item Order-1 values of $c$ in $H_0$-resolving EDE are ruled out by the data. While the SDC does not make any prediction for the fraction of dark matter to which the EDE scalar is coupled, this nonetheless suggests a mild tension between the SDC and EDE resolution of the $H_0$ tension.
\end{enumerate}

There remain many directions for future work. Our analysis is motivated by the SDC, but the latter makes no prediction for fraction of dark matter to which the scalar field couples. Therefore a natural model extension is allow this fraction to vary in the fit to cosmological data sets. Other variations of our analysis would be to consider different choices of $V(\phi)$, such as monomial $\phi^n$ or hyperbolic $\tanh(\phi/f)^n$ potentials, and different choices of the dark matter coupling, such as $m_{\rm DM}=m_0(1+c \phi^2/M_{pl}^2) $. A final possibility is to examine the role of EDE-dark matter interactions in resolving the coincidence problem inherent in early universe resolutions to the Hubble tension, namely, why the new physics becomes transiently relevant around matter-radiation equality, and not in the many decades of redshift before this epoch. We leave these interesting possibilities to future work.\\

{\bf Acknowledgements} The authors thank Mikhail Ivanov, Leah Jenks, Austin Joyce, Hayden Lee, Michael Toomey, and Liantao Wang for insightful comments.
 W.H.\ and M-X.L.\ were supported by U.S.\ Dept.\ of Energy contract DE-FG02-13ER41958 and the Simons Foundation. Computing resources were provided in part by the University of Chicago Research Computing Center through the Kavli Institute for Cosmological Physics at the University of Chicago. J.C.H.\ acknowledges support from NSF grant AST-2108536.  We thank the Scientific Computing Core staff at the Flatiron Institute for computational support.  The Flatiron Institute is supported by the Simons Foundation.

\appendix

\section{Equations of Motion}
\label{app:EOM}

We consider cold dark matter interacting with the EDE scalar field $\phi$. We model dark matter as a population of non-relativistic Dirac fermions. We consider a model with action given by,
\bea
S = \int  d^4 x \sqrt{-g} \left[ \right. \frac{1}{3 \Mpl^2} R - \frac{1}{2}\partial^\mu \phi \partial_\mu \phi - i \bar{\psi} \slashed{D}\psi \\
- V(\phi) - m_{\rm DM}(\phi)\bar{\psi}\psi + h.c. \left. \right].\nonumber
\eea
where $\psi$ is a Dirac fermion, which plays the role of cold dark matter. As such, we take the non-relativistic limit of $\psi$, in which case $\langle \bar{\psi} \psi \rangle \rightarrow n(t)$, where $n(t)$ is the number density, namely, the total number of particles and anti-particles, not to be confused with $\langle \bar{\psi}\gamma^0\psi \rangle$, which is the {\it difference} between the number of particles and antiparticles. In this limit, the dark matter component is described by a stress tensor,
\be
{T^{\rm (DM)}}^{\mu}\, _\nu = n_{\rm DM} m_{\rm DM}(\phi) u^\mu u_\nu
\ee
with $u^\mu = (-1,v^i)$. This comprises only a part of the stress tensor of the full interacting system, which is given by
\be
T_{\mu \nu} = T_{\mu \nu} ^{(\rm DM)}  + T_{\mu \nu} ^{(\phi)},
\ee
where $T_{\mu \nu} ^{(\phi)}$ is the $\phi$ contribution given by
\be
\label{eq:Tmunuphi}
{T^{(\phi)}}^{\mu}\, _\nu = \partial^\mu \phi \partial_\nu \phi - \frac{1}{2}\delta^\mu \, _\nu \partial^\alpha \phi \partial_\alpha \phi - \delta^ \mu \, _\nu V(\phi) .
\ee
The combined stress tensor is covariantly conserved,
\be
\label{eq:nablaT}
\nabla_\mu {T}^{\mu}\, _\nu = 0 ,
\ee
which follows from the contracted Bianchi identities of General Relativity. The equations of motion of the interacting system are dictated by  the conservation equation Eq.~\eqref{eq:nablaT} along with the equations of motion for the scalar field that follow from the variation of the action.

The equations of motion for the scalar field background and perturbations are given by the variation of the action expanded to linear and quadratic order in $\delta \phi$ respectively.  At the background level, where quantities depend only on time, the variation with respect to the scalar field gives,
\be
\ddot{\phi} + 2 a H \dot{\phi} +  a^2 \frac{ d V}{ d \phi}  + a^2 n  \frac{d m_{\rm DM}}{d \phi} = 0,
\label{eq:phiEOM}
\ee
where dot denotes a derivative with respect to conformal time $\tau$, while $H$ is defined with respect to time $t$. This can be expressed in terms of the dark matter energy density as,
\be
\ddot{\phi} + 2 a H \dot{\phi} +  a^2 \frac{d V}{d\phi}  + a^2   \frac{d \log m_{\rm DM}}{d \phi}\rho_{\rm DM} = 0 .
\ee
The equation of motion for the dark matter density is given by
\be
\label{eq:rhoDMwithint}
\dot{\rho}_{\rm DM} + 3 a H \rho_{\rm DM} = \dot{\phi}\frac{d \log m_{\rm DM}}{d \phi}\rho_{\rm DM} .
\ee

We repeat this procedure for the perturbations, working in the synchronous gauge. The metric in synchronous gauge is given in general by
\begin{equation}
\label{eq:synchmetric}
  ds^2 = a^2(\tau)\left(-d\tau^2 + (\delta_{ij} + h_{ij})dx^i dx^j\right)\,.
\end{equation}
The perturbation $h_{ij}$ may be decomposed into two scalar degrees of freedom, $h$ and $\eta$, defined by the decomposition,
\begin{eqnarray}
\label{hijk}
	h_{ij}(\vec{x},\tau) = \int d^3k e^{i\vec{k}\cdot\vec{x}}
\Bigg[  \hat{k}_i\hat{k}_j h(\vec{k},\tau) \\
+	(\hat{k}_i\hat{k}_j - {1 \over 3}\delta_{ij})\,
        6\eta(\vec{k},\tau) \Bigg] \nonumber
\end{eqnarray}
where $\vec{k} = k\hat{k}$.  See, e.g., \cite{Ma:1995ey}, for more details.

The interaction of the scalar field with dark matter generates new terms in the quadratic action for perturbations, which are given by,
\begin{equation}
\delta S_{2}  = - \int d\tau d^3x \, a^4(\tau) \left[ \frac{d^2 m_{\rm DM}}{d\phi^2} \delta \phi^2 n + \frac{d m_{\rm DM}}{d\phi} \delta \phi \delta n  \right] ,
\end{equation}
where $\delta n$ is the perturbation to the dark matter number density.  The resulting equation of motion is,
\begin{eqnarray}
\ddot{\delta \phi}&&+ 2 a H \dot{\delta \phi}+\left(k^{2} + a^{2}\frac{d^{2}V}{d\phi^{2}}\right)\delta \phi+\frac{1}{2}\dot{h}
\dot{\phi}= \nonumber \\
&&-a^2 \left[\frac{d\ln
m}{d\phi}\rho_{\rm DM}\delta_c+\frac{d^{2} \ln
m}{d\phi^{2}}\delta\phi\rho_{\rm DM}\right],
\end{eqnarray} 
where we define the fractional dark matter density perturbation $\delta_c \equiv (\delta \rho_{\rm DM})/\rho_{\rm DM}$.

To derive the equations of motion for the dark matter component, we now explicitly evaluate Eq.~\ref{eq:nablaT}, and apply the scalar field equations of motion. From the $\nu=0$ component, we find the equation of motion for $\delta_c$, given by
\begin{eqnarray}\label{denscont}
\dot{\delta}_c  + \theta + \frac{\dot{h}}{2}=\frac{d \log m}{d \phi}\dot{\delta \phi}+ \frac{d^2 \log m}{d \phi^2}\dot{\phi}\delta\phi,
\end{eqnarray}
while from the $\nu=i$ component we find the equation of motion for the velocity perturbations,
\begin{equation}
\label{eq:theta}
\dot{\theta}  + a H\theta =  + \frac{d \log m}{d \phi} k^{2} \delta \phi-\frac{d \log m}{d \phi}\dot{\phi}\theta .
\end{equation}
where $\theta \equiv \partial_i v^i$.

\section{Scalar-Mediated Force on Dark Matter}\label{app:Geff}

\begin{figure}[!htb]
\centering
\includegraphics[width=\columnwidth]{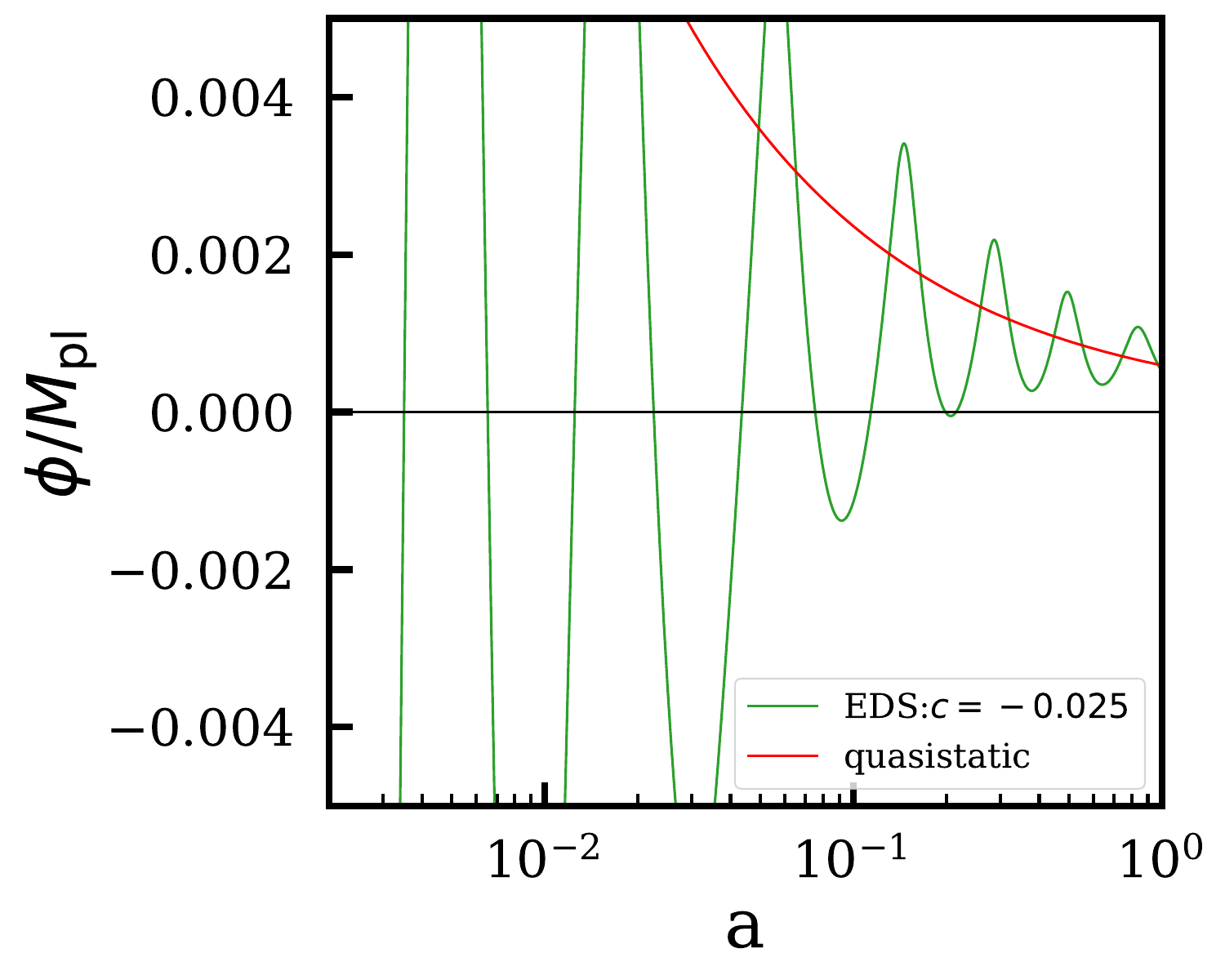}
\caption{\label{fig:phi-quasi}
Scalar field $\phi$ evolution for $c=-0.025$ with fixed $H_0$ and all other parameters (except $\theta_s$) fixed to their values in Eqs.~\eqref{eq:bestfitparams} and \eqref{eq:bestfitparams-2}. The quasistatic estimation Eq.~(\ref{eq:phi-quasi}) is also shown for comparison.
}
\end{figure}

To assess the combined gravitational and scalar mediated forces on the dark matter we start with the time-time and trace of the space-space pieces of the Einstein equation 
with the synchronous metric of Eq.~(\ref{eq:synchmetric})
\begin{equation}
\mathcal{H}\frac{\dot{h}}{2} = k^2\eta+\frac{1}{2\Mpl^2}a^2\delta\rho,
\end{equation}
\begin{equation}
\frac{\ddot{h}}{2}+\mathcal{H}\dot{h} -k^2\eta = -\frac{3}{2\Mpl^2}a^2\delta P
\end{equation}
and combine them
to eliminate $\eta$:
\begin{equation}\label{eq:hddot}
\frac{\ddot{h}}{2}
+\mathcal{H}\frac{\dot{h}}{2} = -\frac{1}{2\Mpl^2}a^2(\delta\rho+3\delta P).
\end{equation}
Taking the derivative of the dark matter continuity equation, Eq.~(\ref{denscont}), and plugging in $\ddot{h}$ from above and $\dot{\theta}$ from the Euler equation (\ref{eq:theta}), we arrive at,
\begin{eqnarray}\label{eq:ddot_delta_c}
\ddot{\delta}_c - \mathcal{H}(\frac{\dot{h}}{2}+\theta_c) =&& \frac{1}{2\Mpl^2}a^2(\delta\rho+3\delta P) -\frac{1}{\Mpl}ck^2\delta\phi \nonumber \\ && +\frac{1}{\Mpl}c\dot{\phi}\theta_c +\frac{1}{\Mpl}c\ddot{\delta\phi}.
\end{eqnarray} 
To gain physical intuition, we consider a quasistatic limit in which the last term in the above can be neglected. 
For small $c$ the second last term is in higher order of $c$. 
Note that there is a non-zero offset value for $\delta\phi$ at late times.
From the equation of motion
\begin{equation}\label{eq:deltaphi0}
	\ddot{\delta\phi} +2aH\dot{\delta\phi} +(k^2+a^2\frac{d^2V}{d\phi^2})\delta\phi +\frac{1}{2}\dot{h}\dot{\phi} = -a^2\frac{1}{M_{\rm pl}}c\delta\rho_{\rm DM},
\end{equation}
we can estimate $\delta\phi$ in the quasistatic limit as
\begin{equation}\label{eq:deltaphi0}
	\delta\phi^{(0)} \approx -a^2\frac{c\delta\rho_{DM}/M_{\rm pl}} {k^2+ a^2 d^2V/d\phi^2},
\end{equation}
We then plug it into Eq.~(\ref{eq:ddot_delta_c}) and, assuming $\delta\rho+\delta P$ is dominated by dark matter, we have,
\begin{equation}
\ddot{\delta}_c +\mathcal{H}\dot{\delta}_c = 4\pi Ga^2\rho_c\delta_c(1+\frac{2c^2k^2}{k^2+a^2d^2V/d\phi^2}).
\end{equation}
From this one may read off an effective gravitational constant,
\begin{equation}
G_{\rm eff} = G_N \left(1+\frac{2c^2k^2}{k^2+a^2d^2V/d\phi^2} \right),
\end{equation}
which is independent of the sign of $c$.

Notice that the modification to $G_{\rm eff}$ appears on scales below the Compton wavelength of the scalar $k/a > d^2V/d\phi^2$ which itself depends on the dark matter density.  The scalar field oscillates around the minimum of the effective potential which evolves quasistatically as
\begin{equation}
\frac{d V}{d\phi} = - \frac{c}{M_{\rm pl}} \rho_{\rm DM}
\end{equation} 
to be
\begin{equation} \label{eq:phi-quasi}
\phi = -{\rm sgn}(c) \frac{ 2^{2/5} c^{1/5} f^{4/5} \rho_{\rm DM}^{1/5}}{3^{1/5} m^{2/5} M_{\rm pl}^{1/5}}.
\end{equation}
At the minimum the scalar mass is
\begin{equation}
m_\phi = \left(\frac{d^2 V}{d\phi^2}\right)^{1/2} = \frac{3^{1/10} 5^{1/2} c^{2/5} m^{1/5} \rho_{\rm DM}^{2/5} }{2^{1/5} f^{2/5} M_{\rm pl}^{2/5}}.
\end{equation}
In Fig.~\ref{fig:phi-quasi} we show the late time evolution of the scalar field for model with $c=-0.025$ with fixed $H_0$ and all other parameters (except $\theta_s$) fixed to their values in Eqs.~\eqref{eq:bestfitparams} and \eqref{eq:bestfitparams-2}. We see that the quasistatic estimation agrees well with the DC offset of the scalar field. The corresponding Compton wavelength at the minimum at $z=0$ is $\sim 1\,$Gpc so that for scales relevant to large-scale structure, $G_{\rm eff} \approx G_N (1+{2c^2})$.   Notice also that the
scaling of the range of the modified force is a fairly mild
$\rho_{\rm DM}^{-2/5}$.   Although EDS admits chameleon screening of the force in a high density environment, even in a virialized halo
where the local density is $\sim 200$ times the background, the range remains large compared with both the scale of the halo and the large-scale structure relevant to $S_8$.

\section{Implementation In \texttt{CLASS}}

We implement the EDS model into the publicly available Boltzmann code \texttt{CLASS} \cite{2011arXiv1104.2932L,2011JCAP...07..034B},\footnote{\url{http://class-code.net}} by modifying the publicly-available \texttt{CLASS\_EDE} \cite{Hill:2020osr}.\footnote{\url{https://github.com/mwt5345/class_ede}}

We use the synchronous gauge functionality of CLASS to solve the Einstein equations, Eq.~(21) of \cite{Ma:1995ey}, given the energy density, pressure, and velocity of the matter content. From the stress tensor Eq.~\eqref{eq:Tmunuphi}, the energy density and pressure of the scalar field are given by,
\begin{eqnarray}
\label{eq:phiback}
    \rho_{\phi} &&= \frac{1}{2 a^2} \dot{\phi}^2 + V(\phi), \\
     p_{\phi} &&= \frac{1}{2 a^2} \dot{\phi}^2 - V(\phi). \nonumber
\end{eqnarray}
The perturbations to the above, along with the scalar field velocity perturbation, are given by,
\begin{eqnarray}
\label{eq:phipert}
    \delta \rho_\phi &&=  \frac{1}{ a^2} \dot{\phi} \dot{\delta \phi} + V'(\phi) \delta \phi. \\
    \delta p_\phi &&=  \frac{1}{ a^2} \dot{\phi} \dot{\delta \phi} - V'(\phi) \delta \phi. \nonumber\\
    (\rho_\phi + p_\phi )v_\phi &&= \frac{1}{a^2}k \dot{\phi} \delta \phi . \nonumber
\end{eqnarray}
We note that \texttt{CLASS} works in units wherein the energy density and pressure are rescaled by $1/3\Mpl^2$, i.e., the stress-energy tensor is rescaled as,
\begin{equation}
T_{\mu \nu} ^{\rm (CLASS)} = \frac{1}{3 \Mpl^2} T_{\mu \nu}   .
\end{equation}
The scalar field retains units of $\Mpl$. The above rescaling manifests in \texttt{CLASS} as a factor of $(1/3)$ in the \texttt{CLASS} definition of $\rho_\phi$, $p_\phi$, etc.~, relative to Eqs.~\eqref{eq:phiback} and \eqref{eq:phipert}.  

The scalar field background equation of motion becomes,
\begin{equation}
\ddot{\phi} + 2 a H \dot{\phi} +  a^2 \frac{d V}{d \phi}  + 3 a^2 \frac{d \log m_{\rm DM}}{d \phi}\rho ^{\rm (CLASS)}_{\rm DM} = 0
\end{equation}
where $\rho ^{\rm (CLASS)}_{\rm DM}$ is in CLASS units. The perturbed Klein-Gordon equation becomes,
\begin{eqnarray}
\ddot{\delta \phi}&&+ 2 a H \dot{\delta \phi}+\left(k^{2} + a^{2}\frac{d^{2}V}{d\phi^{2}}\right)\delta \phi+\frac{1}{2}\dot{h}
\dot{\phi}= \nonumber \\
&&-3 a^2 \left[\frac{d\ln
m}{d\phi}\rho ^{\rm (CLASS)}_{\rm DM}\delta+\frac{d^{2} \ln
m}{d\phi^{2}}\delta\phi\rho ^{\rm (CLASS)}_{\rm DM}\right].
\end{eqnarray} 
The covariant conservation of stress-energy may be expressed as
\begin{equation}
    \nabla^\mu\left(T_{\mu \nu} ^{\rm (DM, CLASS)} +  T_{\mu \nu} ^{(\phi, {\rm CLASS} )}\right)=0 .
\end{equation}
Propagating through the factors of $3$ from the conversion to CLASS units, we find that the equations of motion of dark matter perturbations in CLASS units are unchanged from Eqs.~\eqref{denscont} and \eqref{eq:theta}.

The \texttt{CLASS\_EDE} code \cite{Hill:2020osr} absorbs the cosmological constant $\Lambda$ into the scalar field potential, as
\begin{equation}
    V(\phi)= 3 \Mpl^2\Lambda + m^2 f^2 \left[1 - \cos \frac{\phi}{f} \right]^3
\end{equation}
where $\Lambda$ is a constant. This rewriting utilizes the built-in functionality of CLASS to tune a parameter in $V(\phi)$ in order to satisfy the energy budget equation $\sum \Omega_i = 1$ for arbitrary initial conditions for the scalar field. We impose slow-roll initial conditions on $\phi(t)$ and adiabatic initial conditions on $\delta \phi$, as discussed in \cite{Hill:2020osr}. Finally, in order to sample the EDE parameters $f_{\rm EDE}$ and $\log_{10}(z_c)$, we a shooting method to iteratively determine the corresponding model parameters $f$ and $m$.

In this work we add to \texttt{CLASS\_EDE} \cite{Hill:2020osr} a new cold (pressureless) dark matter component that is coupled to the EDE scalar as discussed above. We retain the \texttt{CLASS} cold dark matter component, with a fixed $\Omega_{\rm cdm}=10^{-5}$, in order to self-consistently define the synchronous gauge.

In order to simultaneously sample the present-day dark matter density and the scalar field initial conditions, we implement a shooting method to determine the initial dark matter density. We impose adiabatic initial conditions for the coupled dark matter component.

\section{Additional Posterior Plots}

The enlarged set of posterior distributions for the analysis with DES-Y3 data and with ACT data are given in 
Fig.~\ref{fig:big-triangle-default-S8}
and \ref{fig:big-triangle-default-ACT}, respectively.

\begin{figure*}
    \centering
    \includegraphics[width=0.99\textwidth]{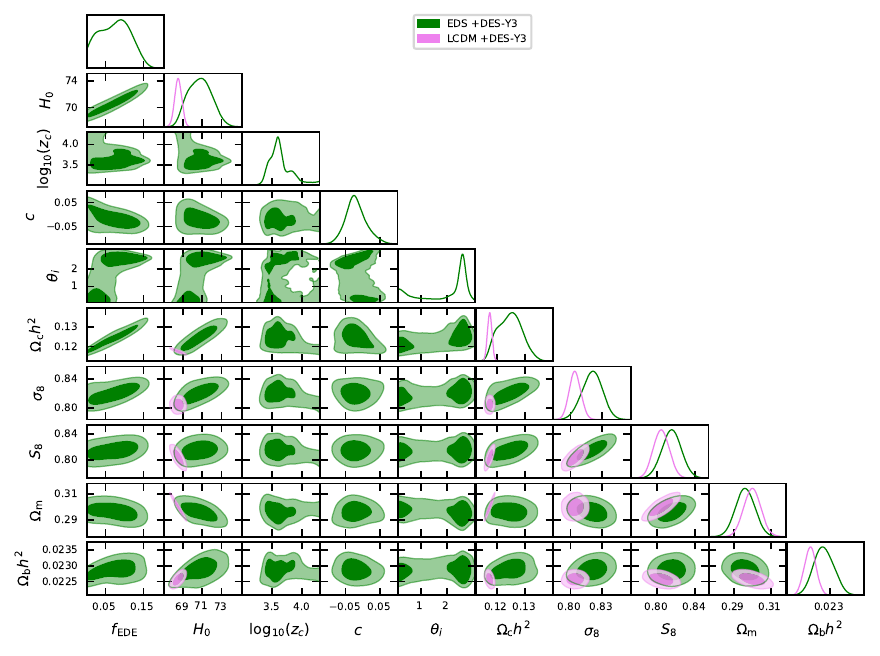}
    \caption{
    Enlarged set of posterior distributions for the fit to the baseline data set (CMB, CMB lensing, BAO, SNIa, and SH0ES) supplemented with DES-Y3 data, approximated by a prior on $S_8$ for $\Lambda$CDM and EDS. }
    \label{fig:big-triangle-default-S8}
\end{figure*}

\begin{figure*}
    \centering
    \includegraphics[width=0.99\textwidth]{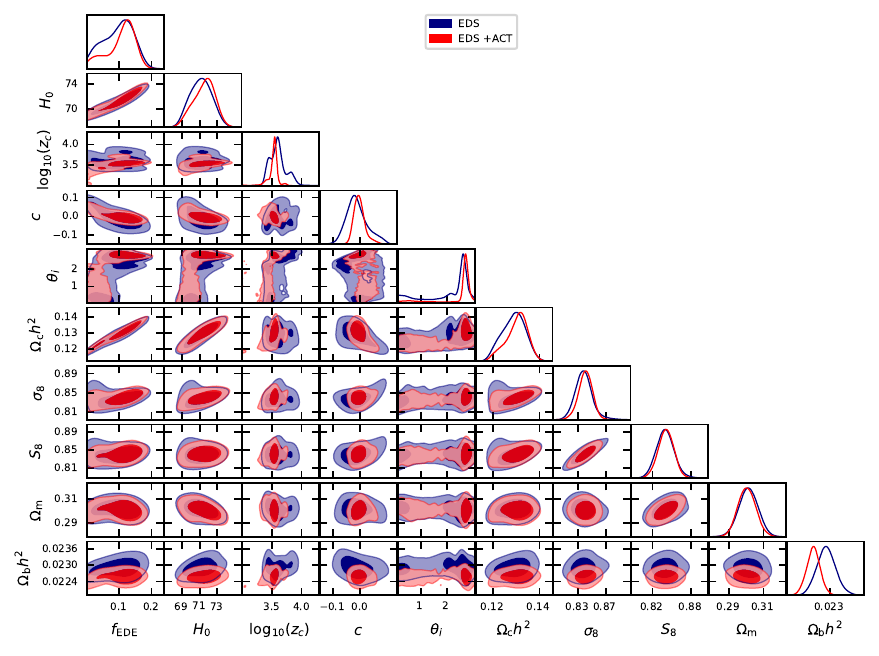}
    \caption{Enlarged set of posterior distributions for the fit of the EDS to the baseline data set (Planck CMB, CMB lensing, BAO, SN, and SH0ES)
with and without the addition of ACT data.  Shaded grey and pink bands denote the SH0ES measurement and the DES-Y3 $S_8$ constraint, respectively. }
    \label{fig:big-triangle-default-ACT}
\end{figure*}

\bibliographystyle{apsrev4-1}

\bibliography{EDS-refs}

%merlin.mbs apsrev4-1.bst 2010-07-25 4.21a (PWD, AO, DPC) hacked
%Control: key (0)
%Control: author (72) initials jnrlst
%Control: editor formatted (1) identically to author
%Control: production of article title (-1) disabled
%Control: page (0) single
%Control: year (1) truncated
%Control: production of eprint (0) enabled
\begin{thebibliography}{88}%
\makeatletter
\providecommand \@ifxundefined [1]{%
 \@ifx{#1\undefined}
}%
\providecommand \@ifnum [1]{%
 \ifnum #1\expandafter \@firstoftwo
 \else \expandafter \@secondoftwo
 \fi
}%
\providecommand \@ifx [1]{%
 \ifx #1\expandafter \@firstoftwo
 \else \expandafter \@secondoftwo
 \fi
}%
\providecommand \natexlab [1]{#1}%
\providecommand \enquote  [1]{``#1''}%
\providecommand \bibnamefont  [1]{#1}%
\providecommand \bibfnamefont [1]{#1}%
\providecommand \citenamefont [1]{#1}%
\providecommand \href@noop [0]{\@secondoftwo}%
\providecommand \href [0]{\begingroup \@sanitize@url \@href}%
\providecommand \@href[1]{\@@startlink{#1}\@@href}%
\providecommand \@@href[1]{\endgroup#1\@@endlink}%
\providecommand \@sanitize@url [0]{\catcode `\\12\catcode `\$12\catcode
  `\&12\catcode `\#12\catcode `\^12\catcode `\_12\catcode `\%12\relax}%
\providecommand \@@startlink[1]{}%
\providecommand \@@endlink[0]{}%
\providecommand \url  [0]{\begingroup\@sanitize@url \@url }%
\providecommand \@url [1]{\endgroup\@href {#1}{\urlprefix }}%
\providecommand \urlprefix  [0]{URL }%
\providecommand \Eprint [0]{\href }%
\providecommand \doibase [0]{http://dx.doi.org/}%
\providecommand \selectlanguage [0]{\@gobble}%
\providecommand \bibinfo  [0]{\@secondoftwo}%
\providecommand \bibfield  [0]{\@secondoftwo}%
\providecommand \translation [1]{[#1]}%
\providecommand \BibitemOpen [0]{}%
\providecommand \bibitemStop [0]{}%
\providecommand \bibitemNoStop [0]{.\EOS\space}%
\providecommand \EOS [0]{\spacefactor3000\relax}%
\providecommand \BibitemShut  [1]{\csname bibitem#1\endcsname}%
\let\auto@bib@innerbib\@empty
%</preamble>
\bibitem [{\citenamefont {Poulin}\ \emph {et~al.}(2019)\citenamefont {Poulin},
  \citenamefont {Smith}, \citenamefont {Karwal},\ and\ \citenamefont
  {Kamionkowski}}]{Poulin:2018cxd}%
  \BibitemOpen
  \bibfield  {author} {\bibinfo {author} {\bibfnamefont {V.}~\bibnamefont
  {Poulin}}, \bibinfo {author} {\bibfnamefont {T.~L.}\ \bibnamefont {Smith}},
  \bibinfo {author} {\bibfnamefont {T.}~\bibnamefont {Karwal}}, \ and\ \bibinfo
  {author} {\bibfnamefont {M.}~\bibnamefont {Kamionkowski}},\ }\href {\doibase
  10.1103/PhysRevLett.122.221301} {\bibfield  {journal} {\bibinfo  {journal}
  {Phys. Rev. Lett.}\ }\textbf {\bibinfo {volume} {122}},\ \bibinfo {pages}
  {221301} (\bibinfo {year} {2019})},\ \Eprint
  {http://arxiv.org/abs/1811.04083} {arXiv:1811.04083 [astro-ph.CO]}
  \BibitemShut {NoStop}%
\bibitem [{\citenamefont {Verde}\ \emph {et~al.}(2019)\citenamefont {Verde},
  \citenamefont {Treu},\ and\ \citenamefont {Riess}}]{Verde:2019ivm}%
  \BibitemOpen
  \bibfield  {author} {\bibinfo {author} {\bibfnamefont {L.}~\bibnamefont
  {Verde}}, \bibinfo {author} {\bibfnamefont {T.}~\bibnamefont {Treu}}, \ and\
  \bibinfo {author} {\bibfnamefont {A.~G.}\ \bibnamefont {Riess}},\ }in\ \href
  {\doibase 10.1038/s41550-019-0902-0} {\emph {\bibinfo {booktitle} {{Nature
  Astronomy 2019}}}}\ (\bibinfo {year} {2019})\ \Eprint
  {http://arxiv.org/abs/1907.10625} {arXiv:1907.10625 [astro-ph.CO]}
  \BibitemShut {NoStop}%
%%CITATION = ARXIV:1907.10625;%%
\bibitem [{\citenamefont {Hill}\ \emph {et~al.}(2020)\citenamefont {Hill},
  \citenamefont {McDonough}, \citenamefont {Toomey},\ and\ \citenamefont
  {Alexander}}]{Hill:2020osr}%
  \BibitemOpen
  \bibfield  {author} {\bibinfo {author} {\bibfnamefont {J.~C.}\ \bibnamefont
  {Hill}}, \bibinfo {author} {\bibfnamefont {E.}~\bibnamefont {McDonough}},
  \bibinfo {author} {\bibfnamefont {M.~W.}\ \bibnamefont {Toomey}}, \ and\
  \bibinfo {author} {\bibfnamefont {S.}~\bibnamefont {Alexander}},\ }\href
  {\doibase 10.1103/PhysRevD.102.043507} {\bibfield  {journal} {\bibinfo
  {journal} {Phys. Rev. D}\ }\textbf {\bibinfo {volume} {102}},\ \bibinfo
  {pages} {043507} (\bibinfo {year} {2020})},\ \Eprint
  {http://arxiv.org/abs/2003.07355} {arXiv:2003.07355 [astro-ph.CO]}
  \BibitemShut {NoStop}%
\bibitem [{\citenamefont {Ivanov}\ \emph {et~al.}(2020)\citenamefont {Ivanov},
  \citenamefont {McDonough}, \citenamefont {Hill}, \citenamefont {Simonovi\'c},
  \citenamefont {Toomey}, \citenamefont {Alexander},\ and\ \citenamefont
  {Zaldarriaga}}]{Ivanov:2020ril}%
  \BibitemOpen
  \bibfield  {author} {\bibinfo {author} {\bibfnamefont {M.~M.}\ \bibnamefont
  {Ivanov}}, \bibinfo {author} {\bibfnamefont {E.}~\bibnamefont {McDonough}},
  \bibinfo {author} {\bibfnamefont {J.~C.}\ \bibnamefont {Hill}}, \bibinfo
  {author} {\bibfnamefont {M.}~\bibnamefont {Simonovi\'c}}, \bibinfo {author}
  {\bibfnamefont {M.~W.}\ \bibnamefont {Toomey}}, \bibinfo {author}
  {\bibfnamefont {S.}~\bibnamefont {Alexander}}, \ and\ \bibinfo {author}
  {\bibfnamefont {M.}~\bibnamefont {Zaldarriaga}},\ }\href {\doibase
  10.1103/PhysRevD.102.103502} {\bibfield  {journal} {\bibinfo  {journal}
  {Phys. Rev. D}\ }\textbf {\bibinfo {volume} {102}},\ \bibinfo {pages}
  {103502} (\bibinfo {year} {2020})},\ \Eprint
  {http://arxiv.org/abs/2006.11235} {arXiv:2006.11235 [astro-ph.CO]}
  \BibitemShut {NoStop}%
\bibitem [{\citenamefont {D'Amico}\ \emph {et~al.}(2021)\citenamefont
  {D'Amico}, \citenamefont {Senatore}, \citenamefont {Zhang},\ and\
  \citenamefont {Zheng}}]{DAmico:2020ods}%
  \BibitemOpen
  \bibfield  {author} {\bibinfo {author} {\bibfnamefont {G.}~\bibnamefont
  {D'Amico}}, \bibinfo {author} {\bibfnamefont {L.}~\bibnamefont {Senatore}},
  \bibinfo {author} {\bibfnamefont {P.}~\bibnamefont {Zhang}}, \ and\ \bibinfo
  {author} {\bibfnamefont {H.}~\bibnamefont {Zheng}},\ }\href {\doibase
  10.1088/1475-7516/2021/05/072} {\bibfield  {journal} {\bibinfo  {journal}
  {JCAP}\ }\textbf {\bibinfo {volume} {05}},\ \bibinfo {pages} {072} (\bibinfo
  {year} {2021})},\ \Eprint {http://arxiv.org/abs/2006.12420} {arXiv:2006.12420
  [astro-ph.CO]} \BibitemShut {NoStop}%
\bibitem [{\citenamefont {Ooguri}\ and\ \citenamefont
  {Vafa}(2007)}]{Ooguri:2006in}%
  \BibitemOpen
  \bibfield  {author} {\bibinfo {author} {\bibfnamefont {H.}~\bibnamefont
  {Ooguri}}\ and\ \bibinfo {author} {\bibfnamefont {C.}~\bibnamefont {Vafa}},\
  }\href {\doibase 10.1016/j.nuclphysb.2006.10.033} {\bibfield  {journal}
  {\bibinfo  {journal} {Nucl. Phys. B}\ }\textbf {\bibinfo {volume} {766}},\
  \bibinfo {pages} {21} (\bibinfo {year} {2007})},\ \Eprint
  {http://arxiv.org/abs/hep-th/0605264} {arXiv:hep-th/0605264} \BibitemShut
  {NoStop}%
\bibitem [{\citenamefont {Baume}\ and\ \citenamefont
  {Palti}(2016)}]{Baume:2016psm}%
  \BibitemOpen
  \bibfield  {author} {\bibinfo {author} {\bibfnamefont {F.}~\bibnamefont
  {Baume}}\ and\ \bibinfo {author} {\bibfnamefont {E.}~\bibnamefont {Palti}},\
  }\href {\doibase 10.1007/JHEP08(2016)043} {\bibfield  {journal} {\bibinfo
  {journal} {JHEP}\ }\textbf {\bibinfo {volume} {08}},\ \bibinfo {pages} {043}
  (\bibinfo {year} {2016})},\ \Eprint {http://arxiv.org/abs/1602.06517}
  {arXiv:1602.06517 [hep-th]} \BibitemShut {NoStop}%
\bibitem [{\citenamefont {Klaewer}\ and\ \citenamefont
  {Palti}(2017)}]{Klaewer:2016kiy}%
  \BibitemOpen
  \bibfield  {author} {\bibinfo {author} {\bibfnamefont {D.}~\bibnamefont
  {Klaewer}}\ and\ \bibinfo {author} {\bibfnamefont {E.}~\bibnamefont
  {Palti}},\ }\href {\doibase 10.1007/JHEP01(2017)088} {\bibfield  {journal}
  {\bibinfo  {journal} {JHEP}\ }\textbf {\bibinfo {volume} {01}},\ \bibinfo
  {pages} {088} (\bibinfo {year} {2017})},\ \Eprint
  {http://arxiv.org/abs/1610.00010} {arXiv:1610.00010 [hep-th]} \BibitemShut
  {NoStop}%
\bibitem [{\citenamefont {Blumenhagen}\ \emph {et~al.}(2017)\citenamefont
  {Blumenhagen}, \citenamefont {Valenzuela},\ and\ \citenamefont
  {Wolf}}]{Blumenhagen:2017cxt}%
  \BibitemOpen
  \bibfield  {author} {\bibinfo {author} {\bibfnamefont {R.}~\bibnamefont
  {Blumenhagen}}, \bibinfo {author} {\bibfnamefont {I.}~\bibnamefont
  {Valenzuela}}, \ and\ \bibinfo {author} {\bibfnamefont {F.}~\bibnamefont
  {Wolf}},\ }\href {\doibase 10.1007/JHEP07(2017)145} {\bibfield  {journal}
  {\bibinfo  {journal} {JHEP}\ }\textbf {\bibinfo {volume} {07}},\ \bibinfo
  {pages} {145} (\bibinfo {year} {2017})},\ \Eprint
  {http://arxiv.org/abs/1703.05776} {arXiv:1703.05776 [hep-th]} \BibitemShut
  {NoStop}%
\bibitem [{\citenamefont {Scalisi}\ and\ \citenamefont
  {Valenzuela}(2019)}]{Scalisi:2018eaz}%
  \BibitemOpen
  \bibfield  {author} {\bibinfo {author} {\bibfnamefont {M.}~\bibnamefont
  {Scalisi}}\ and\ \bibinfo {author} {\bibfnamefont {I.}~\bibnamefont
  {Valenzuela}},\ }\href {\doibase 10.1007/JHEP08(2019)160} {\bibfield
  {journal} {\bibinfo  {journal} {JHEP}\ }\textbf {\bibinfo {volume} {08}},\
  \bibinfo {pages} {160} (\bibinfo {year} {2019})},\ \Eprint
  {http://arxiv.org/abs/1812.07558} {arXiv:1812.07558 [hep-th]} \BibitemShut
  {NoStop}%
\bibitem [{\citenamefont {Riess}\ \emph {et~al.}(2019)\citenamefont {Riess},
  \citenamefont {Casertano}, \citenamefont {Yuan}, \citenamefont {Macri},\ and\
  \citenamefont {Scolnic}}]{Riess:2019cxk}%
  \BibitemOpen
  \bibfield  {author} {\bibinfo {author} {\bibfnamefont {A.~G.}\ \bibnamefont
  {Riess}}, \bibinfo {author} {\bibfnamefont {S.}~\bibnamefont {Casertano}},
  \bibinfo {author} {\bibfnamefont {W.}~\bibnamefont {Yuan}}, \bibinfo {author}
  {\bibfnamefont {L.~M.}\ \bibnamefont {Macri}}, \ and\ \bibinfo {author}
  {\bibfnamefont {D.}~\bibnamefont {Scolnic}},\ }\href {\doibase
  10.3847/1538-4357/ab1422} {\bibfield  {journal} {\bibinfo  {journal}
  {Astrophys. J.}\ }\textbf {\bibinfo {volume} {876}},\ \bibinfo {pages} {85}
  (\bibinfo {year} {2019})},\ \Eprint {http://arxiv.org/abs/1903.07603}
  {arXiv:1903.07603 [astro-ph.CO]} \BibitemShut {NoStop}%
%%CITATION = ARXIV:1903.07603;%%
\bibitem [{\citenamefont {Riess}\ \emph
  {et~al.}(2021{\natexlab{a}})\citenamefont {Riess}, \citenamefont {Casertano},
  \citenamefont {Yuan}, \citenamefont {Bowers}, \citenamefont {Macri},
  \citenamefont {Zinn},\ and\ \citenamefont {Scolnic}}]{Riess:2020fzl}%
  \BibitemOpen
  \bibfield  {author} {\bibinfo {author} {\bibfnamefont {A.~G.}\ \bibnamefont
  {Riess}}, \bibinfo {author} {\bibfnamefont {S.}~\bibnamefont {Casertano}},
  \bibinfo {author} {\bibfnamefont {W.}~\bibnamefont {Yuan}}, \bibinfo {author}
  {\bibfnamefont {J.~B.}\ \bibnamefont {Bowers}}, \bibinfo {author}
  {\bibfnamefont {L.}~\bibnamefont {Macri}}, \bibinfo {author} {\bibfnamefont
  {J.~C.}\ \bibnamefont {Zinn}}, \ and\ \bibinfo {author} {\bibfnamefont
  {D.}~\bibnamefont {Scolnic}},\ }\href {\doibase 10.3847/2041-8213/abdbaf}
  {\bibfield  {journal} {\bibinfo  {journal} {Astrophys. J. Lett.}\ }\textbf
  {\bibinfo {volume} {908}},\ \bibinfo {pages} {L6} (\bibinfo {year}
  {2021}{\natexlab{a}})},\ \Eprint {http://arxiv.org/abs/2012.08534}
  {arXiv:2012.08534 [astro-ph.CO]} \BibitemShut {NoStop}%
\bibitem [{\citenamefont {Aghanim}\ \emph {et~al.}(2018)\citenamefont {Aghanim}
  \emph {et~al.}}]{Aghanim:2018eyx}%
  \BibitemOpen
  \bibfield  {author} {\bibinfo {author} {\bibfnamefont {N.}~\bibnamefont
  {Aghanim}} \emph {et~al.} (\bibinfo {collaboration} {Planck}),\ }\href@noop
  {} {\  (\bibinfo {year} {2018})},\ \Eprint {http://arxiv.org/abs/1807.06209}
  {arXiv:1807.06209 [astro-ph.CO]} \BibitemShut {NoStop}%
%%CITATION = ARXIV:1807.06209;%%
\bibitem [{\citenamefont {{Cooke}}\ \emph {et~al.}(2016)\citenamefont
  {{Cooke}}, \citenamefont {{Pettini}}, \citenamefont {{Nollett}},\ and\
  \citenamefont {{Jorgenson}}}]{2016ApJ...830..148C}%
  \BibitemOpen
  \bibfield  {author} {\bibinfo {author} {\bibfnamefont {R.~J.}\ \bibnamefont
  {{Cooke}}}, \bibinfo {author} {\bibfnamefont {M.}~\bibnamefont {{Pettini}}},
  \bibinfo {author} {\bibfnamefont {K.~M.}\ \bibnamefont {{Nollett}}}, \ and\
  \bibinfo {author} {\bibfnamefont {R.}~\bibnamefont {{Jorgenson}}},\ }\href
  {\doibase 10.3847/0004-637X/830/2/148} {\bibfield  {journal} {\bibinfo
  {journal} {\apj}\ }\textbf {\bibinfo {volume} {830}},\ \bibinfo {eid} {148}
  (\bibinfo {year} {2016})},\ \Eprint {http://arxiv.org/abs/1607.03900}
  {arXiv:1607.03900 [astro-ph.CO]} \BibitemShut {NoStop}%
\bibitem [{\citenamefont {Aubourg}\ \emph {et~al.}(2015)\citenamefont {Aubourg}
  \emph {et~al.}}]{Aubourg:2014yra}%
  \BibitemOpen
  \bibfield  {author} {\bibinfo {author} {\bibfnamefont {E.}~\bibnamefont
  {Aubourg}} \emph {et~al.},\ }\href {\doibase 10.1103/PhysRevD.92.123516}
  {\bibfield  {journal} {\bibinfo  {journal} {Phys. Rev. D}\ }\textbf {\bibinfo
  {volume} {92}},\ \bibinfo {pages} {123516} (\bibinfo {year} {2015})},\
  \Eprint {http://arxiv.org/abs/1411.1074} {arXiv:1411.1074 [astro-ph.CO]}
  \BibitemShut {NoStop}%
\bibitem [{\citenamefont {Cuceu}\ \emph {et~al.}(2019)\citenamefont {Cuceu},
  \citenamefont {Farr}, \citenamefont {Lemos},\ and\ \citenamefont
  {Font-Ribera}}]{Cuceu:2019for}%
  \BibitemOpen
  \bibfield  {author} {\bibinfo {author} {\bibfnamefont {A.}~\bibnamefont
  {Cuceu}}, \bibinfo {author} {\bibfnamefont {J.}~\bibnamefont {Farr}},
  \bibinfo {author} {\bibfnamefont {P.}~\bibnamefont {Lemos}}, \ and\ \bibinfo
  {author} {\bibfnamefont {A.}~\bibnamefont {Font-Ribera}},\ }\href {\doibase
  10.1088/1475-7516/2019/10/044} {\bibfield  {journal} {\bibinfo  {journal}
  {JCAP}\ }\textbf {\bibinfo {volume} {10}},\ \bibinfo {pages} {044} (\bibinfo
  {year} {2019})},\ \Eprint {http://arxiv.org/abs/1906.11628} {arXiv:1906.11628
  [astro-ph.CO]} \BibitemShut {NoStop}%
\bibitem [{\citenamefont {Sch{\"o}neberg}\ \emph {et~al.}(2019)\citenamefont
  {Sch{\"o}neberg}, \citenamefont {Lesgourgues},\ and\ \citenamefont
  {Hooper}}]{Schoneberg:2019wmt}%
  \BibitemOpen
  \bibfield  {author} {\bibinfo {author} {\bibfnamefont {N.}~\bibnamefont
  {Sch{\"o}neberg}}, \bibinfo {author} {\bibfnamefont {J.}~\bibnamefont
  {Lesgourgues}}, \ and\ \bibinfo {author} {\bibfnamefont {D.~C.}\ \bibnamefont
  {Hooper}},\ }\href {\doibase 10.1088/1475-7516/2019/10/029} {\bibfield
  {journal} {\bibinfo  {journal} {JCAP}\ }\textbf {\bibinfo {volume} {10}},\
  \bibinfo {pages} {029} (\bibinfo {year} {2019})},\ \Eprint
  {http://arxiv.org/abs/1907.11594} {arXiv:1907.11594 [astro-ph.CO]}
  \BibitemShut {NoStop}%
\bibitem [{\citenamefont {Abbott}\ \emph {et~al.}(2018)\citenamefont {Abbott}
  \emph {et~al.}}]{Abbott:2017smn}%
  \BibitemOpen
  \bibfield  {author} {\bibinfo {author} {\bibfnamefont {T.~M.~C.}\
  \bibnamefont {Abbott}} \emph {et~al.} (\bibinfo {collaboration} {DES}),\
  }\href {\doibase 10.1093/mnras/sty1939} {\bibfield  {journal} {\bibinfo
  {journal} {Mon. Not. Roy. Astron. Soc.}\ }\textbf {\bibinfo {volume} {480}},\
  \bibinfo {pages} {3879} (\bibinfo {year} {2018})},\ \Eprint
  {http://arxiv.org/abs/1711.00403} {arXiv:1711.00403 [astro-ph.CO]}
  \BibitemShut {NoStop}%
%%CITATION = ARXIV:1711.00403;%%
\bibitem [{\citenamefont {Philcox}\ \emph {et~al.}(2020)\citenamefont
  {Philcox}, \citenamefont {Ivanov}, \citenamefont {Simonovic},\ and\
  \citenamefont {Zaldarriaga}}]{Philcox:2020vvt}%
  \BibitemOpen
  \bibfield  {author} {\bibinfo {author} {\bibfnamefont {O.~H.~E.}\
  \bibnamefont {Philcox}}, \bibinfo {author} {\bibfnamefont {M.~M.}\
  \bibnamefont {Ivanov}}, \bibinfo {author} {\bibfnamefont {M.}~\bibnamefont
  {Simonovic}}, \ and\ \bibinfo {author} {\bibfnamefont {M.}~\bibnamefont
  {Zaldarriaga}},\ }\href@noop {} {\  (\bibinfo {year} {2020})},\ \Eprint
  {http://arxiv.org/abs/2002.04035} {arXiv:2002.04035 [astro-ph.CO]}
  \BibitemShut {NoStop}%
%%CITATION = ARXIV:2002.04035;%%
\bibitem [{\citenamefont {Riess}\ \emph
  {et~al.}(2021{\natexlab{b}})\citenamefont {Riess} \emph
  {et~al.}}]{Riess:2021jrx}%
  \BibitemOpen
  \bibfield  {author} {\bibinfo {author} {\bibfnamefont {A.~G.}\ \bibnamefont
  {Riess}} \emph {et~al.},\ }\href@noop {} {\  (\bibinfo {year}
  {2021}{\natexlab{b}})},\ \Eprint {http://arxiv.org/abs/2112.04510}
  {arXiv:2112.04510 [astro-ph.CO]} \BibitemShut {NoStop}%
\bibitem [{\citenamefont {Freedman}\ \emph {et~al.}(2019)\citenamefont
  {Freedman} \emph {et~al.}}]{Freedman:2019jwv}%
  \BibitemOpen
  \bibfield  {author} {\bibinfo {author} {\bibfnamefont {W.~L.}\ \bibnamefont
  {Freedman}} \emph {et~al.},\ }\href {\doibase 10.3847/1538-4357/ab2f73} {\
  (\bibinfo {year} {2019}),\ 10.3847/1538-4357/ab2f73},\ \Eprint
  {http://arxiv.org/abs/1907.05922} {arXiv:1907.05922 [astro-ph.CO]}
  \BibitemShut {NoStop}%
\bibitem [{\citenamefont {Birrer}\ \emph {et~al.}(2020)\citenamefont {Birrer}
  \emph {et~al.}}]{Birrer:2020tax}%
  \BibitemOpen
  \bibfield  {author} {\bibinfo {author} {\bibfnamefont {S.}~\bibnamefont
  {Birrer}} \emph {et~al.},\ }\href {\doibase 10.1051/0004-6361/202038861}
  {\bibfield  {journal} {\bibinfo  {journal} {Astron. Astrophys.}\ }\textbf
  {\bibinfo {volume} {643}},\ \bibinfo {pages} {A165} (\bibinfo {year}
  {2020})},\ \Eprint {http://arxiv.org/abs/2007.02941} {arXiv:2007.02941
  [astro-ph.CO]} \BibitemShut {NoStop}%
\bibitem [{\citenamefont {Shah}\ \emph {et~al.}(2021)\citenamefont {Shah},
  \citenamefont {Lemos},\ and\ \citenamefont {Lahav}}]{Shah:2021onj}%
  \BibitemOpen
  \bibfield  {author} {\bibinfo {author} {\bibfnamefont {P.}~\bibnamefont
  {Shah}}, \bibinfo {author} {\bibfnamefont {P.}~\bibnamefont {Lemos}}, \ and\
  \bibinfo {author} {\bibfnamefont {O.}~\bibnamefont {Lahav}},\ }\href@noop {}
  {\  (\bibinfo {year} {2021})},\ \Eprint {http://arxiv.org/abs/2109.01161}
  {arXiv:2109.01161 [astro-ph.CO]} \BibitemShut {NoStop}%
\bibitem [{\citenamefont {Efstathiou}(2021)}]{Efstathiou:2021ocp}%
  \BibitemOpen
  \bibfield  {author} {\bibinfo {author} {\bibfnamefont {G.}~\bibnamefont
  {Efstathiou}},\ }\href@noop {} {\  (\bibinfo {year} {2021})},\ \Eprint
  {http://arxiv.org/abs/2103.08723} {arXiv:2103.08723 [astro-ph.CO]}
  \BibitemShut {NoStop}%
\bibitem [{\citenamefont {Jedamzik}\ \emph {et~al.}(2020)\citenamefont
  {Jedamzik}, \citenamefont {Pogosian},\ and\ \citenamefont
  {Zhao}}]{Jedamzik:2020zmd}%
  \BibitemOpen
  \bibfield  {author} {\bibinfo {author} {\bibfnamefont {K.}~\bibnamefont
  {Jedamzik}}, \bibinfo {author} {\bibfnamefont {L.}~\bibnamefont {Pogosian}},
  \ and\ \bibinfo {author} {\bibfnamefont {G.-B.}\ \bibnamefont {Zhao}},\
  }\href@noop {} {\  (\bibinfo {year} {2020})},\ \Eprint
  {http://arxiv.org/abs/2010.04158} {arXiv:2010.04158 [astro-ph.CO]}
  \BibitemShut {NoStop}%
\bibitem [{\citenamefont {Lin}\ \emph {et~al.}(2021)\citenamefont {Lin},
  \citenamefont {Chen},\ and\ \citenamefont {Mack}}]{Lin:2021sfs}%
  \BibitemOpen
  \bibfield  {author} {\bibinfo {author} {\bibfnamefont {W.}~\bibnamefont
  {Lin}}, \bibinfo {author} {\bibfnamefont {X.}~\bibnamefont {Chen}}, \ and\
  \bibinfo {author} {\bibfnamefont {K.~J.}\ \bibnamefont {Mack}},\ }\href
  {\doibase 10.3847/1538-4357/ac12cf} {\bibfield  {journal} {\bibinfo
  {journal} {Astrophys. J.}\ }\textbf {\bibinfo {volume} {920}},\ \bibinfo
  {pages} {159} (\bibinfo {year} {2021})},\ \Eprint
  {http://arxiv.org/abs/2102.05701} {arXiv:2102.05701 [astro-ph.CO]}
  \BibitemShut {NoStop}%
\bibitem [{\citenamefont {Clark}\ \emph {et~al.}(2021)\citenamefont {Clark},
  \citenamefont {Vattis}, \citenamefont {Fan},\ and\ \citenamefont
  {Koushiappas}}]{Clark:2021hlo}%
  \BibitemOpen
  \bibfield  {author} {\bibinfo {author} {\bibfnamefont {S.~J.}\ \bibnamefont
  {Clark}}, \bibinfo {author} {\bibfnamefont {K.}~\bibnamefont {Vattis}},
  \bibinfo {author} {\bibfnamefont {J.}~\bibnamefont {Fan}}, \ and\ \bibinfo
  {author} {\bibfnamefont {S.~M.}\ \bibnamefont {Koushiappas}},\ }\href@noop {}
  {\  (\bibinfo {year} {2021})},\ \Eprint {http://arxiv.org/abs/2110.09562}
  {arXiv:2110.09562 [astro-ph.CO]} \BibitemShut {NoStop}%
\bibitem [{\citenamefont {Allali}\ \emph {et~al.}(2021)\citenamefont {Allali},
  \citenamefont {Hertzberg},\ and\ \citenamefont {Rompineve}}]{Allali:2021azp}%
  \BibitemOpen
  \bibfield  {author} {\bibinfo {author} {\bibfnamefont {I.~J.}\ \bibnamefont
  {Allali}}, \bibinfo {author} {\bibfnamefont {M.~P.}\ \bibnamefont
  {Hertzberg}}, \ and\ \bibinfo {author} {\bibfnamefont {F.}~\bibnamefont
  {Rompineve}},\ }\href {\doibase 10.1103/PhysRevD.104.L081303} {\bibfield
  {journal} {\bibinfo  {journal} {Phys. Rev. D}\ }\textbf {\bibinfo {volume}
  {104}},\ \bibinfo {pages} {L081303} (\bibinfo {year} {2021})},\ \Eprint
  {http://arxiv.org/abs/2104.12798} {arXiv:2104.12798 [astro-ph.CO]}
  \BibitemShut {NoStop}%
\bibitem [{\citenamefont {Karwal}\ \emph {et~al.}(2021)\citenamefont {Karwal},
  \citenamefont {Raveri}, \citenamefont {Jain}, \citenamefont {Khoury},\ and\
  \citenamefont {Trodden}}]{Karwal:2021vpk}%
  \BibitemOpen
  \bibfield  {author} {\bibinfo {author} {\bibfnamefont {T.}~\bibnamefont
  {Karwal}}, \bibinfo {author} {\bibfnamefont {M.}~\bibnamefont {Raveri}},
  \bibinfo {author} {\bibfnamefont {B.}~\bibnamefont {Jain}}, \bibinfo {author}
  {\bibfnamefont {J.}~\bibnamefont {Khoury}}, \ and\ \bibinfo {author}
  {\bibfnamefont {M.}~\bibnamefont {Trodden}},\ }\href@noop {} {\  (\bibinfo
  {year} {2021})},\ \Eprint {http://arxiv.org/abs/2106.13290} {arXiv:2106.13290
  [astro-ph.CO]} \BibitemShut {NoStop}%
\bibitem [{\citenamefont {Hildebrandt}\ \emph {et~al.}(2017)\citenamefont
  {Hildebrandt} \emph {et~al.}}]{Hildebrandt:2016iqg}%
  \BibitemOpen
  \bibfield  {author} {\bibinfo {author} {\bibfnamefont {H.}~\bibnamefont
  {Hildebrandt}} \emph {et~al.},\ }\href {\doibase 10.1093/mnras/stw2805}
  {\bibfield  {journal} {\bibinfo  {journal} {Mon. Not. Roy. Astron. Soc.}\
  }\textbf {\bibinfo {volume} {465}},\ \bibinfo {pages} {1454} (\bibinfo {year}
  {2017})},\ \Eprint {http://arxiv.org/abs/1606.05338} {arXiv:1606.05338
  [astro-ph.CO]} \BibitemShut {NoStop}%
%%CITATION = ARXIV:1606.05338;%%
\bibitem [{\citenamefont {{Hildebrandt}}\ \emph {et~al.}(2020)\citenamefont
  {{Hildebrandt}} \emph {et~al.}}]{2020A&A...633A..69H}%
  \BibitemOpen
  \bibfield  {author} {\bibinfo {author} {\bibfnamefont {H.}~\bibnamefont
  {{Hildebrandt}}} \emph {et~al.},\ }\href {\doibase
  10.1051/0004-6361/201834878} {\bibfield  {journal} {\bibinfo  {journal}
  {\aap}\ }\textbf {\bibinfo {volume} {633}},\ \bibinfo {eid} {A69} (\bibinfo
  {year} {2020})},\ \Eprint {http://arxiv.org/abs/1812.06076} {arXiv:1812.06076
  [astro-ph.CO]} \BibitemShut {NoStop}%
\bibitem [{\citenamefont {Hikage}\ \emph {et~al.}(2019)\citenamefont {Hikage}
  \emph {et~al.}}]{Hikage:2018qbn}%
  \BibitemOpen
  \bibfield  {author} {\bibinfo {author} {\bibfnamefont {C.}~\bibnamefont
  {Hikage}} \emph {et~al.} (\bibinfo {collaboration} {HSC}),\ }\href {\doibase
  10.1093/pasj/psz010} {\bibfield  {journal} {\bibinfo  {journal} {Publ.
  Astron. Soc. Jap.}\ }\textbf {\bibinfo {volume} {71}},\ \bibinfo {pages}
  {Publications of the Astronomical Society of Japan, Volume 71, Issue 2, April
  2019, 43, https://doi.org/10.1093/pasj/psz010} (\bibinfo {year} {2019})},\
  \Eprint {http://arxiv.org/abs/1809.09148} {arXiv:1809.09148 [astro-ph.CO]}
  \BibitemShut {NoStop}%
%%CITATION = ARXIV:1809.09148;%%
\bibitem [{\citenamefont {Smith}\ \emph {et~al.}(2021)\citenamefont {Smith},
  \citenamefont {Poulin}, \citenamefont {Bernal}, \citenamefont {Boddy},
  \citenamefont {Kamionkowski},\ and\ \citenamefont {Murgia}}]{Smith:2020rxx}%
  \BibitemOpen
  \bibfield  {author} {\bibinfo {author} {\bibfnamefont {T.~L.}\ \bibnamefont
  {Smith}}, \bibinfo {author} {\bibfnamefont {V.}~\bibnamefont {Poulin}},
  \bibinfo {author} {\bibfnamefont {J.~L.}\ \bibnamefont {Bernal}}, \bibinfo
  {author} {\bibfnamefont {K.~K.}\ \bibnamefont {Boddy}}, \bibinfo {author}
  {\bibfnamefont {M.}~\bibnamefont {Kamionkowski}}, \ and\ \bibinfo {author}
  {\bibfnamefont {R.}~\bibnamefont {Murgia}},\ }\href {\doibase
  10.1103/PhysRevD.103.123542} {\bibfield  {journal} {\bibinfo  {journal}
  {Phys. Rev. D}\ }\textbf {\bibinfo {volume} {103}},\ \bibinfo {pages}
  {123542} (\bibinfo {year} {2021})},\ \Eprint
  {http://arxiv.org/abs/2009.10740} {arXiv:2009.10740 [astro-ph.CO]}
  \BibitemShut {NoStop}%
\bibitem [{\citenamefont {Kamionkowski}\ \emph {et~al.}(2014)\citenamefont
  {Kamionkowski}, \citenamefont {Pradler},\ and\ \citenamefont
  {Walker}}]{Kamionkowski:2014zda}%
  \BibitemOpen
  \bibfield  {author} {\bibinfo {author} {\bibfnamefont {M.}~\bibnamefont
  {Kamionkowski}}, \bibinfo {author} {\bibfnamefont {J.}~\bibnamefont
  {Pradler}}, \ and\ \bibinfo {author} {\bibfnamefont {D.~G.~E.}\ \bibnamefont
  {Walker}},\ }\href {\doibase 10.1103/PhysRevLett.113.251302} {\bibfield
  {journal} {\bibinfo  {journal} {Phys. Rev. Lett.}\ }\textbf {\bibinfo
  {volume} {113}},\ \bibinfo {pages} {251302} (\bibinfo {year} {2014})},\
  \Eprint {http://arxiv.org/abs/1409.0549} {arXiv:1409.0549 [hep-ph]}
  \BibitemShut {NoStop}%
\bibitem [{\citenamefont {Marsh}(2016)}]{Marsh:2015xka}%
  \BibitemOpen
  \bibfield  {author} {\bibinfo {author} {\bibfnamefont {D.~J.~E.}\
  \bibnamefont {Marsh}},\ }\href {\doibase 10.1016/j.physrep.2016.06.005}
  {\bibfield  {journal} {\bibinfo  {journal} {Phys. Rept.}\ }\textbf {\bibinfo
  {volume} {643}},\ \bibinfo {pages} {1} (\bibinfo {year} {2016})},\ \Eprint
  {http://arxiv.org/abs/1510.07633} {arXiv:1510.07633 [astro-ph.CO]}
  \BibitemShut {NoStop}%
\bibitem [{\citenamefont {Banks}\ \emph {et~al.}(2003)\citenamefont {Banks},
  \citenamefont {Dine}, \citenamefont {Fox},\ and\ \citenamefont
  {Gorbatov}}]{Banks:2003sx}%
  \BibitemOpen
  \bibfield  {author} {\bibinfo {author} {\bibfnamefont {T.}~\bibnamefont
  {Banks}}, \bibinfo {author} {\bibfnamefont {M.}~\bibnamefont {Dine}},
  \bibinfo {author} {\bibfnamefont {P.~J.}\ \bibnamefont {Fox}}, \ and\
  \bibinfo {author} {\bibfnamefont {E.}~\bibnamefont {Gorbatov}},\ }\href
  {\doibase 10.1088/1475-7516/2003/06/001} {\bibfield  {journal} {\bibinfo
  {journal} {JCAP}\ }\textbf {\bibinfo {volume} {0306}},\ \bibinfo {pages}
  {001} (\bibinfo {year} {2003})},\ \Eprint
  {http://arxiv.org/abs/hep-th/0303252} {arXiv:hep-th/0303252 [hep-th]}
  \BibitemShut {NoStop}%
%%CITATION = HEP-TH/0303252;%%
\bibitem [{\citenamefont {Rudelius}(2015)}]{Rudelius:2014wla}%
  \BibitemOpen
  \bibfield  {author} {\bibinfo {author} {\bibfnamefont {T.}~\bibnamefont
  {Rudelius}},\ }\href {\doibase 10.1088/1475-7516/2015/04/049} {\bibfield
  {journal} {\bibinfo  {journal} {JCAP}\ }\textbf {\bibinfo {volume} {1504}},\
  \bibinfo {pages} {049} (\bibinfo {year} {2015})},\ \Eprint
  {http://arxiv.org/abs/1409.5793} {arXiv:1409.5793 [hep-th]} \BibitemShut
  {NoStop}%
%%CITATION = ARXIV:1409.5793;%%
\bibitem [{\citenamefont {Stout}(2020)}]{Stout:2020uaf}%
  \BibitemOpen
  \bibfield  {author} {\bibinfo {author} {\bibfnamefont {J.}~\bibnamefont
  {Stout}},\ }\href@noop {} {\  (\bibinfo {year} {2020})},\ \Eprint
  {http://arxiv.org/abs/2012.11605} {arXiv:2012.11605 [hep-th]} \BibitemShut
  {NoStop}%
\bibitem [{\citenamefont {Bousso}\ and\ \citenamefont
  {Polchinski}(2000)}]{Bousso:2000xa}%
  \BibitemOpen
  \bibfield  {author} {\bibinfo {author} {\bibfnamefont {R.}~\bibnamefont
  {Bousso}}\ and\ \bibinfo {author} {\bibfnamefont {J.}~\bibnamefont
  {Polchinski}},\ }\href {\doibase 10.1088/1126-6708/2000/06/006} {\bibfield
  {journal} {\bibinfo  {journal} {JHEP}\ }\textbf {\bibinfo {volume} {06}},\
  \bibinfo {pages} {006} (\bibinfo {year} {2000})},\ \Eprint
  {http://arxiv.org/abs/hep-th/0004134} {arXiv:hep-th/0004134} \BibitemShut
  {NoStop}%
\bibitem [{\citenamefont {Susskind}(2003)}]{Susskind:2003kw}%
  \BibitemOpen
  \bibfield  {author} {\bibinfo {author} {\bibfnamefont {L.}~\bibnamefont
  {Susskind}},\ }\href@noop {} {\ ,\ \bibinfo {pages} {247} (\bibinfo {year}
  {2003})},\ \Eprint {http://arxiv.org/abs/hep-th/0302219}
  {arXiv:hep-th/0302219} \BibitemShut {NoStop}%
\bibitem [{\citenamefont {Vafa}(2005)}]{Vafa:2005ui}%
  \BibitemOpen
  \bibfield  {author} {\bibinfo {author} {\bibfnamefont {C.}~\bibnamefont
  {Vafa}},\ }\href@noop {} {\  (\bibinfo {year} {2005})},\ \Eprint
  {http://arxiv.org/abs/hep-th/0509212} {arXiv:hep-th/0509212} \BibitemShut
  {NoStop}%
\bibitem [{\citenamefont {Palti}(2019)}]{Palti:2019pca}%
  \BibitemOpen
  \bibfield  {author} {\bibinfo {author} {\bibfnamefont {E.}~\bibnamefont
  {Palti}},\ }\href {\doibase 10.1002/prop.201900037} {\bibfield  {journal}
  {\bibinfo  {journal} {Fortsch. Phys.}\ }\textbf {\bibinfo {volume} {67}},\
  \bibinfo {pages} {1900037} (\bibinfo {year} {2019})},\ \Eprint
  {http://arxiv.org/abs/1903.06239} {arXiv:1903.06239 [hep-th]} \BibitemShut
  {NoStop}%
\bibitem [{\citenamefont {Brennan}\ \emph {et~al.}(2017)\citenamefont
  {Brennan}, \citenamefont {Carta},\ and\ \citenamefont
  {Vafa}}]{Brennan:2017rbf}%
  \BibitemOpen
  \bibfield  {author} {\bibinfo {author} {\bibfnamefont {T.~D.}\ \bibnamefont
  {Brennan}}, \bibinfo {author} {\bibfnamefont {F.}~\bibnamefont {Carta}}, \
  and\ \bibinfo {author} {\bibfnamefont {C.}~\bibnamefont {Vafa}},\ }\href
  {\doibase 10.22323/1.305.0015} {\bibfield  {journal} {\bibinfo  {journal}
  {PoS}\ }\textbf {\bibinfo {volume} {TASI2017}},\ \bibinfo {pages} {015}
  (\bibinfo {year} {2017})},\ \Eprint {http://arxiv.org/abs/1711.00864}
  {arXiv:1711.00864 [hep-th]} \BibitemShut {NoStop}%
\bibitem [{\citenamefont {van Beest}\ \emph {et~al.}(2021)\citenamefont {van
  Beest}, \citenamefont {Calder\'on-Infante}, \citenamefont {Mirfendereski},\
  and\ \citenamefont {Valenzuela}}]{vanBeest:2021lhn}%
  \BibitemOpen
  \bibfield  {author} {\bibinfo {author} {\bibfnamefont {M.}~\bibnamefont {van
  Beest}}, \bibinfo {author} {\bibfnamefont {J.}~\bibnamefont
  {Calder\'on-Infante}}, \bibinfo {author} {\bibfnamefont {D.}~\bibnamefont
  {Mirfendereski}}, \ and\ \bibinfo {author} {\bibfnamefont {I.}~\bibnamefont
  {Valenzuela}},\ }\href@noop {} {\  (\bibinfo {year} {2021})},\ \Eprint
  {http://arxiv.org/abs/2102.01111} {arXiv:2102.01111 [hep-th]} \BibitemShut
  {NoStop}%
\bibitem [{\citenamefont {Agrawal}\ \emph {et~al.}(2021)\citenamefont
  {Agrawal}, \citenamefont {Obied},\ and\ \citenamefont
  {Vafa}}]{Agrawal:2019dlm}%
  \BibitemOpen
  \bibfield  {author} {\bibinfo {author} {\bibfnamefont {P.}~\bibnamefont
  {Agrawal}}, \bibinfo {author} {\bibfnamefont {G.}~\bibnamefont {Obied}}, \
  and\ \bibinfo {author} {\bibfnamefont {C.}~\bibnamefont {Vafa}},\ }\href
  {\doibase 10.1103/PhysRevD.103.043523} {\bibfield  {journal} {\bibinfo
  {journal} {Phys. Rev. D}\ }\textbf {\bibinfo {volume} {103}},\ \bibinfo
  {pages} {043523} (\bibinfo {year} {2021})},\ \Eprint
  {http://arxiv.org/abs/1906.08261} {arXiv:1906.08261 [astro-ph.CO]}
  \BibitemShut {NoStop}%
\bibitem [{\citenamefont {Anchordoqui}\ \emph {et~al.}(2020)\citenamefont
  {Anchordoqui}, \citenamefont {Antoniadis}, \citenamefont {L\"ust},
  \citenamefont {Soriano},\ and\ \citenamefont {Taylor}}]{Anchordoqui:2019amx}%
  \BibitemOpen
  \bibfield  {author} {\bibinfo {author} {\bibfnamefont {L.~A.}\ \bibnamefont
  {Anchordoqui}}, \bibinfo {author} {\bibfnamefont {I.}~\bibnamefont
  {Antoniadis}}, \bibinfo {author} {\bibfnamefont {D.}~\bibnamefont {L\"ust}},
  \bibinfo {author} {\bibfnamefont {J.~F.}\ \bibnamefont {Soriano}}, \ and\
  \bibinfo {author} {\bibfnamefont {T.~R.}\ \bibnamefont {Taylor}},\ }\href
  {\doibase 10.1103/PhysRevD.101.083532} {\bibfield  {journal} {\bibinfo
  {journal} {Phys. Rev. D}\ }\textbf {\bibinfo {volume} {101}},\ \bibinfo
  {pages} {083532} (\bibinfo {year} {2020})},\ \Eprint
  {http://arxiv.org/abs/1912.00242} {arXiv:1912.00242 [hep-th]} \BibitemShut
  {NoStop}%
\bibitem [{\citenamefont {Kim}\ \emph {et~al.}(2021)\citenamefont {Kim},
  \citenamefont {Kim}, \citenamefont {Semertzidis}, \citenamefont {Shin},\ and\
  \citenamefont {Yin}}]{Kim:2021eye}%
  \BibitemOpen
  \bibfield  {author} {\bibinfo {author} {\bibfnamefont {D.}~\bibnamefont
  {Kim}}, \bibinfo {author} {\bibfnamefont {Y.}~\bibnamefont {Kim}}, \bibinfo
  {author} {\bibfnamefont {Y.~K.}\ \bibnamefont {Semertzidis}}, \bibinfo
  {author} {\bibfnamefont {Y.~C.}\ \bibnamefont {Shin}}, \ and\ \bibinfo
  {author} {\bibfnamefont {W.}~\bibnamefont {Yin}},\ }\href {\doibase
  10.1103/PhysRevD.104.095010} {\bibfield  {journal} {\bibinfo  {journal}
  {Phys. Rev. D}\ }\textbf {\bibinfo {volume} {104}},\ \bibinfo {pages}
  {095010} (\bibinfo {year} {2021})},\ \Eprint
  {http://arxiv.org/abs/2105.03422} {arXiv:2105.03422 [hep-ph]} \BibitemShut
  {NoStop}%
\bibitem [{\citenamefont {{Planck
  Collaboration}}(2019)}]{Planck2018likelihood}%
  \BibitemOpen
  \bibfield  {author} {\bibinfo {author} {\bibnamefont {{Planck
  Collaboration}}},\ }\href@noop {} {\bibfield  {journal} {\bibinfo  {journal}
  {arXiv e-prints}\ ,\ \bibinfo {eid} {arXiv:1907.12875}} (\bibinfo {year}
  {2019})},\ \Eprint {http://arxiv.org/abs/1907.12875} {arXiv:1907.12875
  [astro-ph.CO]} \BibitemShut {NoStop}%
\bibitem [{\citenamefont {{Planck Collaboration}}(2018)}]{2018arXiv180706210P}%
  \BibitemOpen
  \bibfield  {author} {\bibinfo {author} {\bibnamefont {{Planck
  Collaboration}}},\ }\href@noop {} {\bibfield  {journal} {\bibinfo  {journal}
  {arXiv e-prints}\ ,\ \bibinfo {eid} {arXiv:1807.06210}} (\bibinfo {year}
  {2018})},\ \Eprint {http://arxiv.org/abs/1807.06210} {arXiv:1807.06210
  [astro-ph.CO]} \BibitemShut {NoStop}%
\bibitem [{\citenamefont {Ross}\ \emph {et~al.}(2015)\citenamefont {Ross},
  \citenamefont {Samushia}, \citenamefont {Howlett}, \citenamefont {Percival},
  \citenamefont {Burden},\ and\ \citenamefont {Manera}}]{Ross:2014qpa}%
  \BibitemOpen
  \bibfield  {author} {\bibinfo {author} {\bibfnamefont {A.~J.}\ \bibnamefont
  {Ross}}, \bibinfo {author} {\bibfnamefont {L.}~\bibnamefont {Samushia}},
  \bibinfo {author} {\bibfnamefont {C.}~\bibnamefont {Howlett}}, \bibinfo
  {author} {\bibfnamefont {W.~J.}\ \bibnamefont {Percival}}, \bibinfo {author}
  {\bibfnamefont {A.}~\bibnamefont {Burden}}, \ and\ \bibinfo {author}
  {\bibfnamefont {M.}~\bibnamefont {Manera}},\ }\href {\doibase
  10.1093/mnras/stv154} {\bibfield  {journal} {\bibinfo  {journal} {Mon. Not.
  Roy. Astron. Soc.}\ }\textbf {\bibinfo {volume} {449}},\ \bibinfo {pages}
  {835} (\bibinfo {year} {2015})},\ \Eprint {http://arxiv.org/abs/1409.3242}
  {arXiv:1409.3242 [astro-ph.CO]} \BibitemShut {NoStop}%
%%CITATION = ARXIV:1409.3242;%%
\bibitem [{\citenamefont {{Beutler}}\ \emph {et~al.}(2011)\citenamefont
  {{Beutler}}, \citenamefont {{Blake}}, \citenamefont {{Colless}},
  \citenamefont {{Jones}}, \citenamefont {{Staveley-Smith}}, \citenamefont
  {{Campbell}}, \citenamefont {{Parker}}, \citenamefont {{Saunders}},\ and\
  \citenamefont {{Watson}}}]{2011MNRAS.416.3017B}%
  \BibitemOpen
  \bibfield  {author} {\bibinfo {author} {\bibfnamefont {F.}~\bibnamefont
  {{Beutler}}}, \bibinfo {author} {\bibfnamefont {C.}~\bibnamefont {{Blake}}},
  \bibinfo {author} {\bibfnamefont {M.}~\bibnamefont {{Colless}}}, \bibinfo
  {author} {\bibfnamefont {D.~H.}\ \bibnamefont {{Jones}}}, \bibinfo {author}
  {\bibfnamefont {L.}~\bibnamefont {{Staveley-Smith}}}, \bibinfo {author}
  {\bibfnamefont {L.}~\bibnamefont {{Campbell}}}, \bibinfo {author}
  {\bibfnamefont {Q.}~\bibnamefont {{Parker}}}, \bibinfo {author}
  {\bibfnamefont {W.}~\bibnamefont {{Saunders}}}, \ and\ \bibinfo {author}
  {\bibfnamefont {F.}~\bibnamefont {{Watson}}},\ }\href {\doibase
  10.1111/j.1365-2966.2011.19250.x} {\bibfield  {journal} {\bibinfo  {journal}
  {\mnras}\ }\textbf {\bibinfo {volume} {416}},\ \bibinfo {pages} {3017}
  (\bibinfo {year} {2011})},\ \Eprint {http://arxiv.org/abs/1106.3366}
  {arXiv:1106.3366 [astro-ph.CO]} \BibitemShut {NoStop}%
\bibitem [{\citenamefont {Alam}\ \emph {et~al.}(2017)\citenamefont {Alam} \emph
  {et~al.}}]{Alam:2016hwk}%
  \BibitemOpen
  \bibfield  {author} {\bibinfo {author} {\bibfnamefont {S.}~\bibnamefont
  {Alam}} \emph {et~al.} (\bibinfo {collaboration} {BOSS}),\ }\href {\doibase
  10.1093/mnras/stx721} {\bibfield  {journal} {\bibinfo  {journal} {Mon. Not.
  Roy. Astron. Soc.}\ }\textbf {\bibinfo {volume} {470}},\ \bibinfo {pages}
  {2617} (\bibinfo {year} {2017})},\ \Eprint {http://arxiv.org/abs/1607.03155}
  {arXiv:1607.03155 [astro-ph.CO]} \BibitemShut {NoStop}%
%%CITATION = ARXIV:1607.03155;%%
\bibitem [{\citenamefont {Scolnic}\ \emph {et~al.}(2018)\citenamefont {Scolnic}
  \emph {et~al.}}]{Scolnic:2017caz}%
  \BibitemOpen
  \bibfield  {author} {\bibinfo {author} {\bibfnamefont {D.~M.}\ \bibnamefont
  {Scolnic}} \emph {et~al.},\ }\href {\doibase 10.3847/1538-4357/aab9bb}
  {\bibfield  {journal} {\bibinfo  {journal} {Astrophys. J.}\ }\textbf
  {\bibinfo {volume} {859}},\ \bibinfo {pages} {101} (\bibinfo {year}
  {2018})},\ \Eprint {http://arxiv.org/abs/1710.00845} {arXiv:1710.00845
  [astro-ph.CO]} \BibitemShut {NoStop}%
%%CITATION = ARXIV:1710.00845;%%
\bibitem [{\citenamefont {Abbott}\ \emph {et~al.}(2021)\citenamefont {Abbott}
  \emph {et~al.}}]{DES:2021wwk}%
  \BibitemOpen
  \bibfield  {author} {\bibinfo {author} {\bibfnamefont {T.~M.~C.}\
  \bibnamefont {Abbott}} \emph {et~al.} (\bibinfo {collaboration} {DES}),\
  }\href@noop {} {\  (\bibinfo {year} {2021})},\ \Eprint
  {http://arxiv.org/abs/2105.13549} {arXiv:2105.13549 [astro-ph.CO]}
  \BibitemShut {NoStop}%
\bibitem [{\citenamefont {Aiola}\ \emph {et~al.}(2020)\citenamefont {Aiola}
  \emph {et~al.}}]{ACT:2020gnv}%
  \BibitemOpen
  \bibfield  {author} {\bibinfo {author} {\bibfnamefont {S.}~\bibnamefont
  {Aiola}} \emph {et~al.} (\bibinfo {collaboration} {ACT}),\ }\href {\doibase
  10.1088/1475-7516/2020/12/047} {\bibfield  {journal} {\bibinfo  {journal}
  {JCAP}\ }\textbf {\bibinfo {volume} {12}},\ \bibinfo {pages} {047} (\bibinfo
  {year} {2020})},\ \Eprint {http://arxiv.org/abs/2007.07288} {arXiv:2007.07288
  [astro-ph.CO]} \BibitemShut {NoStop}%
\bibitem [{\citenamefont {Choi}\ \emph {et~al.}(2020)\citenamefont {Choi} \emph
  {et~al.}}]{ACT:2020frw}%
  \BibitemOpen
  \bibfield  {author} {\bibinfo {author} {\bibfnamefont {S.~K.}\ \bibnamefont
  {Choi}} \emph {et~al.} (\bibinfo {collaboration} {ACT}),\ }\href {\doibase
  10.1088/1475-7516/2020/12/045} {\bibfield  {journal} {\bibinfo  {journal}
  {JCAP}\ }\textbf {\bibinfo {volume} {12}},\ \bibinfo {pages} {045} (\bibinfo
  {year} {2020})},\ \Eprint {http://arxiv.org/abs/2007.07289} {arXiv:2007.07289
  [astro-ph.CO]} \BibitemShut {NoStop}%
\bibitem [{\citenamefont {Hill}\ \emph {et~al.}(2021)\citenamefont {Hill} \emph
  {et~al.}}]{Hill:2021yec}%
  \BibitemOpen
  \bibfield  {author} {\bibinfo {author} {\bibfnamefont {J.~C.}\ \bibnamefont
  {Hill}} \emph {et~al.},\ }\href@noop {} {\  (\bibinfo {year} {2021})},\
  \Eprint {http://arxiv.org/abs/2109.04451} {arXiv:2109.04451 [astro-ph.CO]}
  \BibitemShut {NoStop}%
\bibitem [{\citenamefont {Smith}\ \emph {et~al.}(2020)\citenamefont {Smith},
  \citenamefont {Poulin},\ and\ \citenamefont {Amin}}]{Smith:2019ihp}%
  \BibitemOpen
  \bibfield  {author} {\bibinfo {author} {\bibfnamefont {T.~L.}\ \bibnamefont
  {Smith}}, \bibinfo {author} {\bibfnamefont {V.}~\bibnamefont {Poulin}}, \
  and\ \bibinfo {author} {\bibfnamefont {M.~A.}\ \bibnamefont {Amin}},\ }\href
  {\doibase 10.1103/PhysRevD.101.063523} {\bibfield  {journal} {\bibinfo
  {journal} {Phys. Rev. D}\ }\textbf {\bibinfo {volume} {101}},\ \bibinfo
  {pages} {063523} (\bibinfo {year} {2020})},\ \Eprint
  {http://arxiv.org/abs/1908.06995} {arXiv:1908.06995 [astro-ph.CO]}
  \BibitemShut {NoStop}%
\bibitem [{\citenamefont {Agrawal}\ \emph {et~al.}(2019)\citenamefont
  {Agrawal}, \citenamefont {Cyr-Racine}, \citenamefont {Pinner},\ and\
  \citenamefont {Randall}}]{Agrawal:2019lmo}%
  \BibitemOpen
  \bibfield  {author} {\bibinfo {author} {\bibfnamefont {P.}~\bibnamefont
  {Agrawal}}, \bibinfo {author} {\bibfnamefont {F.-Y.}\ \bibnamefont
  {Cyr-Racine}}, \bibinfo {author} {\bibfnamefont {D.}~\bibnamefont {Pinner}},
  \ and\ \bibinfo {author} {\bibfnamefont {L.}~\bibnamefont {Randall}},\
  }\href@noop {} {\  (\bibinfo {year} {2019})},\ \Eprint
  {http://arxiv.org/abs/1904.01016} {arXiv:1904.01016 [astro-ph.CO]}
  \BibitemShut {NoStop}%
%%CITATION = ARXIV:1904.01016;%%
\bibitem [{\citenamefont {Alexander}\ and\ \citenamefont
  {McDonough}(2019)}]{Alexander:2019rsc}%
  \BibitemOpen
  \bibfield  {author} {\bibinfo {author} {\bibfnamefont {S.}~\bibnamefont
  {Alexander}}\ and\ \bibinfo {author} {\bibfnamefont {E.}~\bibnamefont
  {McDonough}},\ }\href {\doibase 10.1016/j.physletb.2019.134830} {\bibfield
  {journal} {\bibinfo  {journal} {Phys. Lett.}\ }\textbf {\bibinfo {volume}
  {B797}},\ \bibinfo {pages} {134830} (\bibinfo {year} {2019})},\ \Eprint
  {http://arxiv.org/abs/1904.08912} {arXiv:1904.08912 [astro-ph.CO]}
  \BibitemShut {NoStop}%
%%CITATION = ARXIV:1904.08912;%%
\bibitem [{\citenamefont {Lin}\ \emph {et~al.}(2019{\natexlab{a}})\citenamefont
  {Lin}, \citenamefont {Benevento}, \citenamefont {Hu},\ and\ \citenamefont
  {Raveri}}]{Lin:2019qug}%
  \BibitemOpen
  \bibfield  {author} {\bibinfo {author} {\bibfnamefont {M.-X.}\ \bibnamefont
  {Lin}}, \bibinfo {author} {\bibfnamefont {G.}~\bibnamefont {Benevento}},
  \bibinfo {author} {\bibfnamefont {W.}~\bibnamefont {Hu}}, \ and\ \bibinfo
  {author} {\bibfnamefont {M.}~\bibnamefont {Raveri}},\ }\href {\doibase
  10.1103/PhysRevD.100.063542} {\bibfield  {journal} {\bibinfo  {journal}
  {Phys. Rev. D}\ }\textbf {\bibinfo {volume} {100}},\ \bibinfo {pages}
  {063542} (\bibinfo {year} {2019}{\natexlab{a}})},\ \Eprint
  {http://arxiv.org/abs/1905.12618} {arXiv:1905.12618 [astro-ph.CO]}
  \BibitemShut {NoStop}%
\bibitem [{\citenamefont {Sakstein}\ and\ \citenamefont
  {Trodden}(2019)}]{Sakstein:2019fmf}%
  \BibitemOpen
  \bibfield  {author} {\bibinfo {author} {\bibfnamefont {J.}~\bibnamefont
  {Sakstein}}\ and\ \bibinfo {author} {\bibfnamefont {M.}~\bibnamefont
  {Trodden}},\ }\href@noop {} {\  (\bibinfo {year} {2019})},\ \Eprint
  {http://arxiv.org/abs/1911.11760} {arXiv:1911.11760 [astro-ph.CO]}
  \BibitemShut {NoStop}%
%%CITATION = ARXIV:1911.11760;%%
\bibitem [{\citenamefont {Niedermann}\ and\ \citenamefont
  {Sloth}(2021)}]{Niedermann:2019olb}%
  \BibitemOpen
  \bibfield  {author} {\bibinfo {author} {\bibfnamefont {F.}~\bibnamefont
  {Niedermann}}\ and\ \bibinfo {author} {\bibfnamefont {M.~S.}\ \bibnamefont
  {Sloth}},\ }\href {\doibase 10.1103/PhysRevD.103.L041303} {\bibfield
  {journal} {\bibinfo  {journal} {Phys. Rev. D}\ }\textbf {\bibinfo {volume}
  {103}},\ \bibinfo {pages} {L041303} (\bibinfo {year} {2021})},\ \Eprint
  {http://arxiv.org/abs/1910.10739} {arXiv:1910.10739 [astro-ph.CO]}
  \BibitemShut {NoStop}%
\bibitem [{\citenamefont {Niedermann}\ and\ \citenamefont
  {Sloth}(2020)}]{Niedermann:2020dwg}%
  \BibitemOpen
  \bibfield  {author} {\bibinfo {author} {\bibfnamefont {F.}~\bibnamefont
  {Niedermann}}\ and\ \bibinfo {author} {\bibfnamefont {M.~S.}\ \bibnamefont
  {Sloth}},\ }\href {\doibase 10.1103/PhysRevD.102.063527} {\bibfield
  {journal} {\bibinfo  {journal} {Phys. Rev. D}\ }\textbf {\bibinfo {volume}
  {102}},\ \bibinfo {pages} {063527} (\bibinfo {year} {2020})},\ \Eprint
  {http://arxiv.org/abs/2006.06686} {arXiv:2006.06686 [astro-ph.CO]}
  \BibitemShut {NoStop}%
\bibitem [{\citenamefont {Kaloper}(2019)}]{Kaloper:2019lpl}%
  \BibitemOpen
  \bibfield  {author} {\bibinfo {author} {\bibfnamefont {N.}~\bibnamefont
  {Kaloper}},\ }\href {\doibase 10.1142/S0218271819440176} {\bibfield
  {journal} {\bibinfo  {journal} {Int. J. Mod. Phys.}\ }\textbf {\bibinfo
  {volume} {D28}},\ \bibinfo {pages} {1944017} (\bibinfo {year} {2019})},\
  \Eprint {http://arxiv.org/abs/1903.11676} {arXiv:1903.11676 [hep-th]}
  \BibitemShut {NoStop}%
%%CITATION = ARXIV:1903.11676;%%
\bibitem [{\citenamefont {Berghaus}\ and\ \citenamefont
  {Karwal}(2020)}]{Berghaus:2019cls}%
  \BibitemOpen
  \bibfield  {author} {\bibinfo {author} {\bibfnamefont {K.~V.}\ \bibnamefont
  {Berghaus}}\ and\ \bibinfo {author} {\bibfnamefont {T.}~\bibnamefont
  {Karwal}},\ }\href {\doibase 10.1103/PhysRevD.101.083537} {\bibfield
  {journal} {\bibinfo  {journal} {Phys. Rev. D}\ }\textbf {\bibinfo {volume}
  {101}},\ \bibinfo {pages} {083537} (\bibinfo {year} {2020})},\ \Eprint
  {http://arxiv.org/abs/1911.06281} {arXiv:1911.06281 [astro-ph.CO]}
  \BibitemShut {NoStop}%
\bibitem [{\citenamefont {Knox}\ and\ \citenamefont
  {Millea}(2020)}]{Knox:2019rjx}%
  \BibitemOpen
  \bibfield  {author} {\bibinfo {author} {\bibfnamefont {L.}~\bibnamefont
  {Knox}}\ and\ \bibinfo {author} {\bibfnamefont {M.}~\bibnamefont {Millea}},\
  }\href {\doibase 10.1103/PhysRevD.101.043533} {\bibfield  {journal} {\bibinfo
   {journal} {Phys. Rev. D}\ }\textbf {\bibinfo {volume} {101}},\ \bibinfo
  {pages} {043533} (\bibinfo {year} {2020})},\ \Eprint
  {http://arxiv.org/abs/1908.03663} {arXiv:1908.03663 [astro-ph.CO]}
  \BibitemShut {NoStop}%
\bibitem [{\citenamefont {Tsujikawa}(2013)}]{Tsujikawa:2013fta}%
  \BibitemOpen
  \bibfield  {author} {\bibinfo {author} {\bibfnamefont {S.}~\bibnamefont
  {Tsujikawa}},\ }\href {\doibase 10.1088/0264-9381/30/21/214003} {\bibfield
  {journal} {\bibinfo  {journal} {Class. Quant. Grav.}\ }\textbf {\bibinfo
  {volume} {30}},\ \bibinfo {pages} {214003} (\bibinfo {year} {2013})},\
  \Eprint {http://arxiv.org/abs/1304.1961} {arXiv:1304.1961 [gr-qc]}
  \BibitemShut {NoStop}%
\bibitem [{\citenamefont {Lin}\ \emph {et~al.}(2019{\natexlab{b}})\citenamefont
  {Lin}, \citenamefont {Raveri},\ and\ \citenamefont {Hu}}]{Lin:2018nxe}%
  \BibitemOpen
  \bibfield  {author} {\bibinfo {author} {\bibfnamefont {M.-X.}\ \bibnamefont
  {Lin}}, \bibinfo {author} {\bibfnamefont {M.}~\bibnamefont {Raveri}}, \ and\
  \bibinfo {author} {\bibfnamefont {W.}~\bibnamefont {Hu}},\ }\href {\doibase
  10.1103/PhysRevD.99.043514} {\bibfield  {journal} {\bibinfo  {journal} {Phys.
  Rev. D}\ }\textbf {\bibinfo {volume} {99}},\ \bibinfo {pages} {043514}
  (\bibinfo {year} {2019}{\natexlab{b}})},\ \Eprint
  {http://arxiv.org/abs/1810.02333} {arXiv:1810.02333 [astro-ph.CO]}
  \BibitemShut {NoStop}%
\bibitem [{\citenamefont {Poulin}\ \emph {et~al.}(2021)\citenamefont {Poulin},
  \citenamefont {Smith},\ and\ \citenamefont {Bartlett}}]{Poulin:2021bjr}%
  \BibitemOpen
  \bibfield  {author} {\bibinfo {author} {\bibfnamefont {V.}~\bibnamefont
  {Poulin}}, \bibinfo {author} {\bibfnamefont {T.~L.}\ \bibnamefont {Smith}}, \
  and\ \bibinfo {author} {\bibfnamefont {A.}~\bibnamefont {Bartlett}},\
  }\href@noop {} {\  (\bibinfo {year} {2021})},\ \Eprint
  {http://arxiv.org/abs/2109.06229} {arXiv:2109.06229 [astro-ph.CO]}
  \BibitemShut {NoStop}%
\bibitem [{\citenamefont {Lin}\ \emph {et~al.}(2020)\citenamefont {Lin},
  \citenamefont {Hu},\ and\ \citenamefont {Raveri}}]{Lin:2020jcb}%
  \BibitemOpen
  \bibfield  {author} {\bibinfo {author} {\bibfnamefont {M.-X.}\ \bibnamefont
  {Lin}}, \bibinfo {author} {\bibfnamefont {W.}~\bibnamefont {Hu}}, \ and\
  \bibinfo {author} {\bibfnamefont {M.}~\bibnamefont {Raveri}},\ }\href
  {\doibase 10.1103/PhysRevD.102.123523} {\bibfield  {journal} {\bibinfo
  {journal} {Phys. Rev. D}\ }\textbf {\bibinfo {volume} {102}},\ \bibinfo
  {pages} {123523} (\bibinfo {year} {2020})},\ \Eprint
  {http://arxiv.org/abs/2009.08974} {arXiv:2009.08974 [astro-ph.CO]}
  \BibitemShut {NoStop}%
\bibitem [{\citenamefont {McCarthy}\ \emph {et~al.}(2021)\citenamefont
  {McCarthy}, \citenamefont {Hill},\ and\ \citenamefont
  {Madhavacheril}}]{McCarthy:2021lfp}%
  \BibitemOpen
  \bibfield  {author} {\bibinfo {author} {\bibfnamefont {F.}~\bibnamefont
  {McCarthy}}, \bibinfo {author} {\bibfnamefont {J.~C.}\ \bibnamefont {Hill}},
  \ and\ \bibinfo {author} {\bibfnamefont {M.~S.}\ \bibnamefont
  {Madhavacheril}},\ }\href@noop {} {\  (\bibinfo {year} {2021})},\ \Eprint
  {http://arxiv.org/abs/2103.05582} {arXiv:2103.05582 [astro-ph.CO]}
  \BibitemShut {NoStop}%
\bibitem [{\citenamefont {{Lesgourgues}}(2011)}]{2011arXiv1104.2932L}%
  \BibitemOpen
  \bibfield  {author} {\bibinfo {author} {\bibfnamefont {J.}~\bibnamefont
  {{Lesgourgues}}},\ }\href@noop {} {\bibfield  {journal} {\bibinfo  {journal}
  {arXiv e-prints}\ ,\ \bibinfo {eid} {arXiv:1104.2932}} (\bibinfo {year}
  {2011})},\ \Eprint {http://arxiv.org/abs/1104.2932} {arXiv:1104.2932
  [astro-ph.IM]} \BibitemShut {NoStop}%
\bibitem [{\citenamefont {{Blas}}\ \emph {et~al.}(2011)\citenamefont {{Blas}},
  \citenamefont {{Lesgourgues}},\ and\ \citenamefont
  {{Tram}}}]{2011JCAP...07..034B}%
  \BibitemOpen
  \bibfield  {author} {\bibinfo {author} {\bibfnamefont {D.}~\bibnamefont
  {{Blas}}}, \bibinfo {author} {\bibfnamefont {J.}~\bibnamefont
  {{Lesgourgues}}}, \ and\ \bibinfo {author} {\bibfnamefont {T.}~\bibnamefont
  {{Tram}}},\ }\href {\doibase 10.1088/1475-7516/2011/07/034} {\bibfield
  {journal} {\bibinfo  {journal} {\jcap}\ }\textbf {\bibinfo {volume} {2011}},\
  \bibinfo {eid} {034} (\bibinfo {year} {2011})},\ \Eprint
  {http://arxiv.org/abs/1104.2933} {arXiv:1104.2933 [astro-ph.CO]} \BibitemShut
  {NoStop}%
\bibitem [{\citenamefont {Torrado}\ and\ \citenamefont
  {Lewis}(2019)}]{torrado_lewis_2019}%
  \BibitemOpen
  \bibfield  {author} {\bibinfo {author} {\bibfnamefont {J.}~\bibnamefont
  {Torrado}}\ and\ \bibinfo {author} {\bibfnamefont {A.}~\bibnamefont
  {Lewis}},\ }\href {https://github.com/CobayaSampler/cobaya} {\enquote
  {\bibinfo {title} {Cobaya},}\ } (\bibinfo {year} {2019})\BibitemShut
  {NoStop}%
\bibitem [{\citenamefont {{Lewis}}(2019)}]{GetDist}%
  \BibitemOpen
  \bibfield  {author} {\bibinfo {author} {\bibfnamefont {A.}~\bibnamefont
  {{Lewis}}},\ }\href@noop {} {\bibfield  {journal} {\bibinfo  {journal} {arXiv
  e-prints}\ ,\ \bibinfo {eid} {arXiv:1910.13970}} (\bibinfo {year} {2019})},\
  \Eprint {http://arxiv.org/abs/1910.13970} {arXiv:1910.13970 [astro-ph.IM]}
  \BibitemShut {NoStop}%
\bibitem [{\citenamefont {Gelman}\ and\ \citenamefont
  {Rubin}(1992)}]{Gelman:1992zz}%
  \BibitemOpen
  \bibfield  {author} {\bibinfo {author} {\bibfnamefont {A.}~\bibnamefont
  {Gelman}}\ and\ \bibinfo {author} {\bibfnamefont {D.~B.}\ \bibnamefont
  {Rubin}},\ }\href {\doibase 10.1214/ss/1177011136} {\bibfield  {journal}
  {\bibinfo  {journal} {Statist. Sci.}\ }\textbf {\bibinfo {volume} {7}},\
  \bibinfo {pages} {457} (\bibinfo {year} {1992})}\BibitemShut {NoStop}%
%%CITATION = STSCE,7,457;%%
\bibitem [{\citenamefont {Powell}(2009)}]{Powell2009}%
  \BibitemOpen
  \bibfield  {author} {\bibinfo {author} {\bibfnamefont {M.}~\bibnamefont
  {Powell}},\ }\href@noop {} {\bibfield  {journal} {\bibinfo  {journal}
  {Technical Report, Department of Applied Mathematics and Theoretical
  Physics}\ } (\bibinfo {year} {2009})}\BibitemShut {NoStop}%
\bibitem [{\citenamefont {{Cartis}}\ \emph
  {et~al.}(2018{\natexlab{a}})\citenamefont {{Cartis}}, \citenamefont
  {{Fiala}}, \citenamefont {{Marteau}},\ and\ \citenamefont
  {{Roberts}}}]{Cartis2018a}%
  \BibitemOpen
  \bibfield  {author} {\bibinfo {author} {\bibfnamefont {C.}~\bibnamefont
  {{Cartis}}}, \bibinfo {author} {\bibfnamefont {J.}~\bibnamefont {{Fiala}}},
  \bibinfo {author} {\bibfnamefont {B.}~\bibnamefont {{Marteau}}}, \ and\
  \bibinfo {author} {\bibfnamefont {L.}~\bibnamefont {{Roberts}}},\ }\href@noop
  {} {\bibfield  {journal} {\bibinfo  {journal} {arXiv e-prints}\ ,\ \bibinfo
  {eid} {arXiv:1804.00154}} (\bibinfo {year} {2018}{\natexlab{a}})},\ \Eprint
  {http://arxiv.org/abs/1804.00154} {arXiv:1804.00154 [math.OC]} \BibitemShut
  {NoStop}%
\bibitem [{\citenamefont {{Cartis}}\ \emph
  {et~al.}(2018{\natexlab{b}})\citenamefont {{Cartis}}, \citenamefont
  {{Roberts}},\ and\ \citenamefont {{Sheridan-Methven}}}]{Cartis2018b}%
  \BibitemOpen
  \bibfield  {author} {\bibinfo {author} {\bibfnamefont {C.}~\bibnamefont
  {{Cartis}}}, \bibinfo {author} {\bibfnamefont {L.}~\bibnamefont {{Roberts}}},
  \ and\ \bibinfo {author} {\bibfnamefont {O.}~\bibnamefont
  {{Sheridan-Methven}}},\ }\href@noop {} {\bibfield  {journal} {\bibinfo
  {journal} {arXiv e-prints}\ ,\ \bibinfo {eid} {arXiv:1812.11343}} (\bibinfo
  {year} {2018}{\natexlab{b}})},\ \Eprint {http://arxiv.org/abs/1812.11343}
  {arXiv:1812.11343 [math.OC]} \BibitemShut {NoStop}%
\bibitem [{\citenamefont {Spurio~Mancini}\ \emph {et~al.}(2021)\citenamefont
  {Spurio~Mancini}, \citenamefont {Piras}, \citenamefont {Alsing},
  \citenamefont {Joachimi},\ and\ \citenamefont
  {Hobson}}]{SpurioMancini:2021ppk}%
  \BibitemOpen
  \bibfield  {author} {\bibinfo {author} {\bibfnamefont {A.}~\bibnamefont
  {Spurio~Mancini}}, \bibinfo {author} {\bibfnamefont {D.}~\bibnamefont
  {Piras}}, \bibinfo {author} {\bibfnamefont {J.}~\bibnamefont {Alsing}},
  \bibinfo {author} {\bibfnamefont {B.}~\bibnamefont {Joachimi}}, \ and\
  \bibinfo {author} {\bibfnamefont {M.~P.}\ \bibnamefont {Hobson}},\
  }\href@noop {} {\  (\bibinfo {year} {2021})},\ \Eprint
  {http://arxiv.org/abs/2106.03846} {arXiv:2106.03846 [astro-ph.CO]}
  \BibitemShut {NoStop}%
\bibitem [{\citenamefont {Moss}\ \emph {et~al.}(2021)\citenamefont {Moss},
  \citenamefont {Copeland}, \citenamefont {Bamford},\ and\ \citenamefont
  {Clarke}}]{Moss:2021obd}%
  \BibitemOpen
  \bibfield  {author} {\bibinfo {author} {\bibfnamefont {A.}~\bibnamefont
  {Moss}}, \bibinfo {author} {\bibfnamefont {E.}~\bibnamefont {Copeland}},
  \bibinfo {author} {\bibfnamefont {S.}~\bibnamefont {Bamford}}, \ and\
  \bibinfo {author} {\bibfnamefont {T.}~\bibnamefont {Clarke}},\ }\href@noop {}
  {\  (\bibinfo {year} {2021})},\ \Eprint {http://arxiv.org/abs/2109.14848}
  {arXiv:2109.14848 [astro-ph.CO]} \BibitemShut {NoStop}%
\bibitem [{\citenamefont {Obied}\ \emph {et~al.}(2018)\citenamefont {Obied},
  \citenamefont {Ooguri}, \citenamefont {Spodyneiko},\ and\ \citenamefont
  {Vafa}}]{Obied:2018sgi}%
  \BibitemOpen
  \bibfield  {author} {\bibinfo {author} {\bibfnamefont {G.}~\bibnamefont
  {Obied}}, \bibinfo {author} {\bibfnamefont {H.}~\bibnamefont {Ooguri}},
  \bibinfo {author} {\bibfnamefont {L.}~\bibnamefont {Spodyneiko}}, \ and\
  \bibinfo {author} {\bibfnamefont {C.}~\bibnamefont {Vafa}},\ }\href@noop {}
  {\  (\bibinfo {year} {2018})},\ \Eprint {http://arxiv.org/abs/1806.08362}
  {arXiv:1806.08362 [hep-th]} \BibitemShut {NoStop}%
\bibitem [{\citenamefont {Raveri}\ \emph {et~al.}(2019)\citenamefont {Raveri},
  \citenamefont {Hu},\ and\ \citenamefont {Sethi}}]{Raveri:2018ddi}%
  \BibitemOpen
  \bibfield  {author} {\bibinfo {author} {\bibfnamefont {M.}~\bibnamefont
  {Raveri}}, \bibinfo {author} {\bibfnamefont {W.}~\bibnamefont {Hu}}, \ and\
  \bibinfo {author} {\bibfnamefont {S.}~\bibnamefont {Sethi}},\ }\href
  {\doibase 10.1103/PhysRevD.99.083518} {\bibfield  {journal} {\bibinfo
  {journal} {Phys. Rev. D}\ }\textbf {\bibinfo {volume} {99}},\ \bibinfo
  {pages} {083518} (\bibinfo {year} {2019})},\ \Eprint
  {http://arxiv.org/abs/1812.10448} {arXiv:1812.10448 [hep-th]} \BibitemShut
  {NoStop}%
\bibitem [{\citenamefont {Heisenberg}\ \emph {et~al.}(2018)\citenamefont
  {Heisenberg}, \citenamefont {Bartelmann}, \citenamefont {Brandenberger},\
  and\ \citenamefont {Refregier}}]{Heisenberg:2018yae}%
  \BibitemOpen
  \bibfield  {author} {\bibinfo {author} {\bibfnamefont {L.}~\bibnamefont
  {Heisenberg}}, \bibinfo {author} {\bibfnamefont {M.}~\bibnamefont
  {Bartelmann}}, \bibinfo {author} {\bibfnamefont {R.}~\bibnamefont
  {Brandenberger}}, \ and\ \bibinfo {author} {\bibfnamefont {A.}~\bibnamefont
  {Refregier}},\ }\href {\doibase 10.1103/PhysRevD.98.123502} {\bibfield
  {journal} {\bibinfo  {journal} {Phys. Rev. D}\ }\textbf {\bibinfo {volume}
  {98}},\ \bibinfo {pages} {123502} (\bibinfo {year} {2018})},\ \Eprint
  {http://arxiv.org/abs/1808.02877} {arXiv:1808.02877 [astro-ph.CO]}
  \BibitemShut {NoStop}%
\bibitem [{\citenamefont {Akrami}\ \emph {et~al.}(2019)\citenamefont {Akrami},
  \citenamefont {Kallosh}, \citenamefont {Linde},\ and\ \citenamefont
  {Vardanyan}}]{Akrami:2018ylq}%
  \BibitemOpen
  \bibfield  {author} {\bibinfo {author} {\bibfnamefont {Y.}~\bibnamefont
  {Akrami}}, \bibinfo {author} {\bibfnamefont {R.}~\bibnamefont {Kallosh}},
  \bibinfo {author} {\bibfnamefont {A.}~\bibnamefont {Linde}}, \ and\ \bibinfo
  {author} {\bibfnamefont {V.}~\bibnamefont {Vardanyan}},\ }\href {\doibase
  10.1002/prop.201800075} {\bibfield  {journal} {\bibinfo  {journal} {Fortsch.
  Phys.}\ }\textbf {\bibinfo {volume} {67}},\ \bibinfo {pages} {1800075}
  (\bibinfo {year} {2019})},\ \Eprint {http://arxiv.org/abs/1808.09440}
  {arXiv:1808.09440 [hep-th]} \BibitemShut {NoStop}%
\bibitem [{\citenamefont {Akaike}(1973)}]{akaike1973information}%
  \BibitemOpen
  \bibfield  {author} {\bibinfo {author} {\bibfnamefont {H.}~\bibnamefont
  {Akaike}},\ }\enquote {\bibinfo {title} {Information theory and an extension
  of the maximum likelihood principle},}\ in\ \href@noop {} {\emph {\bibinfo
  {booktitle} {Selected Papers of Hirotugu Akaike}}}\ (\bibinfo  {publisher}
  {Springer New York},\ \bibinfo {address} {New York, NY},\ \bibinfo {year}
  {1973})\ pp.\ \bibinfo {pages} {199--213}\BibitemShut {NoStop}%
\bibitem [{\citenamefont {Ma}\ and\ \citenamefont
  {Bertschinger}(1995)}]{Ma:1995ey}%
  \BibitemOpen
  \bibfield  {author} {\bibinfo {author} {\bibfnamefont {C.-P.}\ \bibnamefont
  {Ma}}\ and\ \bibinfo {author} {\bibfnamefont {E.}~\bibnamefont
  {Bertschinger}},\ }\href {\doibase 10.1086/176550} {\bibfield  {journal}
  {\bibinfo  {journal} {Astrophys. J.}\ }\textbf {\bibinfo {volume} {455}},\
  \bibinfo {pages} {7} (\bibinfo {year} {1995})},\ \Eprint
  {http://arxiv.org/abs/astro-ph/9506072} {arXiv:astro-ph/9506072} \BibitemShut
  {NoStop}%
\end{thebibliography}%

\end{document}